%% file: main.tex
\documentclass[acmsmall]{acmart}

\usepackage[english]{babel}
\usepackage{blindtext}
\usepackage{subcaption}
\usepackage{tabularx}
\usepackage{amsmath}
\usepackage{enumitem}
\usepackage{pbox}
\usepackage{multirow}
\usepackage{tabularx}
\usepackage{makecell}
\usepackage{mathtools}
\usepackage{caption}
\usepackage{float}
\usepackage{booktabs}
\usepackage{xurl}
\usepackage{hyperref}
\usepackage{url}

\newcommand{\ashwin}[1]{\textcolor{black}{#1}}
\usepackage[normalem]{ulem}

\usepackage{tikz}
\usepackage{amsmath}
\newcommand*\circledsmall[1]{\tikz[baseline=(char.base)]{
            \node[shape=circle,draw,inner sep=1pt] (char) {#1};}}
\newcommand{\parab}[1]{\vspace{0.05in}\noindent\textbf{#1}}
\renewcommand\footnotetextcopyrightpermission[1]{}
\begin{document}
\title[Nautilus]{Nautilus: A Framework for Cross-Layer Cartography of Submarine Cables and IP Links}

\author{Alagappan Ramanathan}
\orcid{0009-0003-3293-1790}
\affiliation{%
   \institution{University of California, Irvine}
   \country{USA}
}
\email{alagappr@uci.edu}

\author{Sangeetha Abdu Jyothi   }
\orcid{0009-0000-0503-4478}
\affiliation{%
   \institution{University of California, Irvine \& VMware Research}
   \country{USA}
}
\email{sangeetha.aj@uci.edu}

\input{sections/abstract}

\maketitle

\input{sections/intro_current}

\input{sections/related_current}

\input{sections/design_current}

\input{sections/datasets_current}

\input{sections/evaluation_current}

\input{sections/validation}

\input{sections/discussion}

\input{sections/conclusion}

\bibliographystyle{ACM-Reference-Format}
\bibliography{main}

\clearpage
\input{sections/appendix}

\end{document}

%% file: sections/abstract.tex
\begin{abstract}

Submarine cables constitute the backbone of the Internet. However, these critical infrastructure components are vulnerable to several natural and man-made threats, and during failures, are difficult to repair in remote oceans. In spite of their crucial role, we have a limited understanding of the impact of submarine cable failures on global connectivity, particularly on the higher layers of the Internet.

In this paper, we present Nautilus, a framework for cross-layer cartography of submarine cables and IP links. Using a corpus of public datasets and Internet cartographic techniques, Nautilus identifies IP links that are likely traversing submarine cables and maps them to one or more potential cables. Nautilus also gives each IP to cable assignment a prediction score that reflects the confidence in the mapping. Nautilus generates a mapping for 3.05 million and 1.43 million IPv4 and IPv6 links, respectively, spanning 91\% of all active cables. In the absence of ground truth data, we validate Nautilus mapping using three techniques: analyzing past cable failures, using targeted traceroute measurements, and comparing with public network maps of two operators.

\end{abstract}

%% file: sections/intro_current.tex
\section{Introduction}

Submarine cables form the backbone of the Internet infrastructure. Today, fewer than 500 undersea cables interconnect various landmasses, carrying over 99\% of intercontinental traffic~\cite{99_traffic}. Nearly \$10 trillion worth of financial transactions are transmitted over transoceanic cables every day~\cite{cable-10-tril}. By virtue of their role in international connectivity and the global economy, submarine cables are deemed critical components of the Internet infrastructure. There is a rallying cry for protecting submarine cables as a national security priority due to their strategic relevance~\cite{cable-10-tril}. However, these cables are vulnerable to several threats~\cite{earthquake_1,typhoon-1,solarstorm-cable,anchor-1,tonga_cable_issue}.

Submarine cables are often damaged by natural disasters such as earthquakes~\cite{earthquake_1,earthquake_2,earthquake_3,earthquake_3,earthquake-1,earthquake-2,earthquake-3}, typhoons~\cite{typhoon-1,typhoon-2}, and underwater volcanic eruptions~\cite{tonga_cable_issue}. In addition, they are susceptible to accidental physical damages from ship anchors~\cite{anchor-1,anchor-2} and various marine activities of humans. These cables could also be used as a target by malicious states and non-state terrorist organizations~\cite{cable-10-tril, cable-security}. On average, there are over 100 submarine cable failures each year~\cite{submarine_faq}. In the recent past, there have been multiple reports of cable failures and repairs around the globe~\cite{cable_failures_1, cable_failures_2, cable_failures_3, cable_failures_4, cable_failures_5, cable_failures_6, multiple_providers_aae1_outage}.   Unfortunately, repair of submarine cables, particularly when the damage is located in the deep sea, could take anywhere from a few weeks to months~\cite{tonga_cable_repair}. 

The significance of submarine cables in the global economy, coupled with the challenges in their restoration during failures, underscores the need for a better understanding of the role of submarine cables in end-to-end Internet connectivity. The impact of submarine cable failures is further exacerbated at higher layers of the Internet due to the sharing of each cable by multiple Autonomous Systems (ASes)~\cite{submarine}. When one cable fails, all ASes that have IP links on the cable are affected~\cite{multiple_providers_aae1_outage}. Today, while the physical map of submarine cables is publicly available, we have a limited understanding of the IP links and Internet paths traversing these cables. To bridge this gap, we need a cross-layer map of submarine cables and IP links.

Past work on Internet mapping was largely confined to specific layers of the Internet and did not support cross-layer mapping~\cite{internet_mapping_survey_1, internet_mapping_survey_2}; for example, mapping of physical cables~\cite{physical_1, physical_2, physical_3}, interfaces~\cite{subnet_1, subnet_2, subnet_3, subnet_4}, routers~\cite{router_paper_4, router_paper_8, router_paper_10, router_paper_11}, PoPs~\cite{pop_2, pop_3, pop_6, pop_7}, and ASes~\cite{AS_1, AS_3, AS_8}. Recently, iGDB~\cite{igdb} put forward the first cross-layer mapping framework for the Internet. But iGDB does not currently support mapping over transoceanic paths. We find that the mapping techniques used by iGDB, particularly Thiessen polygons around urban regions that aggregate traffic to the area, align well with the layout of long-haul cables on land but are not well-suited for submarine cable mapping. Most submarine cable landing points are located farther away from urban areas. Moreover, accounting for geolocation inaccuracies during mapping, which iGDB does not support, is critical for cross-layer mapping on submarine cables.

In this paper, we present Nautilus, a framework for cross-layer cartography of submarine cables and IP links, using a wide range of publicly available datasets and tools as well as custom mapping techniques. Specifically, Nautilus uses traceroutes and the global submarine cable map from TeleGeography~\cite{submarine_cable_map} as key data sources. Nautilus also leverages eleven geolocation services, speed-of-light-based geolocation validation, and custom mapping techniques. Using these datasets and tools, Nautilus first identifies IP links that are traversing submarine cables and then maps them to one or more potential cables. Nautilus also assigns a prediction score to each IP link to submarine cable mapping, which denotes the degree of certainty in mapping.

Nautilus leverages long-running RIPE~\cite{ripe_atlas} and CAIDA~\cite{CAIDA} traceroute measurements to extract IP links (\S~\ref{ip_link_extraction}). Nautilus then classifies a link into definitely submarine, definitely terrestrial, or potentially/unconfirmed submarine categories based on the geolocation of IP link endpoints and various geolocation-based tests. Nautilus relies on eleven geolocation services~\cite{ripe_ipmap,caida_itdk,maxmind,ip2location,ipinfo,ipapi,db-ip,ipregistry,ipgeolocation,ipdata, ipapi_co} in this stage to improve the accuracy of geolocation.  %

After link classification, Nautilus employs a two-pronged approach for mapping. The two submodules in Nautilus relies on disjoint data sources and techniques to identify IP link to cable mapping independently, which are then aggregated to obtain the final mapping. The \textit{Geolocation module} of Nautilus uses IP geolocation and Speed of Light (SoL) based geolocation validation to identify a set of potential cables (\S~\ref{geolocation_module}). The \textit{Cable owner module} leverages IP to AS mapping of link endpoints and AS information of cable owners to generate another set of potential cable mappings (\S~\ref{organization_module}). Finally, the aggregation phase combines the mapping generated by the two modules to generate a set of candidate cables with a prediction score per cable (\S~\ref{final_mapping}). The prediction scores express the degree of certainty associated with mapping.

Nautilus leverages a dataset of over 100 million IPv4 traceroutes, from which it extracts 5.8 million links. Speed-of-Light (SoL) validation of geolocation data on both endpoints of these links leaves us with 5.11 million valid links. Within this set, based on the location of endpoints, Nautilus classifies 17.5\% of the links as definitely submarine, while 55\% fall into the potential submarine category. The remaining are categorized as definitely terrestrial. Out of the 3.73 million submarine IPv4 links (combining definitely and potentially submarine), Nautilus generates a mapping for 82\% of them. Similarly, from the IPv6 traceroutes, Nautilus extracts 3.08 million links, with 2.34 million links passing the SoL validation. Among these, 17\% of the links are classified as definitely submarine, and 59\% fall into the potential submarine category. Nautilus generates a mapping for 80\% of the 1.79 million submarine IPv6 links. For the mapped submarine links (IPv4 and IPv6), Nautilus predicts five or fewer submarine cables for 93\% of the cases, with 54\% of the links mapped to a single cable. The IP link to cable mapping generated by Nautilus identifies at least one mapping on 91\% of active cables and 90\% of submarine cable landing points.

As there is no directly related work for comparison, we evaluate Nautilus mapping against the cable predictions generated by Liu et al. (referred to as SCN-Crit)~\cite{submarine_drivability} and an extension of iGDB~\cite{igdb} to include submarine cables. Our analysis reveals that Nautilus generates a more accurate mapping, predicting 35\% fewer cables than SCN-Crit and 75\% fewer cables than iGDB per IP link, respectively. Note that an IP link should ideally be mapped to a single cable that it traverses. When a link is mapped to a set of potential cables, it indicates a higher degree of uncertainty in mapping. Hence, mapping to fewer cables is better.

Unfortunately, there are no public datasets available for ground truth validation of Nautilus mapping. We contacted over a dozen large ISPs around the globe and received a response from only two tier-1 providers in the US; they pointed out the highly sensitive nature of the data we requested and declined to share it. Hence, we perform validation of Nautilus mapping with three measurement techniques: (i) analysis of traceroutes collected before, during, and after cable failures to test the presence of links mapped to the failed cables, (ii) targeted traceroute measurements between probes located near submarine cable landing points to test the presence of IP links mapped to the cable, and (iii) comparisons with publicly available physical layer network maps of two providers. 

Our validation experiments show that (i) links mapped by Nautilus to a specific cable are missing during an outage of the cable while being present before and after the event, (ii) Nautilus' top cable prediction accurately matches the expected cable for 77\% of the links, and is a secondary (non-top) choice for 19\% of the links; only 4\% of the links do not have a match, and (iii) for two providers, there is a significant overlap between Nautilus' cable predictions for the operator and the operator's ground-truth physical map.

In summary, we make the following contribution:
\begin{itemize}[leftmargin=*,nolistsep]
    \item We present Nautilus, a framework for cross-layer mapping of submarine cables and IP links. Nautilus uses publicly available datasets and a range of well-known and custom Internet cartography techniques to generate an IP link to cable mapping with an assigned prediction score (\S~\ref{design}). \ashwin{Nautilus is open source}~\footnote{\ashwin{Code and data available at \href{https://gitlab.com/netsail-uci/nautilus}{https://gitlab.com/netsail-uci/nautilus}}}.
    \item Nautilus generates a mapping for 82\% and 80\% of all identified (definitely and potentially) submarine links for IPv4 and IPv6, respectively, spanning 91\% of all active cables (\S~\ref{cma}).
    \item We demonstrate that Nautilus is more accurate and predicts fewer submarine cables for an IP link when compared against related work---SCN-Crit and iGDB (\S~\ref{cpw}).
    \item We validate Nautilus' mapping by performing a wide range of validation experiments (\S~\ref{validation}).
\end{itemize}

%% file: sections/related_current.tex
\section{Related Work}

\parab{Cross-layer mapping on submarine cables:} 
Liu et al. (SCN-Crit~\cite{submarine_drivability}) estimated the fraction of paths traversing subsea cables using traceroute measurements to top 50 websites from 63 countries. Note that cable mapping was not the primary goal of this work. However, this work identified a set of potential submarine cables that an IP link could be traversing by (i) geolocating all routers on the path, (ii) identifying potential submarine hops using drivability criteria, and (iii) assigning all cables between the two countries of the submarine endpoints as the potential set. We observe that this subset is much larger and coarse-grained compared to the mapping generated by Nautilus. 

iGDB~\cite{igdb} is the first framework that supports cross-layer mapping of the Internet, but it is restricted to terrestrial cables. iGDB maps potential land cable endpoints by geolocating IP link endpoints, mapping these geolocations to urban areas using Thiessen polygons, and identifying potential cables between these urban areas. While this approach may work for long-haul land cables, due to limited geolocation and restrictive mapping techniques, iGDB does not work well when extended to submarine cables.

\parab{Single layer mapping:} Past work in Internet mapping mainly focused on individual layers and can be broadly classified as mapping at the (i) the physical layer~\cite{physical_1, physical_2, physical_3} where each link represents a physical cable between two locations, (ii) the interface level~\cite{subnet_1, subnet_2, subnet_3, subnet_4}, (iii) the router level~\cite{router_paper_1, router_paper_2, router_paper_4, router_paper_6, router_paper_8, router_paper_14, router_hot, router_recursive_2}, (iv) the PoP level~\cite{pop_1, pop_2, pop_3, pop_4, pop_5, pop_6, pop_7}, and (v) the AS level~\cite{AS_1, AS_2, AS_3, AS_4, AS_7, AS_8, AS_9}. The common methods employed by various mapping architectures can be broadly classified into aggregation-based and constraint-based approaches. Examples of aggregation-based solutions include physical layer techniques~\cite{physical_1, physical_2, physical_3} that collect data from various public and private sources and aggregate them to generate the mapping. An example of a constraint-based approach is the router-level mapping solution, HOT~\cite{router_hot}, which uses various network constraint parameters and optimizes an objective function to generate a mapping. These solutions designed for single-layer mapping cannot be employed directly for cross-layer mapping, which requires correlating maps at two different layers of the network.

%% file: sections/design_current.tex
\section{Design} \label{design}

In this section, we present the design of Nautilus, a framework for cross-layer IP link to submarine cable mapping (Figure~\ref{system_diagram}). The high-level workflow of Nautilus is below. A running example of Nautilus outputs on a single link at various stages of the workflow is in Figure~\ref{running_example}. \ashwin{A detailed workflow for each stage in Nautilus is presented in Appendix~\ref{nautilus_workflow_pieces}}.

\begin{itemize}[leftmargin=*,nolistsep]
    \item First, Nautilus extracts IP links from traceroutes (\circledsmall{1}).
    \item Next, the geolocation module (i) geolocates IP addresses of link endpoints and performs Speed-of-Light (SoL) validation (\circledsmall{2}), (ii) classifies the given IP link to definitely submarine, definitely terrestrial, or potentially submarine (\circledsmall{3}), and (iii) generates a cable mapping for the given IP link based on geolocation information (\circledsmall{4}).
    \item Third, the cable owner module (i) gets the AS number corresponding to IP addresses of link endpoints and retrieves all cables belonging to those ASes (\circledsmall{5}), and (ii) generates a cable mapping based on this cable ownership information (\circledsmall{6}).
    \item Finally, the aggregation phase combines the mappings generated by the geolocation and cable owner modules to generate a final cable mapping with a prediction score (\circledsmall{7}).
\end{itemize}

\subsection{IP \& Link Extraction} \label{ip_link_extraction}

\begin{figure*}
  \centering
  \includegraphics[width=\textwidth]{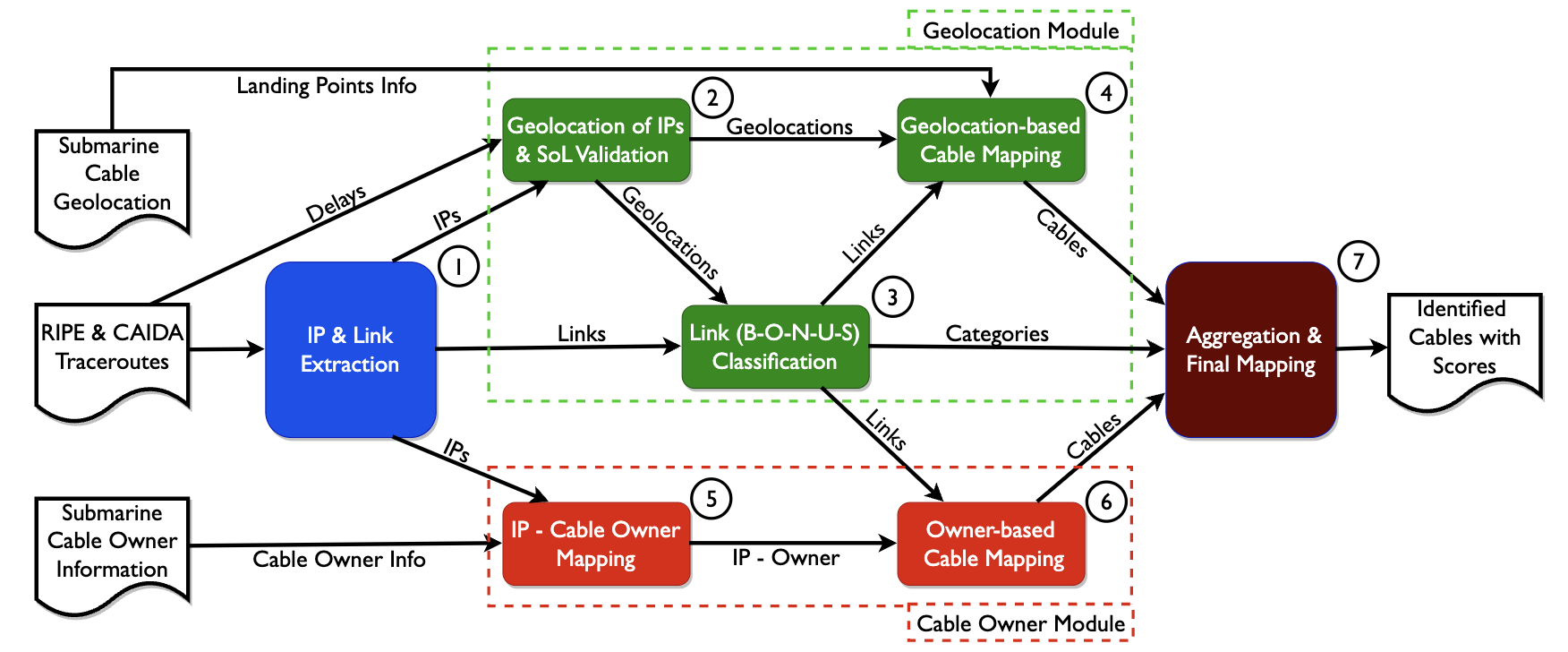}
  \caption{Nautilus System Architecture. Each stage in the pipeline is earmarked with a number that indicates the order of processing information. \ashwin{A detailed workflow for each stage is presented in Appendix~\ref{nautilus_workflow_pieces}}.}
  \label{system_diagram}
\end{figure*}

\begin{figure*}
  \centering
  \includegraphics[width=\textwidth]{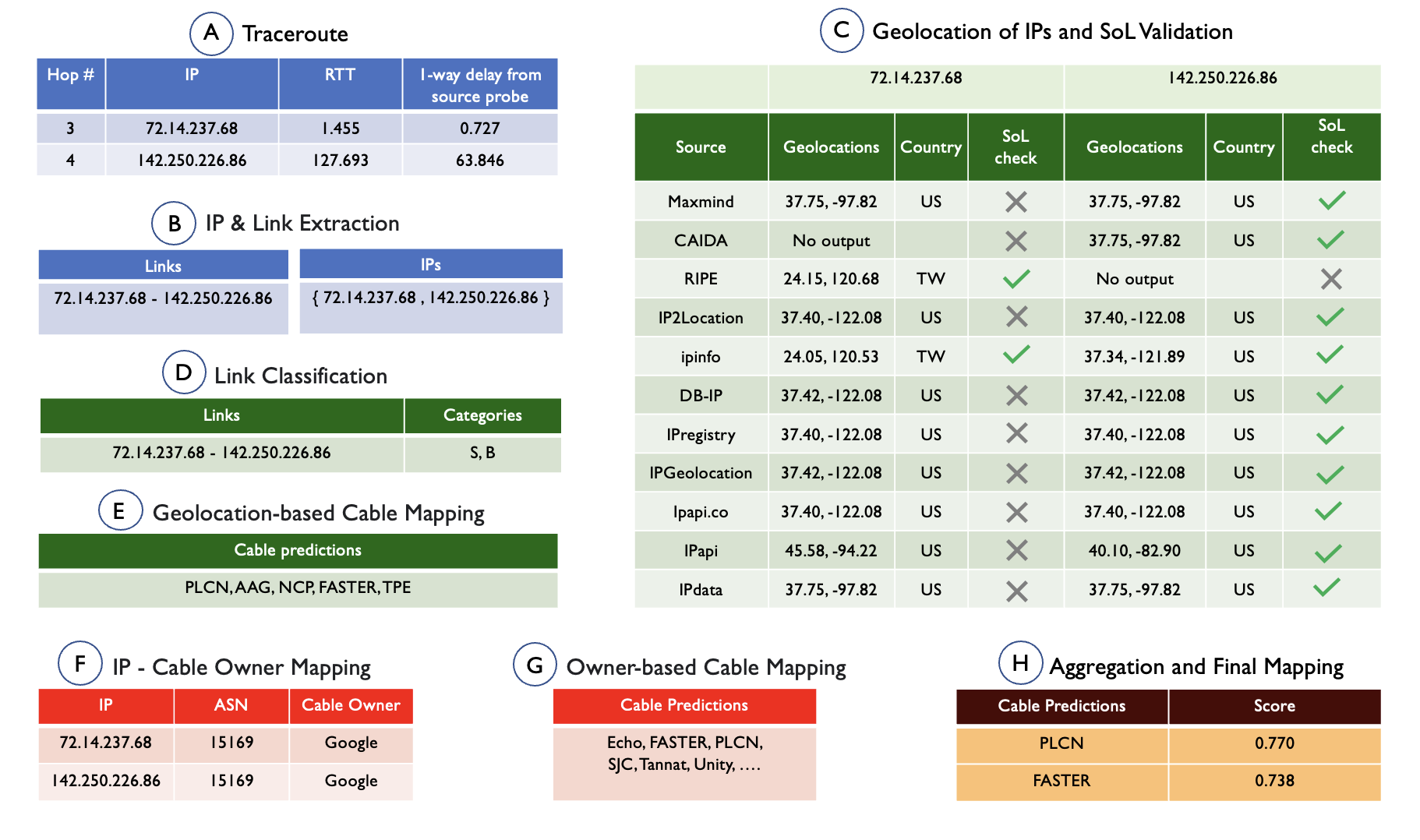}
  \caption{A running example to illustrate the outputs at each stage of Nautilus. These tables show the outputs for a single link extracted from a traceroute between RIPE probes in Taiwan and Los Angeles, US.}
  \label{running_example}
\end{figure*}

The first step in the Nautilus pipeline involves extracting IP links from traceroutes, as represented by \circledsmall{1} in Figure~\ref{system_diagram}. Nautilus uses long-running traceroutes from RIPE Atlas~\cite{ripe_atlas} and CAIDA~\cite{CAIDA} (\S~\ref{subsubsec:traceroutes}), to ensure extensive coverage of submarine links. Nautilus then eliminates traceroutes labeled as invalid or loop traceroutes. Since a single traceroute could potentially contain multiple submarine links, we process each IP link individually.
Nautilus extracts IP links from the traceroute that satisfy the following criteria: (i) both endpoints are non-private and valid, and (ii) the IP link endpoints are consecutive hops in a given traceroute (some routers disable ping and appear as * in traceroutes; such non-consecutive hops are filtered out). Based on the extracted IP links, we generate a list of unique IPs. Thus from a single traceroute, multiple valid IP links (and corresponding IPs) will be extracted and sent to further stages of the pipeline. Figure~\ref{running_example} A shows two hops in a traceroute, and Figure~\ref{running_example} B shows the IPs and the link extracted from it. 

\vspace{1mm}

\subsection{Geolocation Module} \label{geolocation_module}

Nautilus leverages geolocation as a crucial tool for establishing an IP link to submarine cable mapping. To generate this mapping, Nautilus first collects and aggregates the geolocation information from eleven services (\S~\ref{geolocation_sol_validation}), then classifies each IP link into one of the seven categories based on the aggregated geolocation information (\S~\ref{bonus_classification}), and finally correlates the IP geolocation with the submarine cable endpoint information from the Telegeography map~\cite{submarine_cable_map} to generate an IP link to submarine cable mapping (\S~\ref{geolocation_cable_mapping}).

\subsubsection{Collection \& Aggregation of IP Geolocations} \label{geolocation_sol_validation}

In this segment, we focus on IP geolocation (\circledsmall{2} in Fig.~\ref{system_diagram}). Sample output is shown in Figure~\ref{running_example} C, and we demonstrate the overall accuracy of the IP geolocation used by Nautilus in \S~\ref{multiple_geolocation_performance_gains}.

\parab{Multi-Source Approach for Geolocation of IPs.} Prior work on mapping relied on one or a few geolocation services. However, IP geolocation, especially for routers at the city level, is known to be inaccurate~\cite{geolocation_problem}. Additionally, the coverage of different geolocation services varies significantly, with no single geolocation source achieving high coverage across all IPs. To overcome these limitations, we use eleven geolocation services in Nautilus and employ aggregation techniques to combine the results. This multi-source approach aims to enhance the fidelity of geolocation. 

\parab{Speed-of-Light (SoL) Validation.} Nautilus employs a Speed-of-Light (SoL) based validation technique to eliminate inaccurate geolocations and improve fidelity. SoL dictates the minimum delay theoretically possible between two geolocations in fiber optic networks. \ashwin{The theoretical minimum delay is computed as the Haversine distance between the geolocation of two endpoints divided by the speed of light in fiber ($2*10^8 
 m/s$).} As locations of RIPE and CAIDA probes are well-known, Nautilus computes the upper bound of distance to all IPs in the traceroute from the source probes based on measured latencies. For a given IP, if a geolocation fails SoL check \ashwin{(i.e., the measured latencies are less than the lowest possible delay)} in at least 5\% of the traceroutes that the IP belongs to, we invalidate the corresponding geolocation. We incorporate a 5\% margin of error to account for issues with delay annotation in traceroutes and inaccuracies in probe locations. In Figure~\ref{running_example} C, for the IP 72.14.237.68, which is based in Taiwan, most geolocation services predominantly predict a location in the US, while only two services predict Taiwan accurately. During SoL validation, all geolocations in the US in Figure~\ref{running_example} C are discarded due to a distance estimate from the well-known location of the source probe that violates the speed-of-light constraint.

\parab{Multi-Source Geolocation Aggregation.} After SoL validation, the validated geolocations are likely to be in close proximity to each other. Hence, Nautilus aggregates the geolocations into clusters by employing an auto-clustering algorithm, DBSCAN~\cite{dbscan}. DBSCAN is a density-based clustering algorithm that merges points into a cluster if two conditions are met: (i) there are at least a minimum number of points (\textit{minPoints}) in the cluster, and (ii) all points within the cluster are within a threshold ($\epsilon$) computed using a specific distance metric. In Nautilus, we set the \textit{minPoints} parameter to 1 and define the $\epsilon$ threshold as 20 km computed based on the Haversine distance. Haversine distance is defined as the great circle distance between two points on a sphere given their latitudes and longitudes~\cite{haversine}. The purpose of this $\epsilon$ threshold is to help cluster proximal IP geolocations and ensure that the generated clusters are close to a smaller subset of landing points. We chose a 20 km threshold because fewer than 25\% of the landing points are within this distance of each other, striking a balance between clustering of geolocations and proximity to a small subset of landing points.

For each resulting cluster, Nautilus assigns a cluster score, which is computed as the ratio of the number of points in the cluster to the number of valid geolocation results. For example, in Figure~\ref{running_example} C, the IP 142.250.226.86 has 10 valid geolocations that pass the SoL check. Hence after applying DBSCAN, three clusters are generated with 6, 3, and 1 points in it, with a cluster score of 0.6, 0.3, and 0.1, respectively. While not all services perform equally, estimating the accuracy of a geolocation service is challenging as it depends on various factors like geography and AS. Thus, as a starting point, we cluster by assigning equal weights to all geolocation services. Organizing the geolocation results into clusters using DBSCAN also streamlines downstream analysis \ashwin{(details on how the clusters are leveraged, including an example, is presented in Appendix~\ref{need_for_all_clusters_retention}).}

Note that we are \textit{not} clustering the landing points. When two landing points are less than 20 km apart, there are two potential scenarios. If the geolocation is accurate across all sources, the IP cluster will have a much shorter distance to the correct landing point. If the geolocation accuracy is low, which is typical at the city scale, both landing points will be considered potential candidates.

\subsubsection{Link (B-O-N-U-S) Classification} \label{bonus_classification}

A crucial step in the Nautilus pipeline is the identification of IP links that potentially traverse submarine cables. Given considerable variations in submarine cable lengths, determining an optimal latency threshold to do so is challenging. Hence, Nautilus adopts a geolocation-based approach known as B-O-N-U-S classification (\circledsmall{3} in Figure~\ref{system_diagram}). Nautilus categorizes IP links into seven distinct categories based on two criteria: (i) confidence in the accuracy of geolocation and (ii) the potential for a link to traverse a submarine cable. Table~\ref{bonus_classification_table} provides an overview of the various categories resulting from the B-O-N-U-S classification. An example output is given in Figure~\ref{running_example} D. Note that all the links that have a valid geolocation and have passed the SoL validation will be mapped to one of the seven categories detailed in Table~\ref{bonus_classification_table}.

\begin{table}[h]
    \centering
    \small
    \begin{tabularx}{\linewidth}{
     >{\centering\arraybackslash}X |
    >{\centering\arraybackslash}X  
   >{\centering\arraybackslash}X  
   >{\centering\arraybackslash}X  }
        \textit{{\# endpoints with good geolocation}} & \emph{Definitely \textbf{S}ubmarine} & \emph{\textbf{U}nconfirmed Submarine} & \emph{Definitely \textbf{T}errestrial} \\ \hline
        \textit{\textbf{B}oth} & S, B & U, B & T \\ %
        \textit{\textbf{O}ne} & S, O & U, O & T \\ %
        \textit{\textbf{N}one} & S, N & U, N & T \\ %
    \end{tabularx}
    \caption{B-O-N-U-S Classification categories}
    \label{bonus_classification_table}
\end{table}

\parab{Confidence in Geolocation (B-O-N Classification).} Post DBSCAN clustering, Nautilus obtains valid geolocation clusters, each with an associated cluster score. Nautilus considers the geolocation of an endpoint to be reliable if a cluster has a score greater than 0.6. In the running example (Figure~\ref{running_example} C), the IP 142.250.226.86 is classified as having good geolocation because the cluster centered around (37.4, -122.08) has a cluster score of 0.6. We empirically converged on this threshold since (i) for each endpoint, at least half of the valid geolocation services have a consensus, indicating good geolocation accuracy, and (ii) this threshold offers the best-normalized geolocation accuracy when evaluated against an IP geolocation ground-truth dataset~\cite{geolocation_problem} (\S~\ref{multiple_geolocation_performance_gains}). Using this threshold, Nautilus classifies each IP link into three categories based on the number of endpoints with good geolocation: B (Both), O (One), or N (None), as outlined in Table~\ref{bonus_classification_table}. Finally, Nautilus computes the link \emph{geolocation cluster score} by adding the cluster scores of the two endpoints of the IP link. This measure provides an overall indication of the accuracy of the geolocation for the entire link (more details in \S~\ref{final_mapping}).

\begin{figure}
    \centering
    \includegraphics[width=0.8\columnwidth]{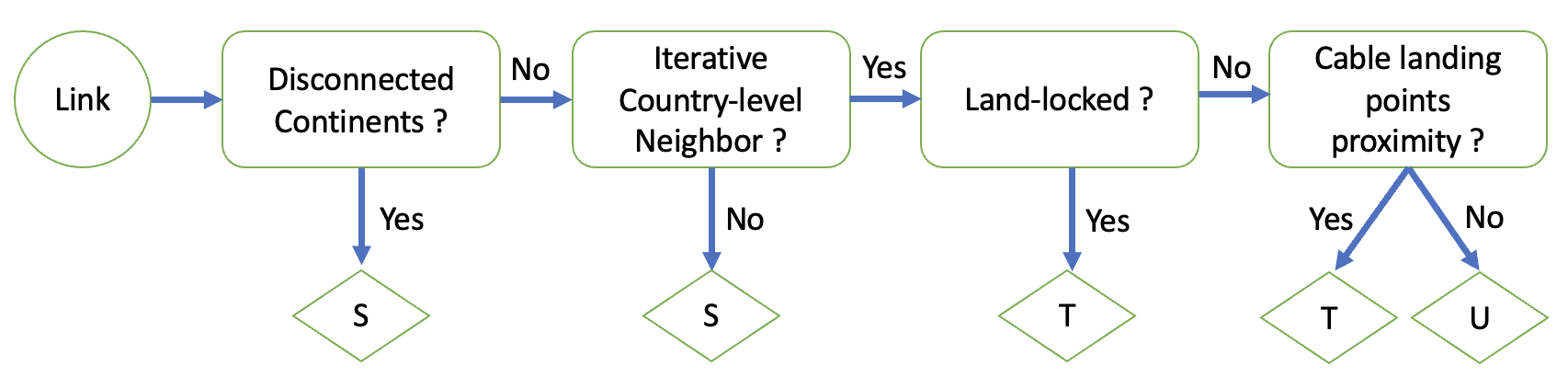}
    \caption{The workflow depicting the U-S link classification. The rectangular boxes represent the various tests on the geolocation of link endpoints to classify links into one of three categories: S (Definitely Submarine), U (Unconfirmed/Potential Submarine), and T (Definitely Terrestrial).}
    \label{fig:us_classification}
\end{figure}

\parab{Submarine Cable Potential (U-S Classification).} Nautilus employs a four-step process to determine the potential of an IP link to traverse a submarine cable as shown in Figure~\ref{fig:us_classification}. The classification of IP links is based on whether they are \emph{definitely submarine} (S), \emph{potential/unconfirmed submarine} (U), or \emph{definitely terrestrial} (T) categories. Despite SoL validation and clustering, each IP may still have clusters across different countries or continents. Hence to handle this, Nautilus examines all combinations of clusters for the endpoints of each IP link rather than selecting the cluster with the highest cluster score. 

The first step, the disconnected continents test, is based on the analysis of the continent information of the IP link endpoints. IP links connecting non-connected continents, such as Asia and North America, are tagged as \emph{definitely submarine}. \ashwin{All continent pairs except Europe-Asia and Asia-Africa are considered as disconnected.} The links not tagged in this phase are examined in the second step, the iterative country-level neighbor test. For this test, Nautilus employs an iterative neighbor search where Nautilus labels an IP link temporarily as \emph{potential submarine} if any combination of country pairs for the given endpoints either border each other or share a common land neighbor (e.g., Netherlands and Denmark, both of which share a border with Germany). If no such combination is found, the IP link is labeled as \emph{definitely submarine}. \ashwin{For instance, applying the iterative country-level neighbor test, Nautilus tags country pairs such as Denmark and Spain as definitely submarine. Nautilus employs a one-level search for land neighbors because a single terrestrial cable spanning multiple countries without an intermediate termination point or gateway is rare. Additionally, geopolitical situations in many regions, such as Africa and South Asia, preclude such deployments.}

In the final two steps, Nautilus captures the IP links that are \emph{definitely terrestrial}. For this, first Nautilus employs the land-locked test, which examines if all country combinations are land-locked (from the perspective of submarine cables). For links that fail this test, Nautilus employs a cable landing point proximity check, where links with a straight-line distance between the endpoints significantly shorter than the distance to the nearest submarine cable landing points (terrestrial termination points of submarine cables) are classified as \emph{definitely terrestrial} while tagging links that fail this check as the \emph{potentially submarine}. \ashwin{For example, a link connecting Austria and Hungary is categorized as definitely terrestrial during the land-locked test. In contrast, links within the same countries, such as those linking California and Hawaii or cities within Papua New Guinea, are classified as potentially submarine links during the cable landing point proximity test. This classification occurs after these links have failed the disconnected continent tests, the iterative country-level neighbor test, and the land-locked test.}

\subsubsection{Geolocation-based cable mapping} \label{geolocation_cable_mapping}

To identify the relevant landing points and cables for an IP link, Nautilus employs a recursive radius-based search (\circledsmall{4} in Figure~\ref{system_diagram}). To efficiently determine nearby landing points, Nautilus constructs a BallTree~\cite{balltree} using the coordinates of the submarine cable landing points. The BallTree partitions the coordinate space into intersecting hyper-spheres, enabling efficient identification of landing points within a given radius.

A key challenge is determining the optimal radius for querying. Nautilus initiates the search with a radius of 500 km and progressively expands it by 50 km until a cable match is found or the query radius reaches the threshold of 1000 km. \ashwin{Our first candidate, $k$-nearest neighbors, proved to be a poor choice due to the varying landing point density across regions and the ongoing expansion of new cables, making it difficult to estimate an appropriate value for $k$}. In Figure~\ref{running_example}, the IP 142.250.226.86, which connects to an IP in Taiwan, is geolocated at (37.4, -122.08), which has no nearby landing points. A large radius helps in such cases. This threshold value is determined based on the average country diameter of $\approx$ 920 km\ashwin{, with the intention of largely restricting the search to within a country or its neighbors}. For IP links with an endpoint located more than 1000 km away from any landing point, Nautilus does not generate any mapping, as the number of potential landing points and cables becomes excessively large and could lead to poor mapping accuracy. We empirically observe that the fraction of such links is small (11\% of all links). %

Furthermore, Nautilus calculates a \emph{distance-based score} by using the Haversine distance between the IP link endpoints and the identified submarine cable landing points, which helps prune irrelevant cables (more details in \S~\ref{final_mapping}). The cable set generated by the geolocation module for the running example is shown in Figure~\ref{running_example} E.

\subsection{Cable Owner Module} \label{organization_module}

Cable owner-based inference can help in regions with several parallel submarine cables (e.g., US-UK). Nautilus relies on the idea that links of an AS are more likely to be on a submarine cable owned by the same AS. The cable owner module in Nautilus employs a two-step process. The first step is the IP to Cable owner mapping wherein we generate an IP to ASN mapping from multiple sources and then employ Natural Language Processing (NLP) techniques to match a cable owner to a list of ASN, and finally merge these results to generate an IP to cable owner map (\S~\ref{ip_to_cable_owner}). The second step utilizes the IP to cable owner map to generate a submarine cable mapping (\S~\ref{cable_owner_based_mapping}).

\subsubsection{IP to Cable Owner Mapping} \label{ip_to_cable_owner}

This stage (\circledsmall{5} in Figure~\ref{system_diagram}) maps an IP address to a potential submarine cable owner using three steps. An example is given in Figure~\ref{running_example} F, where both the examined IP addresses in the link are mapped to Google as the link's submarine cable owner.

\parab{(i) IP to ASN mapping:} Nautilus uses data from four distinct IP to ASN mapping sources to identify the Autonomous System Number (ASN) of an IP address. The aggregation process is based on the principle of maximum agreement, wherein the mappings are combined through a voting mechanism. This approach improves coverage while instilling confidence in the accuracy of the final mapping results. For instance, one of the sources used by Nautilus, CAIDA~\cite{caida_as_to_org}, is based on bdrmapIT~\cite{router_paper_4}, which enhances the accuracy of effectively mapping routers located at the boundaries of ASes.

\parab{(ii) Cable Owner to ASN mapping:} To generate the mapping of submarine cable owners to ASNs, Nautilus combines information from the submarine cable map and the CAIDA ASRank database~\cite{ASRank}. Nautilus primarily relies on two attributes from the ASRank database: (i) organization and (ii) AS name. For example, AS10753 has the organization attribute as "Level 3 Parent, LLC" and the AS name as "LUMEN-LEGACY-L3-CUSTOMER-SHARED-USE". These attributes hence help us to correlate the relationship between Lumen and Level 3 (Lumen's acquisition). As there is no standardized convention with the organization names across the two sources (TeleGeography cable map and ASRank), particularly with acquisitions, mergers, abbreviation usage, and other variations in organization names, Nautilus employs a search along three dimensions on these two attributes to address the naming complexity and identify ASNs corresponding to cable owners. 

The first dimension is a partial or complete match with the AS organization name. The second dimension involves searching for an abbreviation that matches the AS organization name, such as "KT" for Korea Telecom. The third dimension is a match with the standard AS name. We observe that AS names often reflect alternative names of an AS after acquisition, e.g., Lumen and Level 3 (Lumen's acquisition). As the search across these dimensions yields multiple results, Nautilus employs a pruning mechanism to identify relevant matches. The organization with the highest AS rank is selected as the primary match since submarine cables are typically owned by large organizations. For the remaining matches, customers of the selected AS are prioritized. \ashwin{Although we perform manual validation for a limited number of cable owners, resulting in accurate AS matches by Nautilus, there is a potential for false positives. However, these false positives typically have minimal impact on the mapping process for two key reasons: First, the pruning mechanism selects the AS with the highest AS rank and its customers from among the matches, reducing the likelihood of matching false positives. Second, to reduce false positives involving high-ranked ASNs, we validate them by comparing their country of operation with the countries of the respective submarine cable owners. Some examples from the manual validation are presented in Appendix~\ref{cable_owner_to_as_examples}.}

\parab{(iii) IP to cable owner mapping:} 
In this step, Nautilus integrates the mapping of submarine cable owners to lists of ASNs generated in step (ii) with the IP to ASN map from step (i) to establish the mapping of IP addresses to submarine cable owners. By employing this structured approach and utilizing multiple dimensions of search, Nautilus accurately maps IPs to submarine cable owners.

\subsubsection{Cable owner-based mapping} \label{cable_owner_based_mapping}

The cable owner-based mapping plays a crucial role in the final step of Nautilus, as it helps to identify and prioritize the valid submarine cable from a set of parallel cables. Examples of parallel cables can be seen in Figure~\ref{fig:parallel_cables} of Appendix~\ref{parallel_cables_analysis}. Using the mapping of IP addresses to submarine cable owners, Nautilus generates a mapping to include all cables owned by the identified submarine cable owner for that specific IP link. Nautilus assigns an \emph{ownership score} to each of these cables. This step corresponds to \circledsmall{6} in Figure~\ref{system_diagram}.

The ownership score is determined based on the comparison between the ASN of an IP endpoint and the list of ASNs associated with cable owners. If the ASN of one IP endpoint matches with any ASN in the cable owner list, a score of 0.5 is added. This indicates that the IP is owned by one of the identified submarine cable owners. Thus this ownership score is always from the set \{0, 0.5, 1\} that corresponds to no match, one endpoint match, and both endpoints match, respectively.  

Note that in the final stage, cable-based mapping only accounts for 10\% of the final score, with geolocation-based mapping given a higher weightage (90\%\ashwin{---50\% and 40\% for cluster scores and distance scores respectively, both of which are computed based on geolocations}). We made this design choice since the ownership information on the submarine cable map is incomplete. Additionally, some large ASes are known to have leased cables instead of owning them. Hence we assign a higher weight to the geolocation-based mapping. The ownership score primarily helps in distinguishing between potential parallel cables.

\subsection{Aggregation \& Final mapping} \label{final_mapping}

To generate the final mapping of IP links to submarine cables, Nautilus combines the outputs of the geolocation module and the cable owner module (\circledsmall{7} in Figure~\ref{system_diagram}). In this module, we first combine the cluster score and distance-based score from the geolocation module and the ownership score from the cable owner module to derive the final prediction score for each cable mapped to an IP link. Then, we prune the unlikely predicted cables by using a threshold defined as the PaCT (Parallel Cable Threshold).

\parab{Prediction score: } The final prediction score, $S$, is calculated as a weighted sum of the cluster score ($C_i$) (fraction of elements in an IP's geolocation cluster), distance-based score ($d_i$) (normalized distance between IP geolocation at endpoint $i$ and the corresponding submarine landing point location), and ownership score ($O_i$) (1 if ASN of IP address at endpoint $i$ and ASN of any of cables owners match, 0 otherwise), for each endpoint $i$ of the IP link. The weighted sum is scaled by the category factor, $f$ (0.5 for definitely submarine links and 0.25 for potentially submarine links from B-O-N-U-S classification). Thus, prediction score, $S$ $\epsilon$ [0,1], is computed as:

\begin{equation}
S = f * [0.5 * (C_1 + C_2) + 0.4 * (2 - d_{1} - d_{2}) + 0.1 * (O_1 + O_2)]
\end{equation}

Note that geolocation carries the highest weightage (50\%), followed by distance to landing points (40\%) and cable ownership (10\%). Additionally, the distance scores $d_i$ are subtracted from a constant to ensure that we prioritize IPs geolocated closer to the landing points. Once all the predicted cables for an IP link are generated, we prioritize the likely cables by applying a Parallel Cable Threshold (PaCT). Thus, cables that are closer to the landing point and potentially have a match with a submarine cable owner are given a higher weightage. In Figure~\ref{running_example} H, the final mapping includes the prediction score for each potential submarine cable. In this example, the IP link with endpoints belonging to Google's ASN is associated with the PLCN and FASTER cables, which are also owned by Google.

The determination of the 5:4:1 weight ratio and the Parallel Cable Threshold (PaCT) value of 0.05 in Nautilus is based on empirical evaluation using validation experiments. Through extensive testing and analysis, the values that yielded the highest accuracy across all validation experiments were selected. For instance, the weight of 0.1 assigned to the ownership score was chosen after considering various factors. The lower weight was primarily due to observations such as the absence of certain major ISPs like Cogent from the submarine cable map, despite evidence suggesting otherwise~\cite{cogent_network_map}. Moreover, most ASes lease their cables and do not own them.

%% file: sections/datasets_current.tex
\section{Datasets} \label{data_sources}
\label{subsubsec:traceroutes}

In this section, we present a brief overview of the data sources used by Nautilus and their characteristics. \ashwin{A complete list of all data sources used in Nautilus is given in Table~\ref{nautilus_data_sources} in the Appendix.}

\begin{table*}
    \centering
    \resizebox{\textwidth}{!}{%
    \begin{tabular}{lcccccccc}
    \toprule
    \multirow{2}{*}{ } &
        \multicolumn{4}{c}{\textbf{IPv4}} &
        \multicolumn{4}{c}{\textbf{IPv6}} \\
        \cmidrule(lr){2-5}
        \cmidrule(lr){6-9}
        & RIPE 5051 & RIPE 5151 & CAIDA (v4) & \textbf{Total} & RIPE 6052 & RIPE 6152 & CAIDA (v6) & \textbf{Total} \\
        \midrule
    Total traceroutes & 15.3 M & 15.3 M & 86.2 M & \textbf{116.8 M} & 7.13 M & 7.13 M & 105 M & \textbf{119.3 M}\\
    \# links & 3.4 M & 3.9 M & 2.1 M & \textbf{5.8 M} & 1.83 M & 1.23 M & 1.17 M & \textbf{3.08 M} \\
    \# unique links & 1.09 M & 1.23 M & 724 K & & 441 K & 345 K & 1.42 M & \\
    \# link endpoint & 790 K & 937 K & 709 K & \textbf{1.11 M} & 424 K & 337 K & 1.38 M & \textbf{1.64 M} \\
    \bottomrule
    \end{tabular}%
    }
    \caption{RIPE Atlas and CAIDA traceroute dataset characteristics.}
    \label{traceroute_table}
\end{table*}

\parab{Traceroutes: } Nautilus relies on traceroutes collected by RIPE Atlas~\cite{ripe_atlas} and CAIDA's Ark~\cite{CAIDA}. From RIPE Atlas, Nautilus uses long-running traceroute measurements---(5051 \& 5151) and (6052 \& 6152) for IPv4 and IPv6, respectively. These measurements execute traceroutes between a subset of $\approx$ 1000 RIPE probes every 15 minutes. The CAIDA dataset has /24 (IPv4) and /48 (IPv6) prefix measurements~\cite{caida_24_probing, caida_48_probing}. CAIDA initiates ICMP-based Paris traceroute from a collection of 15-30 anchor probes to a random IP address in every /24 and /48 address space. We use RIPE Atlas traceroutes over the period of two weeks, March 15-29, 2022, with approximately 30M (30 million) IPv4 and 14M IPv6 traceroutes. We use CAIDA traceroutes collected over a single cycle (1647) corresponding to about ten days, March 13-23, 2022, containing 86M IPv4 and 105M IPv6 traceroutes. We chose two weeks, as the number of unique IPs and links went up only marginally, at around 3-4\% per day, past two weeks. Table~\ref{traceroute_table} gives a detailed breakdown of the datasets obtained post-processing based on extraction of IPs and links (\S~\ref{ip_link_extraction}). The final dataset thus includes 5.8 million IPv4 and 3.08M IPv6 unique links obtained from $\approx$ 120M traceroutes in IPv4 and IPv6, respectively. The number of unique IPs (link endpoints) is relatively small in comparison to the links due to the dense concentration of fewer IPs being involved with multiple IP links using submarine cables.

\parab{Submarine Cable Information: } Nautilus uses the Submarine Cable map provided by Telegeography~\cite{submarine_cable_map}. This map offers information such as landing points (terrestrial termination points of submarine cables), cable owners, Ready For Service (RFS) year, and the length of $\approx$ 480 submarine cables. In Nautilus, we only consider 450 cables with an RFS earlier than 2022. Additionally, there are $\approx$ 1200 valid landing points and 407 unique cable owners.

\parab{Geolocation: } For the 1.11M IPv4 and 1.64M IPv6 unique addresses identified from traceroutes, we use the following 11 geolocation sources: RIPE IPMap~\cite{ripe_ipmap}, CAIDA geolocation~\cite{caida_itdk}, Maxmind~\cite{maxmind}, IP2Location~\cite{ip2location}, IPinfo~\cite{ipinfo}, DB-IP~\cite{db-ip}, IPregistry~\cite{ipregistry}, IPGeolocation~\cite{ipgeolocation}, IPapi~\cite{ipapi}, IPapi.co~\cite{ipapi_co}, and IPdata~\cite{ipdata} to generate a geolocation output for a given IP. For the 1.11 million IPv4 addresses, we obtain geolocation results for 90K IPs from RIPE IPMap, 560K IPs from CAIDA geolocation, 1.07M IPs from Maxmind, and 1.1M IPs from all other private databases. Similarly, for 1.64M IPv6 addresses, we have the results for 18K IPs from RIPE IPMap, 1.64M IPs from Maxmind, and 1.62M IPs from other private databases. When combining the geolocation from multiple sources, we give equal weightage to all sources. Following the SoL validation process (\S~\ref{geolocation_sol_validation}), we determine that only 5.11M out of 5.8M IPv4 links and 2.34M out of 3.08M IPv6 links have valid geolocation information for both endpoints, satisfying the SoL validation criteria.

Note that the CAIDA dataset relies on multiple geolocation sources, including one of the sources we use, and requires combining the IP-to-node map (IP aliasing) and the node-to-geolocation map to generate the geolocation. We treat the CAIDA dataset as one of the unique sources and give it an equal weightage since its coverage and accuracy diverged significantly from the other individual sources (details in \S~\ref{sec:geolocation_validation}).

\parab{IP to ASN mapping: } For generating an IP to AS mapping, we use four sources: RADB servers~\cite{radb_server}, Routinator RPKI validator~\cite{routinator_whois}, Cymru WhoIS~\cite{cymru_whois} and CAIDA AS2Org~\cite{caida_as_to_org}. While we use the APIs for RADB and Cymru, for the RPKI validator we use the web utility provided by RIPE, which auto-resolves IPs into prefixes before generating an IP to ASN mapping. For the 1.11 M IPv4 addresses, we obtain matches for 470 K IPs from Whois RADB servers, 1.05 M IPs from routinator servers, 391 K IPs from CAIDA AS2Org mapping, and 1.11 M IPs from Cymru whois. Similarly, for 1.64 million IPv6 addresses, we obtain matches for 511 K IPs from Whois RADB servers, 1.56 M IPs from routinator servers, and 1.64 million IPs from Cymru whois. Upon merging the results from all sources, we obtain the mapping for all 1.11 M IPs and 1.64 million IPs in IPv4 and IPv6, respectively. As we use the maximum agreement (voting) mechanism, the impact of wrong predictions of a single source goes down significantly. \ashwin{The average agreement amongst the valid sources per IP is 90\%. Additionally, we have a valid output from all 4 sources for only 10\% of IPs, with 79\% of the IPs having a mapping from two or three valid sources.}

In summary, we gather 5.8M IPv4 and 3.08M IPv6 unique links with 1.11M IPv4 and 1.64M IPv6 unique IPs from the traceroutes collected from RIPE Atlas and CAIDA over a period of approximately 15 days. We obtain geolocation information for $\approx$ 99\% of all these unique IPs and a valid AS number mapping for all the IPs. Among these links, only 5.11M IPv4 and 2.34M IPv6 links satisfy the SoL constraint.

%% file: sections/evaluation_current.tex
\section{Mapping Analysis} 
\label{mapping_analysis}

In this section, we analyze the mapping generated by Nautilus. In particular, we answer the following questions:
\begin{itemize}[leftmargin=*,nolistsep]
    \item How effectively can Nautilus generate a mapping for a given link? (\S~\ref{cma})
    \item How does Nautilus mapping compare with prior work: SCN-Crit~\cite{submarine_drivability} and iGDB~\cite{igdb}? (\S~\ref{cpw})
\end{itemize}

\subsection{Geolocation Validation}
\label{sec:geolocation_validation}

\subsubsection{Performance with multiple geolocation sources} \label{multiple_geolocation_performance_gains}

We conduct geolocation accuracy validation on eleven geolocation sources, against Nautilus, using ground-truth data generated by Gharaibeh et al.~\cite{geolocation_problem}. The ground-truth data\ashwin{, containing $\approx$16500 data points,} was created through DNS-based and delay-based approaches and underwent rigorous filtering and validation to identify a subset of IPs with accurate geolocation~\cite{geolocation_problem}. 

\begin{figure}[!ht]
    \centering
    \begin{subfigure}{0.47\textwidth}
        \centering
        \includegraphics[width=\textwidth]{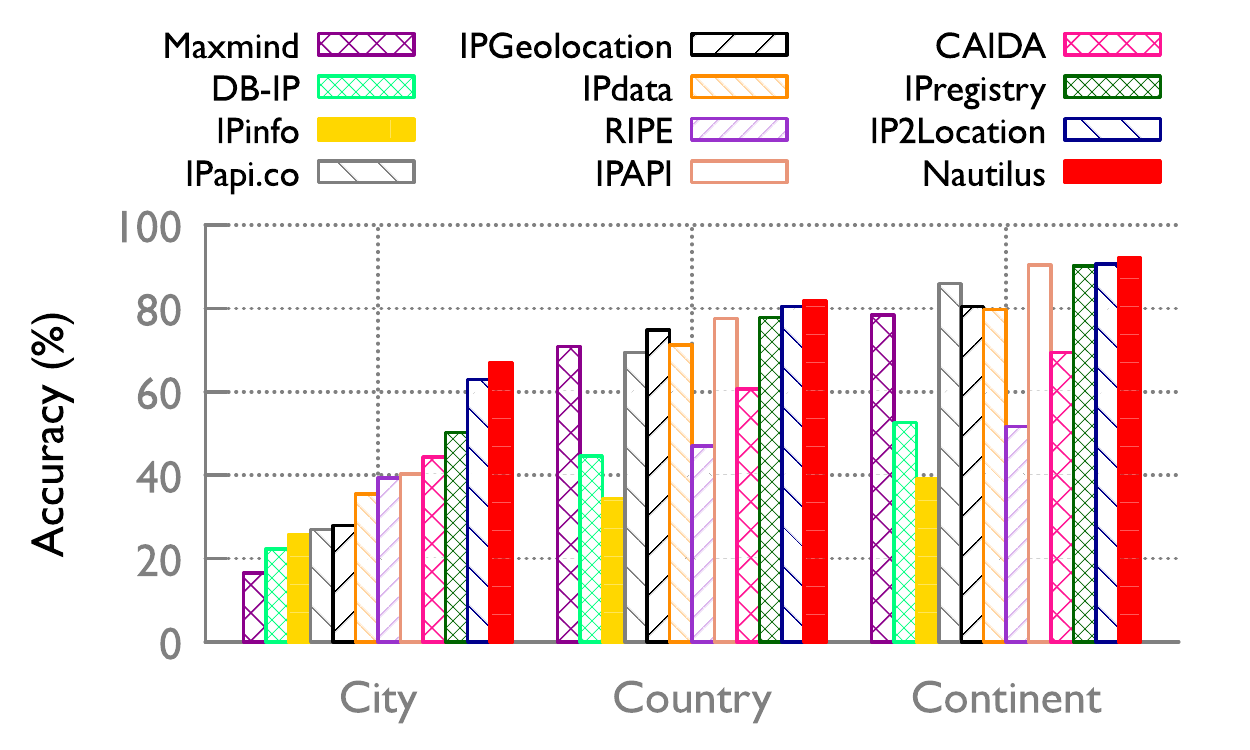}
        \caption[]%
        {{ The accuracy of geolocation sources at City, Country, and Continent granularity (normalized based on the total number of IPs in the ground truth data). Nautilus geolocation is determined by the cluster with the largest cluster score.}}
        \label{fig:geolocation_accuracies}
    \end{subfigure}
    \hfill
    \begin{subfigure}{0.48\textwidth}
        \centering
        \includegraphics[width=\textwidth]{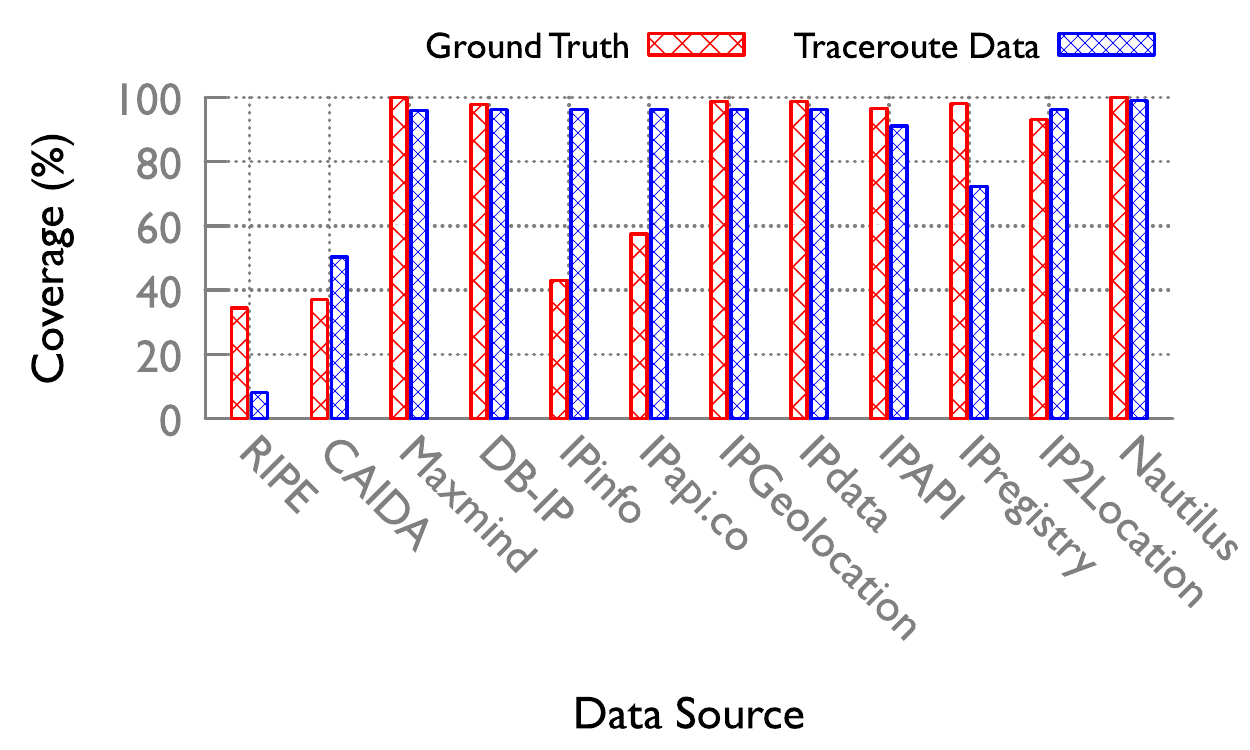}
        \caption[]%
        {{ The coverage of geolocation sources for (i) the ground truth data and (ii) IPs gathered from traceroutes. RIPE and CAIDA have low coverage across both datasets.}%
        }
        \label{fig:geolocation_coverage}
    \end{subfigure}
    \label{fig:geolocation_data}
    \caption{Validation experiment for geolocation sources. (a) Comparison of accuracy of the eleven geolocation sources against Nautilus. (b) Fraction of IPs in the ground truth data and traceroute data successfully geolocated by various sources. }
\end{figure}

We evaluate the accuracy of each source at three levels: city, country, and continent. To ensure a fair comparison, we normalize the results by dividing them by the total number of IPs in the ground truth data, rather than the number of IPs geolocated by each service. For Nautilus, we select the cluster with the highest cluster score from the geolocation validation process described in \S~\ref{geolocation_sol_validation} as an IP's geolocation. \ashwin{On average, we observe a 75\% agreement among sources at the city level and a 97\% agreement at the country level. Furthermore, approximately 87\% of the IPs have valid geolocation data from at least 8 sources, while only 7\% of IPs have data from fewer than 5 sources.} Figure~\ref{fig:geolocation_accuracies} presents the results of this validation. Nautilus exhibits $\approx$ 5\% improvement in geolocation accuracy at the city level compared to individual geolocation sources. At the country and continent levels, Nautilus shows a gain of $\approx$ 3\% in accuracy. The lower city-level accuracy observed for RIPE and CAIDA can be attributed to the lack of geolocation data for many IP addresses in their datasets. 

\ashwin{ While the accuracy improvements with Nautilus may appear modest, it is important to consider two key factors. First, even a minor improvement in accuracy can result in the accurate mapping of a large number of links in our dataset with millions of links. Second, when compared to the closest single source (in terms of accuracy), Nautilus has $\approx$ 6\% improvement in coverage.} It is also worth noting that although the ground truth data predominantly includes IPs from North America and Europe (90\%), Nautilus outperforms other sources even in this region, and the trend remains for other regions in the dataset as well. The coverage for each geolocation source is shown in Figure~\ref{fig:geolocation_coverage}. RIPE and CAIDA datasets have the lowest coverage across both the ground truth dataset and the traceroute dataset \footnote{While the ground truth data was published in 2017 and the geolocations for the sources and Nautilus were estimated in 2022, the overall trend is expected to remain largely the same, as all geolocation services are subject to similar changes in geolocation over time. It is also important to highlight that various sources show consistent coverage for the ground truth data and the IPs extracted from traceroute data. However, a few geolocation sources exhibit variations due to changes in the inherent mechanisms employed by each of these sources over time.}.

\subsubsection{SoL validation threshold} \label{sol_validation_threshold}

Nautilus incorporates the Speed-of-Light (SoL) validation to improve geolocation accuracy. To assess the effectiveness of SoL validation, we use the same ground-truth data~\cite{geolocation_problem} and follow the SoL validation criteria outlined in \S~\ref{geolocation_sol_validation}. We vary the SoL threshold and evaluate the geolocation accuracy at each threshold. Through empirical analysis, we determine that an SoL threshold of 0.05 (equivalent to a 5\% error margin) yields the highest geolocation accuracy, surpassing the accuracy achieved without SoL validation by 2\%. The results of this validation can be found in Figure~\ref{fig:sol_validation} in Appendix~\ref{sol_validation_analysis}.

\begin{table}[]
    \parbox{.45\columnwidth}{
    \centering
    \begin{tabular}{ c c c }
    \hline
    \textbf{Stage} & \textbf{Count (v4)} & \textbf{Count (v6)} \\
    \hline
    Total & 5.8 M & 3.08 M \\
    Classified & 5.11 M & 2.34 M \\
    Submarine & 3.73 M & 1.79 M \\
    Mapped & 3.05 M & 1.43 M \\
    \hline
    \end{tabular}
    \vspace{2mm}
    \caption{The count of links at the end of each stage. Classified refers to the links that receive a category from the BONUS classification. Submarine indicates the links that receive either a definite or potential submarine category (S+U). Mapped indicates the number of submarine links for which Nautilus generates a mapping.}
    \label{tab:simplified_res}
    }
    \hfill
    \parbox{.45\columnwidth}{
    \centering
    \begin{tabular}{ c c c }
    \hline
    \textbf{Category} & \textbf{Count (v4)} & \textbf{Count (v6)} \\
    \hline
    S, B & 511 K & 200 K \\
    S, O & 342 K & 172 K \\
    S, N & 42 K & 30 K \\
    U, B & 1261 K & 490 K \\
    U, O & 1112 K & 497 K \\
    U, N & 469 K & 406 K \\
    Terrestrial & 1377 K & 564 K \\
    Unclassified & 672 K & 712 K \\\hline
    Total & 5.8 M & 3.08 M \\
    \hline
    \end{tabular}
    \vspace{2mm}
    \caption{The count of links under each category of B-O-N-U-S classification. Links that lack geolocation data or fail SoL validation are in the Unclassified category.}
    \label{link_classification_results}
    }
\end{table}

\begin{table}[]
    \parbox{.45\columnwidth}{
    \centering
    \begin{tabular}{ c c }
    \hline
    \textbf{Country Pairs (v4)} & \textbf{Count} \\
    \hline
        US-DE & 40 K \\
        US-UK & 30 K  \\
        US-FR & 18.5 K \\
        SG-IN & 18.2 K  \\
        US-NL & 16.9 K  \\
        UK-DE & 12.9 K  \\
        US-SW & 10 K \\
        US-AU & 9.9 K  \\
        US-BR & 9.8 K  \\
        UK-NL & 8.9 K \\
    \hline
    \end{tabular}
    \vspace{2mm}
    \caption{Top 10 country pairs based on definitely submarine links (S category) of IPv4 (using ISO 2-alpha code for countries).}
    \label{tab:country_count_v4}
    }
    \hfill
     \parbox{.45\columnwidth}{
    \centering
    \begin{tabular}{ c c }
    \hline
    \textbf{Country Pairs (v6)} & \textbf{Count} \\
    \hline
        US-CL & 23 K \\
        US-DE & 10.2 K \\
        UK-RU & 7.4 K \\
        US-HK & 7.3 K \\
        DE-UK & 6.3 K \\
        US-UK & 6.1 K \\
        UK-NL & 5.6 K \\
        US-JP & 5.2 K \\
        US-BR & 5 K \\
        SW-NL & 4.5 K \\
    \hline
    \end{tabular}
    \vspace{2mm}
    \caption{Top 10 country pairs based on definitely submarine links (S category) of IPv6 (using ISO 2-alpha code for countries).}
    \label{tab:country_count_v6}
    }
\end{table}

\begin{figure}[!ht]
  \begin{minipage}[b]{0.47\linewidth}
    \centering
    \includegraphics[width=\linewidth]{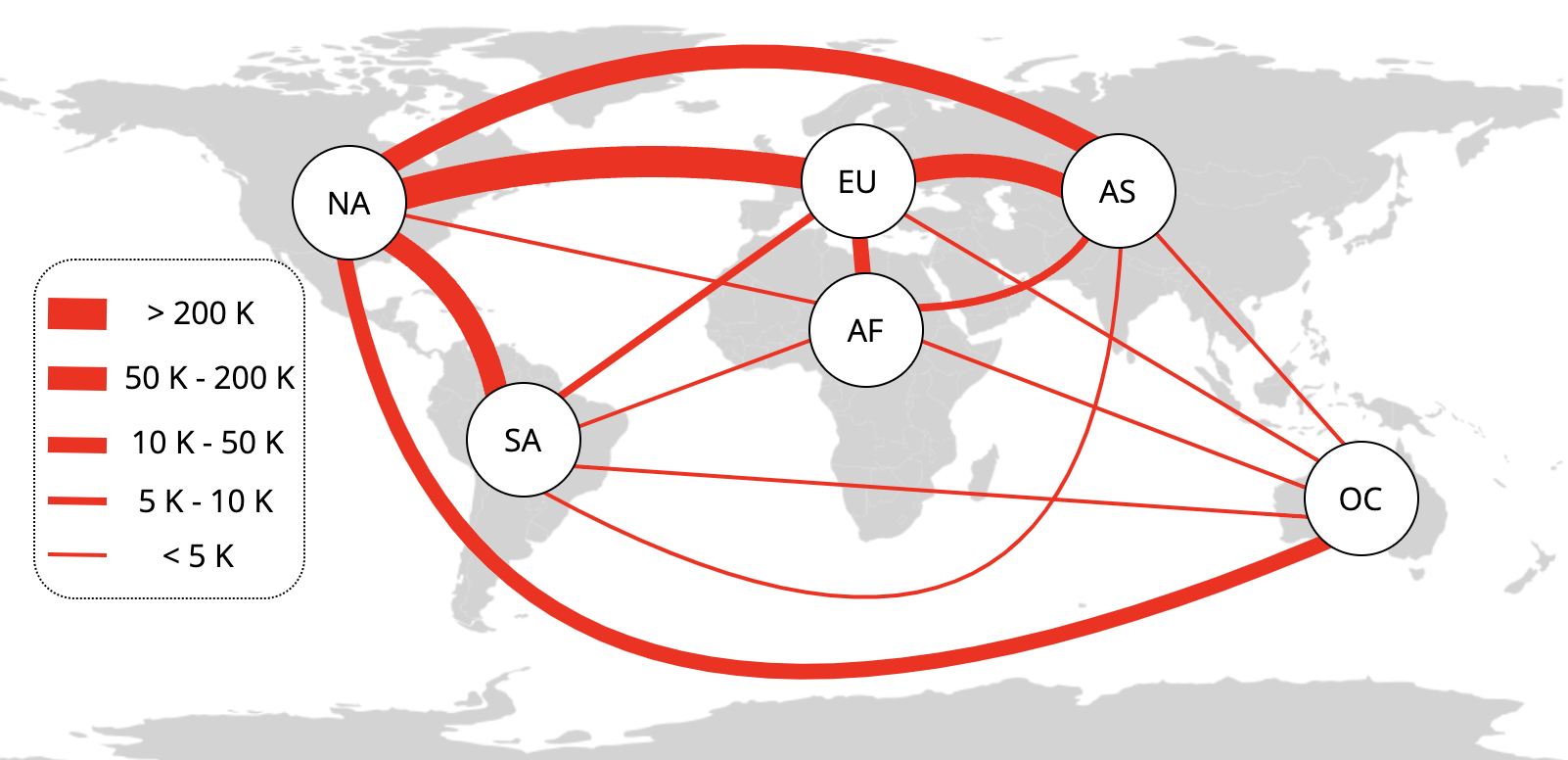}
    \caption{Inter-continental link distribution (IPv4 and IPv6 combined). NA-EU makes up around 205 K links, followed by EU-EU and AS-AS (not depicted in the figure) which make up at least 90 K links each.}
    \label{continent_map}
  \end{minipage}%
  \hfill
  \begin{minipage}[b]{0.48\linewidth}
    \centering
    \includegraphics[width=\linewidth]{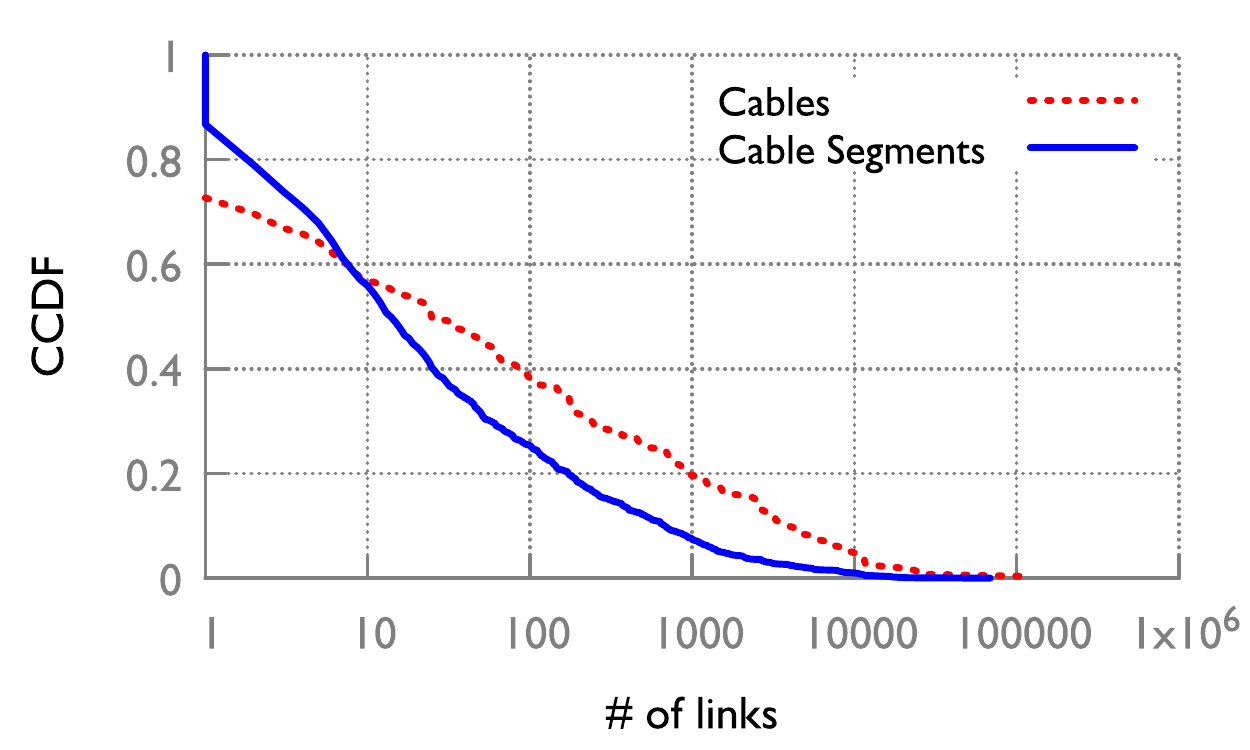}
    \caption{\ashwin{The distribution of the number of links per cable/cable segment (in log scale). The average number of links per cable segment and cable are $\approx$ 447 and 2174, respectively. Both distributions have a long tail, with a maximum of 68,000 and 120,000 links per cable segment and cable, respectively.}}
    \label{link_per_cable}
  \end{minipage}
\end{figure}

\subsection{B-O-N-U-S classification analysis}

Of the 5.8M IPv4 and 3.08M IPv6 \ashwin{links extracted from all traceroutes}, we generate a classification for 5.11M IPv4 and 2.34M IPv6 links. 15\% of links are not classified due to either invalidation of geolocation during SoL tests (14\%) or lack of geolocation data (1\%) \ashwin{for all 11 sources}. The number of links at the end of each stage of the pipeline is given in Table~\ref{tab:simplified_res}. For the classified links, Nautilus labels approximately 17.5\% IPv4 and 17\% IPv6 links as \emph{definitely submarine}, and 27\% IPv4 and 24\% IPv6 links as \emph{definitely terrestrial}. The remaining are tagged as \emph{potential submarine} (detailed breakdown in Table~\ref{link_classification_results}).

\parab{Top Country Pairs:} We analyze the top 10 country pairs for \emph{definitely submarine} links in Table~\ref{tab:country_count_v4} (IPv4) and Table~\ref{tab:country_count_v6} (IPv6). There is a high overlap between the top country pairs of IPv4 and IPv6. Note that these tables exclude intra-country links. An observation from Tables~\ref{tab:country_count_v4} and \ref{tab:country_count_v6} is that most of the countries in top pairs, especially in Europe, house some of the largest IXPs in the world, e.g., Germany (DE-CIX), Netherlands (AMS-IX), UK (LINX), etc. We analyze the IPv4 links between the US and Germany (US-DE) to validate our findings. We observe that 81\% ($\approx$32.2K) of the 40K links have endpoints in Germany within 10 km of cities with known IXPs \ashwin{compiled based on the IXP information from PeeringDB~\cite{PeeringDB}.} These link endpoints located farther inland (away from the coast) in Germany also confirm the presence of MPLS tunnels in long-haul paths.

\parab{Top Continent Pairs:} Figure~\ref{continent_map} shows that NA-EU accounts for 205K (or $\approx$38.5\%) of inter-continental links, while EU-AS and NA-AS account for $\approx$70K links each. Almost 41\% of all links in the Nautilus map in the \emph{definitely submarine} category are intra-continental links, under which EU and AS account for at least 90K links each.

\subsection{Cable Mapping Analysis} \label{cma}

We first analyze the coverage of Nautilus' mapping with respect to submarine cables and IP links. We then analyze the distributions for all categories.

\begin{figure}[!ht]
    \centering
    \begin{subfigure}{0.47\textwidth}
        \centering
        \includegraphics[width=\textwidth]{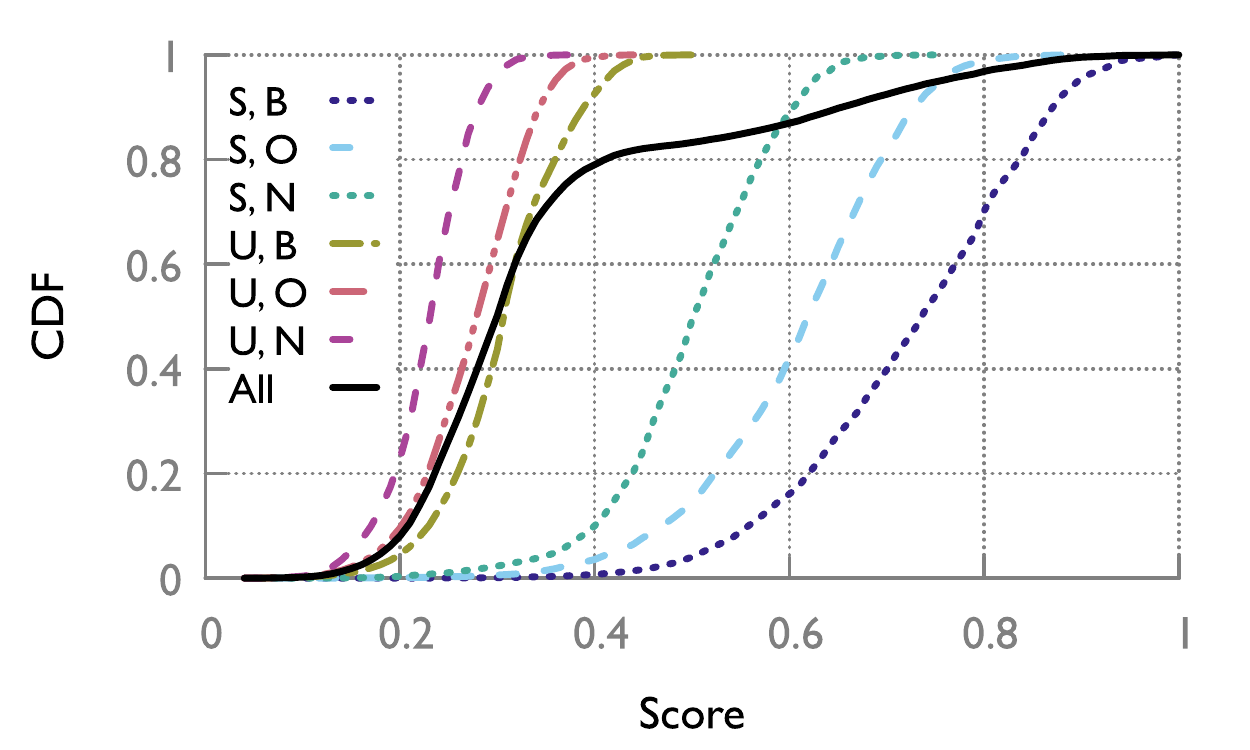}
        \caption[]%
        {{\small Prediction scores for various categories in IPv4}}
        \label{scores_v4}
    \end{subfigure}
    \hfill
    \begin{subfigure}{0.47\textwidth}
        \centering
        \includegraphics[width=\textwidth]{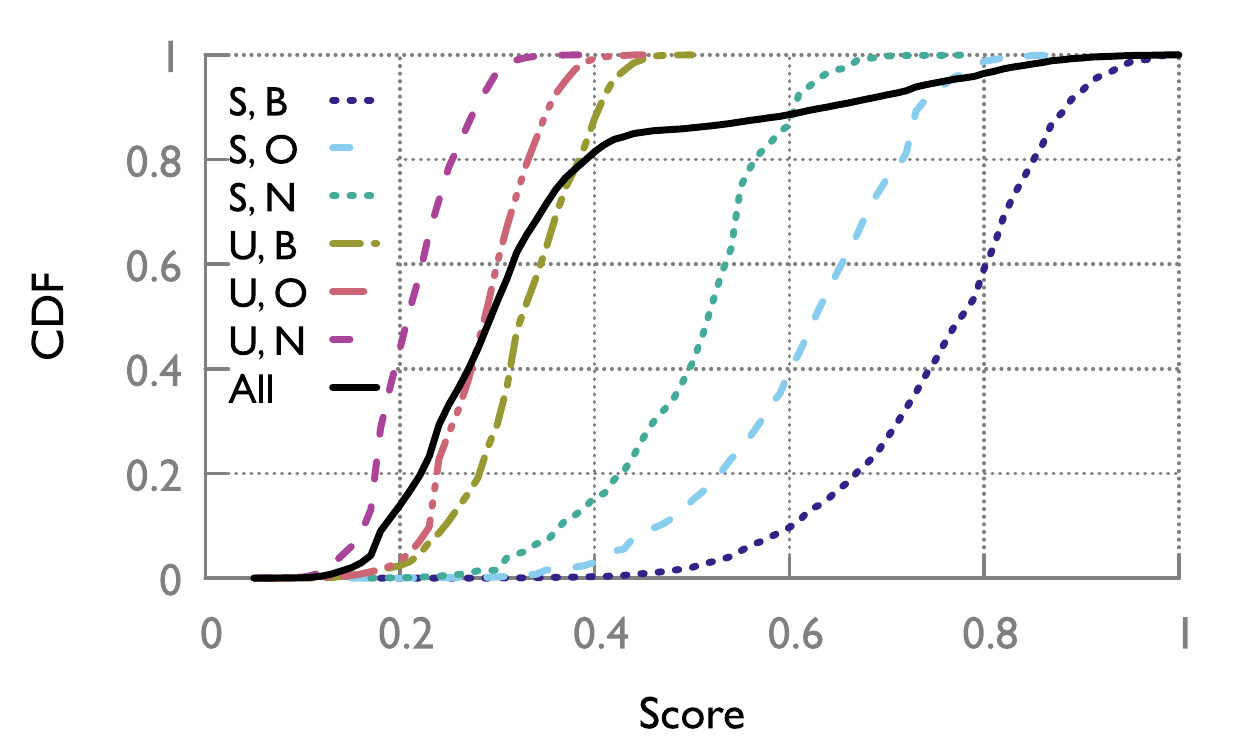}
        \caption[]%
        {{\small Prediction scores for various categories in IPv6}}
        \label{scores_v6}
    \end{subfigure}
    \label{score_result}
    \caption{The distribution of prediction scores for IPv4 and IPv6}
\end{figure}

\begin{figure}[ht]
    \centering
    \begin{subfigure}{0.47\textwidth}
        \centering
        \includegraphics[width=\textwidth]{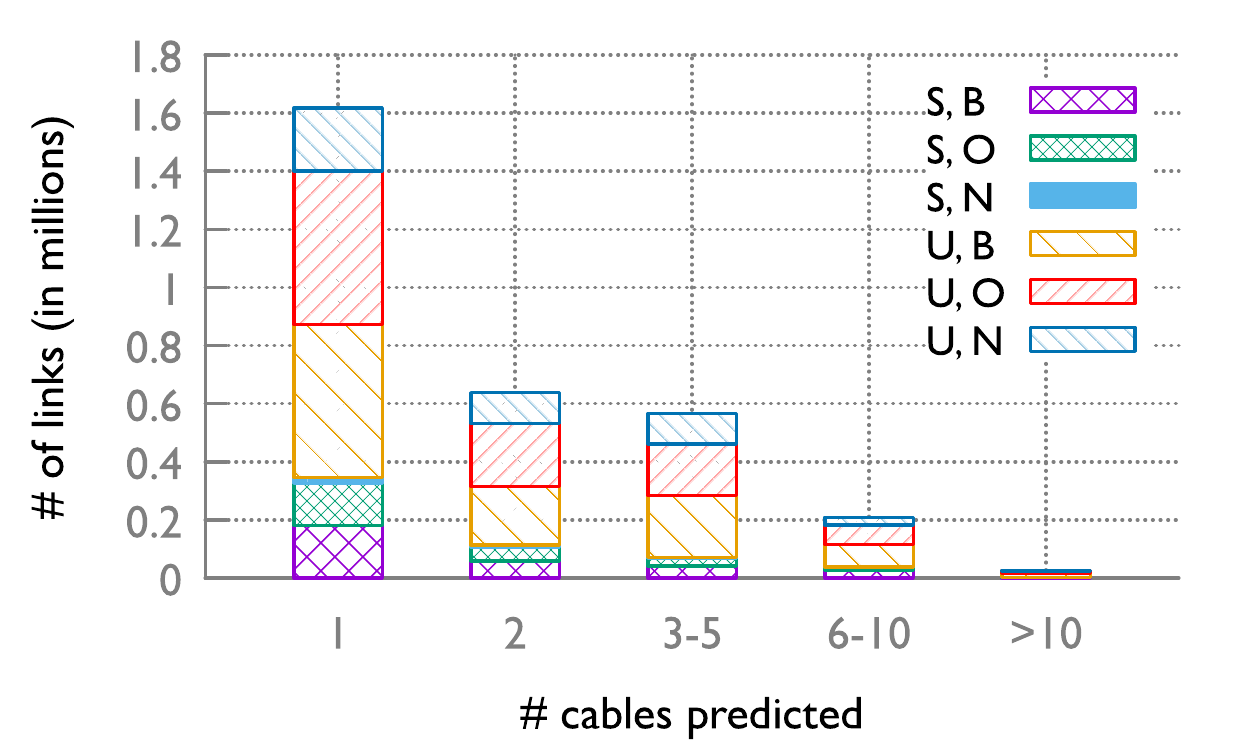}
        \caption[]%
        {{\small Cable count for various link categories in IPv4}}
        \label{count_v4}
    \end{subfigure}
    \hfill
    \begin{subfigure}{0.47\textwidth}
        \centering
        \includegraphics[width=\textwidth]{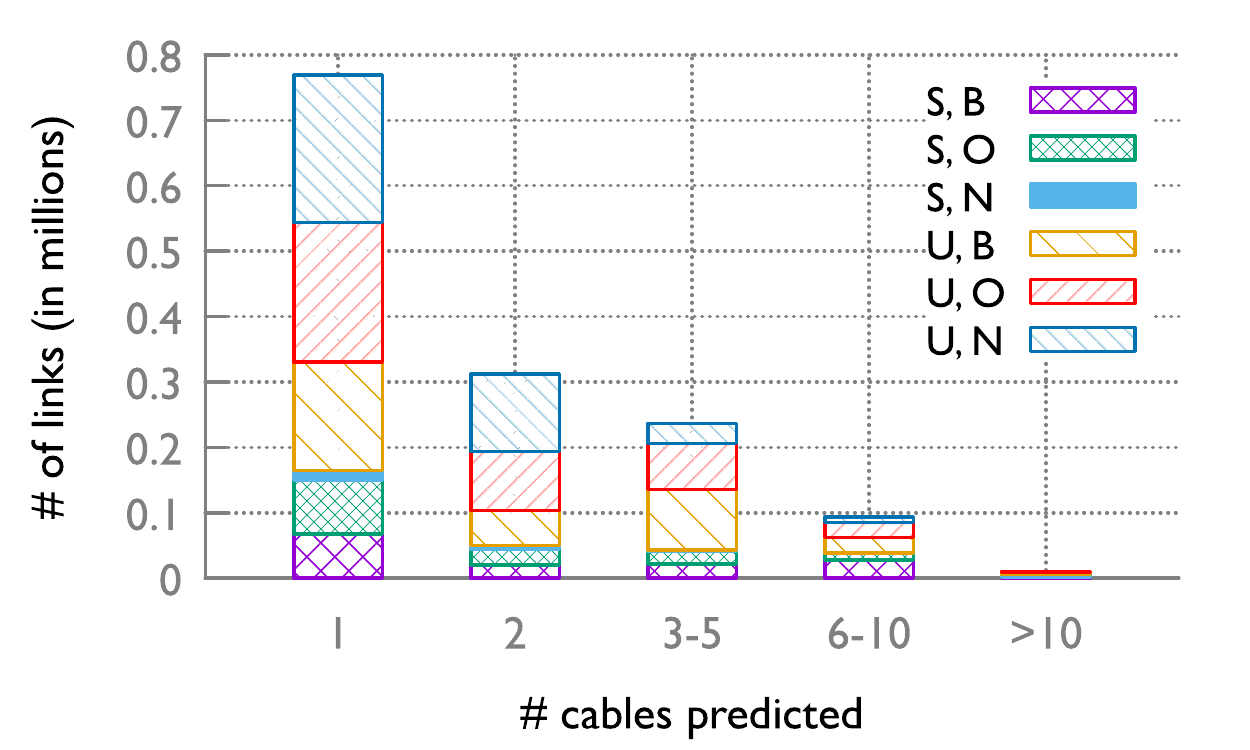}
        \caption[]%
        {{\small Cable count for various link categories in IPv6}}
        \label{count_v6}
    \end{subfigure}
    \caption{The distribution of the number of cables predicted for links belonging to each category. Most links are mapped to a single cable.}
    \label{count_result}
\end{figure}

\parab{Cable and link coverage:} Based on the generated cable mapping, of the 448 active cables, we find at least one IP link in 91\% of the submarine cables. For the $\approx$ 1200 landing points, we identify at least a single IP link that passes through 1074 landing points, thus accounting for 90\% of all landing points. Most of the missed submarine cables and landing points are part of small or regional cables with a short geographic spread. We generate a mapping for 82\% and 80\% of submarine links (definitely and potential submarine combined) in IPv4 and IPv6, respectively. Mapping fails when (i) geolocation or SoL validation is unsuccessful (incorrect geolocation) or (ii) link endpoints are more than 1000 km away from the nearest cable landing point. \ashwin{We also evaluate the distribution of IP links per cable and cable segment, shown in Figure~\ref{link_per_cable}. About 40\% of cables and cable segments have fewer than ten links mapped to them. Over 80\% and 90\% of cables and cable segments, respectively, have less than 1000 links. Additionally, both distributions have a long tail, with a maximum of about 68,000 and 120,000 IP links per cable segment and cable, respectively.}

\parab{Predicted Cable Distribution:} The prediction score for each cable mapping determines the confidence in the mapping. Hence, we analyze the scores in each category by plotting a CDF of the \ashwin{maximum} prediction score for each IP link. Figures~\ref{scores_v4} and~\ref{scores_v6} shows the results for IPv4 and IPv6 respectively. The CDF plots for IPv4 and IPv6 look similar but have subtle differences. Particularly, IPv6 has slightly higher scores in comparison to IPv4. From the scores, we observe that in the best scenario of S, B (definitely submarine links with good geolocation accuracy), Nautilus predicts cables with a prediction score higher than 0.8 for at least 30\% of the links and a score higher than 0.6 for more than 85\% of the links in both IPv4 and IPv6. The other categories have a similar distribution but with lower prediction score bounds. Finally, considering all the links, we observe that only 20\% of the links have a score greater than 0.4. This is expected as most of the links fall under the unconfirmed category, which is assigned a lower prediction score by Nautilus.

Due to the possibility of parallel cables, Nautilus predicts greater than one cable in some scenarios. Using the PaCT value of 0.05 (detailed in \S \ref{final_mapping}), the stacked bar plot  on the number of cables predicted per link in each category is shown in Figures~\ref{count_v4} and \ref{count_v6} for IPv4 and IPv6, respectively. From these plots, we observe that for all links with a mapping, $\approx$54\% of the links have exactly one cable mapped to them for both IPv4 and IPv6 (the best-case scenario). We have more than 76\% of the links with a mapping of $\leq$ two cables. Finally, only $\approx$7\% of the links have a cable mapping with more than five cables. On average, Nautilus maps 2.04 cables per link and prunes 65\% of the potential cables in the final aggregation step.

\subsection{Comparison with Prior Work} \label{cpw}

As there exists no prior work on uniquely mapping IP links to submarine cables, we compare Nautilus against the two closest related works: (i) Liu et. al. (which we call SCN-Crit)~\cite{submarine_drivability} and (ii) iGDB~\cite{igdb}. Since there are no agreed-upon validation criteria for these prior works, we use the number of cable predictions as our benchmark for comparison, with fewer cables indicating a better mapping. Moreover, prior works do not validate their mapping, but Nautilus attempts to validate its mapping efficacy in \S~\ref{validation}.

\parab{SCN-Crit.} SCN-Crit~\cite{submarine_drivability} ran traceroute measurements from RIPE probes to the top 50 websites from 63 countries. Its primary goal was to evaluate the impact of submarine cable networks on end users. SCN-Crit used the drivability metric (no drivable route between ends of a given path) to determine the presence of a submarine cable in any given path. While analyzing the impact on end users, SCN-Crit also generated a cable mapping based on the submarine cable overlap between the countries in the path.

We reproduce a part of this work by initiating  traceroute measurements to the top 50 websites from a RIPE probe within each country for 63 countries, thus generating 3150 traceroutes (one measurement per website-probe pair). Though this does not capture the entirety of measurements carried out in SCN-Crit which includes multiple locations for the same resource, this acts as a good proxy. For the $3150$ traceroutes, Nautilus identifies a cable match for $\approx$70\% of all the websites. On average for all links found in both the mappings, Nautilus predicts 35\% fewer cables per link than SCN-Crit, due to Nautilus' fine-grained geolocation. The number of cables predicted by SCN-Crit is higher primarily due to the policy of cable mapping at the country level. For example, between the US and the UK, SCN-Crit considers a large set of cables between the countries, while Nautilus, which operates at a finer granularity, will only predict the subset of cables in proximity to the relevant landing points.

Finally, as SCN-Crit takes a conservative approach to identifying paths with submarine cables (based on its drivability metric), this approach misses many paths that could potentially utilize a submarine cable. For example, in our validation experiments, we observe that cities within the same country might sometimes use a submarine cable, as seen in the case of an IP link between two cities in Yemen that relies on a submarine cable (\S~\ref{subsec:failures}). SCN-Crit ignores this possibility as these cities are drivable and hence would fail the drivability constraint. Such intra-country links are also common in large countries such as the US.

\parab{iGDB.} iGDB~\cite{igdb} is a framework for mapping between terrestrial cables and the IP layer. iGDB, like Nautilus, relies on geolocation as the critical piece to infer the potential terrestrial cable for a given IP link. iGDB partitions the Earth into Thessian polygons based on the nearest city and then overlays this information with the road/rail network to predict potential routes. The paper shows mapping in specific use cases, but as confirmed with the authors of iGDB, it does not support mapping IP links to submarine cables.

We extend the iGDB framework for generating a mapping over submarine cables. As geolocation scripts are missing in iGDB codebase, we use the CAIDA's geolocation data as a proxy. CAIDA also employs geolocation based on DNS hostname resolutions and IXP locations that iGDB relies on. From our traceroutes, we extract links where both endpoints have a geolocation in CAIDA and merge this with the iGDB framework to map each IP endpoint to the nearest city cluster. Next, to identify the submarine cables, we use an intersection of submarine cables across landing points within each city cluster for a given IP link. We observe that Nautilus predicts $\approx$ 75\% fewer cables than iGDB, indicating better accuracy. The higher number of cable predictions in iGDB is due to the higher concentration of submarine cable landing points within a few city clusters (ex: New York). More importantly, iGDB generates a mapping for only 11\% of the extracted IP links. iGDB's strict geolocation constraints and its reliance on limited geolocation sources contribute to its limited coverage. For example, an IP link between France and the US with an endpoint geolocated at Paris (owing to either geolocation inaccuracies or the use of MPLS tunnels) would not find a cable match under iGDB mapping since no landing points fall within the Paris city cluster. But Nautilus succeeds in mapping such links due to the increasing radius-based queries. 

The main conclusion drawn from the comparison is that, while there are no directly comparable works, previous studies that share similarities with Nautilus predict a significantly larger number of cables per IP link. In contrast, Nautilus provides a \textit{more accurate} and refined mapping. More importantly, these prior works exhibit a notable limitation: they fail to generate any mapping for a substantial number of IP links. Nautilus thus achieves both higher coverage of IP links and higher accuracy, generating a more reliable, comprehensive submarine cable mapping dataset.

%% file: sections/validation.tex
\section{Validation} \label{validation}

In this section, we provide an overview of the techniques we use to validate the map generated by Nautilus. We initially attempted to validate Nautilus mapping by requesting ground truth information from nearly a dozen large ISPs across the globe. Only two ISPs responded; they were unable to share the data due to its sensitive nature. Hence, we use the following validation techniques:
\begin{itemize}[leftmargin=*,nolistsep]
    \item Validation using submarine cable failures: When a submarine cable fails, all links mapped to the failed cable segment should disappear. Hence, inspired by Bischoff et. al.~\cite{submarine}, we evaluate the mapping accuracy of Nautilus by examining two specific submarine cable failure scenarios.
    \item Validation using targeted measurements: We run targeted traceroute measurements using RIPE probes located close to submarine cable landing points. We then extract the potential submarine hop (high latency IP link) and compare it against the mapping generated by Nautilus.
    \item Validation using operator network maps: A few network operators have made their global network map images publicly available. To validate Nautilus, we predict the likely cables each of these network operators use and compare this against the operator's network map. 
\end{itemize}

\subsection{Using Submarine Cable Failures}
\label{subsec:failures}

We use publicly available information about cable failures to validate the accuracy of Nautilus mapping. When a cable fails, all links on the failed segment should disappear during the outage, while being active before and after the outage. We leverage this idea to perform validation of our mapping.

In step 1, depending on the nature of the failure, we either classify a failure as a landing point failure (all cables at a specific landing point are affected) or a cable failure (a single cable affected). Once the landing point or the submarine cable is selected, we extract all IP links from Nautilus' mapping that traverse the failed location/cable. In step 2, two days worth of RIPE traceroute data (IPv4 5051 and 5151 measurements) is collected before, during, and after an outage (a total of 6 days' data for a given failure). Finally, we perform an intersection of the IP links from step 1 and step 2 for three periods (\textit{Before\_Failure}, \textit{During\_Failure}, \textit{After\_Failure}). We present this analysis on two recent submarine cable failures.

\parab{Yemen Outage:} The FALCON and SeaMeWe-5 cables experienced a 4-day outage in Yemen in January 2022 due to an airstrike at Al Hudayah, one of the cable landing points. Both cables connect the Arabian peninsula, with no parallel cables in the exact cable path, except for a short overlap of FALCON and SeaMeWe-5 cables between Egypt and Yemen (Figure~\ref{fig:yemen_cables} in Appendix~\ref{cable_failures_appendix}). This example is selected to display Nautilus' performance with non-parallel, single landing point, and multiple cable failure scenarios. We analyze the traceroutes collected between 19-21 January (\textit{Before\_Failure}), 22-24 January (\textit{During\_Failure}), and 25-27 January (\textit{After\_Failure}). Analyzing the IP links mapped to FALCON \& SeaMeWe-5 cables in Nautilus' mapping, we observe an overlap of (i) 106 links in \textit{Before\_Failure} traceroutes, (ii) 5 links in \textit{During\_Failure} traceroutes, and (iii) 93 links in \textit{After\_Failure} traceroutes collected after the outage.

This drastic fall in the number of links only during the outage indicates the accurate predictions made by Nautilus. As there are no parallel cables, during an outage we should observe 0 links, but we observe 5 active links. Upon closer examination, we find that all these links were classified as \emph{potential submarine}, and Nautilus had assigned a low prediction score for all these 5 links. Another interesting observation is that there was a potential submarine link mapped to the FALCON cable between two points within Yemen (Al Hudaydah and Al Gaydah), which was active in \textit{Before\_Failure} and \textit{After\_Failure} but was missing in \textit{During\_Failure}. Its disappearance indicates that the link marked as potentially submarine was indeed a submarine cable. While SCN-Crit does not map such links as the endpoints are connected by land and drivable, the potential submarine category in Nautilus allows us to identify and map them.

\parab{Papua New Guinea Earthquake:} During the Sep' 2022 earthquake in Papua New Guinea, the KDSC cable system experienced a failure near Madang (Figure~\ref{fig:kdcn} in Appendix~\ref{cable_failures_appendix}). Traceroutes were collected before the failure (5-7 September, labeled as \textit{Before\_Failure}) and during the failure (12-14 September, labeled as \textit{During\_Failure}). Since the cable was not fully operational during the experiment, no traceroutes were available after the failure (\textit{After\_Failure}).

Upon analysis, it was observed that there were 100 IP links from Nautilus' mapping in the \textit{Before\_Failure} dataset. However, during the failure (\textit{During\_Failure}), only 61 links were active. We further discovered that all IP links between islands disappeared during the outage. The 61 active IP links belonged to the potential submarine category, indicating the possibility of \ashwin{these links were using a terrestrial cable instead}. As a result, the impact of the outage was not as pronounced as the Yemen outage example. Classifying connections between two endpoints that could potentially be connected by both land and submarine cables is challenging. It is difficult to determine whether those links are truly submarine or terrestrial unless the possibility of a terrestrial cable between those points can be completely ruled out. Furthermore, some potential submarine links, such as the one between Madang and Lae, disappeared during the outage, confirming that they indeed traverse the submarine cable, despite passing the drivability test between endpoints.

\subsection{Using targeted traceroutes}

We run targeted traceroutes between a pair of RIPE probes located near cable landing points to validate Nautilus mapping. Specifically, we consider probes that meet two key requirements: (i) their proximity to the desired landing points, and (ii) the ownership of the probe's ISP aligning with the submarine cable owner. By selecting specific probe pairs that meet these criteria, we can direct the traceroutes to target a particular cable that we intend to test. In this process, we focus only on the link with the highest latency within these traceroutes. Our aim is to evaluate the accuracy of Nautilus mapping for these targeted links.

First, we eliminate all measurements with high-latency links that are not present in Nautilus mapping, i.e., those traversing a different path (Ex: Link between France and Singapore would be expected to take a cable connecting Europe to Singapore, but due to paths chosen by BGP, the traceroute might follow a circuitous path from France to the US, and then to Singapore). After this elimination, we identify 328 IP links that satisfy our criteria and compare them with Nautilus mapping. For $\approx$77\% (or 252 links) of all such IP links, the top cable prediction from Nautilus matches the expected cable. For $\approx$19\% (or 63 links) of the IP links, Nautilus predicts the expected cable, but not as the top prediction \ashwin{(but within the top 3 predictions for 80\% of links and within top 5 predictions for 100\% of links)}. Finally, for only $\approx$ 4\% (or 13 links), the expected cable does not feature in the predicted cable set.

\subsection{Using operator network maps}

Some large ISPs and network operators make their submarine cable network map images publicly available. We attempt to construct a similar map using Nautilus mapping and compare the results. We perform validation using network maps from Tata Communications~\cite{tata_map} and Vodafone~\cite{vodafone_map}. \ashwin{First, we create a mapping of IP endpoints to their respective organization names by combining the IP-to-ASN mapping from Nautilus with the ASN-to-Organization names from CAIDA ASRank. Then, for a specific network operator like Tata Communications, we locate IPs whose organization names closely correspond to the network operator and extract the IP links associated with these identified IPs.} Next, we generate the list of cables that these IP links were mapped to (by Nautilus). We then compare the potential submarine cables identified from the publicly available network map of operators \ashwin{(transcribed manually)} with the Nautilus' cable list. We observe that of the 34 cables present in the Tata Communications network map, 31 cables are present in Nautilus' predicted cable list. Similarly, for the Vodafone Network Map, Nautilus correctly identifies 52 of the 58 cables present in Vodafone's network map. We observe that the missed cables are from regions where Nautilus data sources do not have sufficient coverage.

%% file: sections/discussion.tex
\section{Discussion}

Nautilus is by no means a perfect solution for generating an IP link to submarine cable mapping, but rather a first attempt. In this section, we discuss the inaccuracies in underlying datasets used by Nautilus, identify special case scenarios in submarine mapping, and present suggestions for improving mapping workflows in the future.

\parab{Inaccuracies in datasets } \textit{(i) Geolocation:} Nautilus relies on geolocation as a key factor in determining the IP link to submarine cable mapping. However, geolocation services are known to be imperfect, with only 50-60\% accuracy at the city level~\cite{geolocation_problem}. Nautilus employs multiple geolocation services to improve accuracy and coverage as shown by our experiments. But despite this, there is still scope for improvement, especially at the city level, and this is a key future direction that we envision to improve the accuracy of cross-layer mapping. (\textit{ii) IP to Owner Mapping:} Telegeography has the list of owners for a subset of cables that Nautilus leverages. The accuracy of cross-layer mapping can be significantly improved if the set of ASes that own or lease a cable are known.

\parab{Special Cases} \textit{(i) Boomerang paths: } Boomerang behavior is sometimes observed where a traceroute path may traverse a submarine link and return to the same landmass using another submarine link. In such cases, Nautilus considers these as two separate links at the IP layer: one going from country A to B and the other from B to A or a nearby country C. An example of boomerang behavior would be a traceroute originating and terminating in adjoining nations within Africa but transiting via Europe for connectivity. Since Nautilus evaluates IP links rather than paths, such scenarios can be handled by the framework. 

\textit{(ii) MPLS: } Nautilus is currently oblivious to the presence of MPLS tunnels. Nautilus observes an MPLS tunnel as an IP link with endpoints at the tunnel points and attempts to identify a submarine cable potentially used by this link. However, this approach may result in some links having no mapped cables or incorrect mappings. Prior work~\cite{mpls_issues,mpls_city} showed that MPLS tunnels are present in approximately 30\% of experimental data, especially at the core of the Internet and in cities. These results also align with our observations of submarine links with endpoints in inland cities like Frankfurt that host large Internet Exchange Points (IXPs), which are most likely MPLS tunnels. Unfortunately, cross-layer mapping of private MPLS tunnels is a much harder problem. 

\parab{Improving Cross-Layer Mapping: } In addition to improving the accuracy of underlying datasets, we identify several opportunities for improving the accuracy and coverage of cross-layer mapping, particularly for submarine links. 

\textit{(i) Improving the mapping confidence for potential submarine links. } Ideally, we want to distinctly identify and map IP links that traverse a submarine cable. However, Nautilus categorizes more than 55\% of the IP links as "potential submarine links" due to the presence of a viable terrestrial path between the link endpoints. Failure-based validation technique (\S~\ref{subsec:failures}) enables us to identify potential submarine links that disappeared during failures and reclassify them as definitely submarine. Thus, the cross-layer mapping can be further improved by leveraging traceroutes collected during failures. 

\textit{(ii) Considering paths and alternative infrastructure: } Nautilus focuses solely on submarine links currently. An Internet path may traverse terrestrial or Low Earth Orbit (LEO) satellite links. An MPLS tunnel could be a combination of terrestrial, submarine, and LEO links. Considering \ashwin{sub-paths (to account for links missed due to a * in traceroutes),} end-to-end paths and all types of possible links could potentially improve the cross-layer mapping accuracy.  

\textit{(iii) Prediction of multiple cables for a single IP link: } Ideally, a single transoceanic IP link should use a single submarine cable. Nautilus currently predicts on an average 2.04 cables per IP link, with around 76\% of the links mapped to $\leq$ 2 cables and 93\% of links mapped to $\leq$ 5 cables. \ashwin{Moreover, when Nautilus predicts multiple cables for a single IP link, they often represent parallel cables. However, disagreements between data sources at different stages may result in predicted cables being associated with different regions.} While Nautilus has better accuracy and coverage compared to prior work, there is scope for improvement. As Nautilus relies on multiple important but partially complete public datasets, improving these datasets can help reduce the average number of cables predicted per IP link. 

\textit{ (iv) Considerations of proximity to landing points: } Nautilus predicts cables based on the assumption that links using a submarine cable generally have endpoints close to the cable landing points. While this assumption is valid in a large fraction of cases, scenarios such as MPLS leads to deviations. Nautilus addresses such corner cases with its recursive radius search extending up to 1000km from the landing point. More sophisticated search and mapping techniques could be devised for such links with endpoints located further inland.

\parab{Comparison with prior work } Due to a lack of standard metrics for measuring the accuracy of cross-layer mapping, we report link coverage and the number of cables predicted per link during comparison with SCN-Crit and iGDB. One may argue that the number of cables predicted per link is a weak proxy for accuracy. Hence, to further strengthen our comparison, we attempt to validate prior work against one of the submarine cable failure scenarios used in Nautilus validation, the Yemen outage of 2022. 

Since SCN-Crit analysis was limited to 63 countries and did not cover any of the failure regions we evaluated, including Yemen, validation was not feasible. However, given the drivability-based link elimination criterion, SCN-Crit will not map any intra-country links in Yemen that Nautilus mapped. More surprisingly, iGDB also fails in mapping links in this region. We surmise that this is due to the inherent limitations of iGDB, namely: (i) use of a single geolocation source, (ii) mapping geolocations to prominent cities using Thessian polygons, and (iii) attempting to generate a map based on the cables between these prominent cities. \ashwin{Interestingly, integrating Nautilus geolocation into iGDB for this validation did not yield any improvement in the results.} A detailed description of this experiment is presented in Appendix~\ref{validating_prior_work}. Note that these prior works did not validate their mapping, while Nautilus was validated through multiple experiments.

\parab{Use Cases } \ashwin{Mapping IP links to submarine cables offers valuable insights for network design and operations. It helps in failure planning and recovery, enabling rapid assessment of infrastructure impact during cable failures. Additionally, it supports basic routing optimizations, such as leveraging information on cables with lower latency for specific destinations, to make more informed routing decisions at the operator level. Augmenting Nautilus' map with terrestrial cross-layer maps can also open avenues for end-to-end Internet connectivity analysis.}

%% file: sections/conclusion.tex
\section{Conclusion}
While the submarine cable network is a critical component of the Internet infrastructure, we have a limited understanding of how higher layers of the Internet map onto these cables. We design Nautilus, a framework for cross-layer mapping on submarine cables that relies on publicly available datasets and a corpus of techniques to generate a mapping of IP links to submarine cables with a confidence score for each prediction. Nautilus generates a mapping for 81\% of all identified submarine (definite and unconfirmed) links. We validate Nautilus using a variety of techniques and demonstrate that Nautilus achieves high accuracy. By filling a critical gap in Internet cartography, Nautilus opens an avenue for research on end-to-end analysis of the Internet that stitches together both terrestrial and submarine segments.

%% file: sections/appendix.tex
\appendix

\section{Ethics}

In our research work, we have taken into consideration any potential ethical concerns and have implemented measures to ensure the responsible handling of sensitive information. While our work involves generating the cross-layer map that identifies IPs, their geolocations, and corresponding submarine cables and has been identified as sensitive information by network operators, we would like to emphasize the following points:

\paragraph{Use of Publicly Available Information:} Our research is based solely on publicly available information and data sources. We have not engaged in any activities that involve unauthorized access to private or confidential information. All data used in our study has been obtained from sources that are freely accessible to the public.

\paragraph{Prevalence of Existing Risks:} It is important to acknowledge that the risks associated with identifying IPs, geolocations, and submarine cables already exist in various domains. Malicious attacks, unauthorized access, and misuse of sensitive information are unfortunate realities in today's interconnected world. Our work does not introduce any new risks or vulnerabilities beyond what already exists.

\paragraph{Absence of Harmful Intent:} Our research work is conducted with the intention of contributing to knowledge and understanding in the field of network infrastructure. We do not have any harmful intent or malicious agenda associated with the generated information. Our aim is to provide insights that can support network management, improve decision-making processes, and enhance communication infrastructure.

\paragraph{Ethical Awareness and Reflection:} While our work may involve sensitive information, we understand the potential ethical implications and have reflected upon these throughout the research process. We have considered the possible risks, societal impact, and the need for responsible data handling. Our intention is to contribute to the knowledge base while ensuring ethical conduct.

\paragraph{Consideration of Ethical Concerns:} We encourage open dialogue and critical discussion around the ethical concerns associated with our research. We are receptive to feedback and willing to address any ethical concerns raised by the research community, reviewers, or stakeholders before making this work open-source. We are committed to continuous improvement and will take the necessary steps to rectify any inadvertent oversights.

In summary, our research work adheres to ethical principles, complies with legal requirements, and aims to contribute to knowledge without introducing any new risks. We have conducted our study with transparency and responsibility. We remain open to discussions regarding ethical considerations and are committed to upholding the highest standards of integrity and accountability in our research endeavors.

\section{Detailed workflow representation of all stages in Nautilus pipeline} ~\label{nautilus_workflow_pieces}

\begin{figure}[h]
    \centering
    \includegraphics[width=\columnwidth]{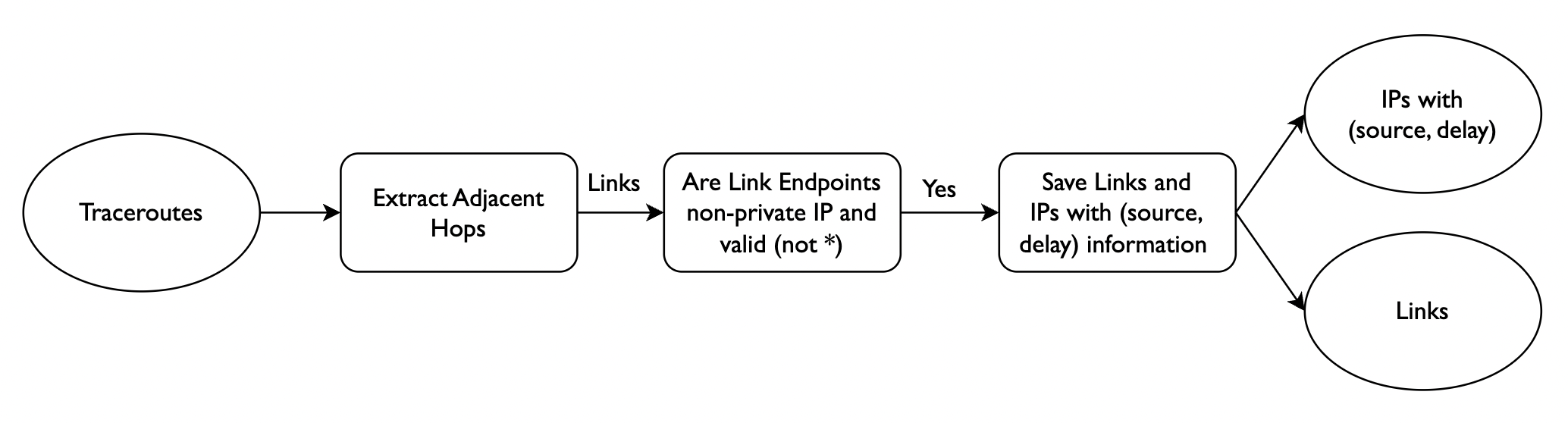}
    \caption{\ashwin{The workflow of IP and Link Extraction stage.}}
    \label{fig:ip_and_link_extraction}
\end{figure}

\begin{figure}[h]
    \centering
    \includegraphics[width=0.75\columnwidth]{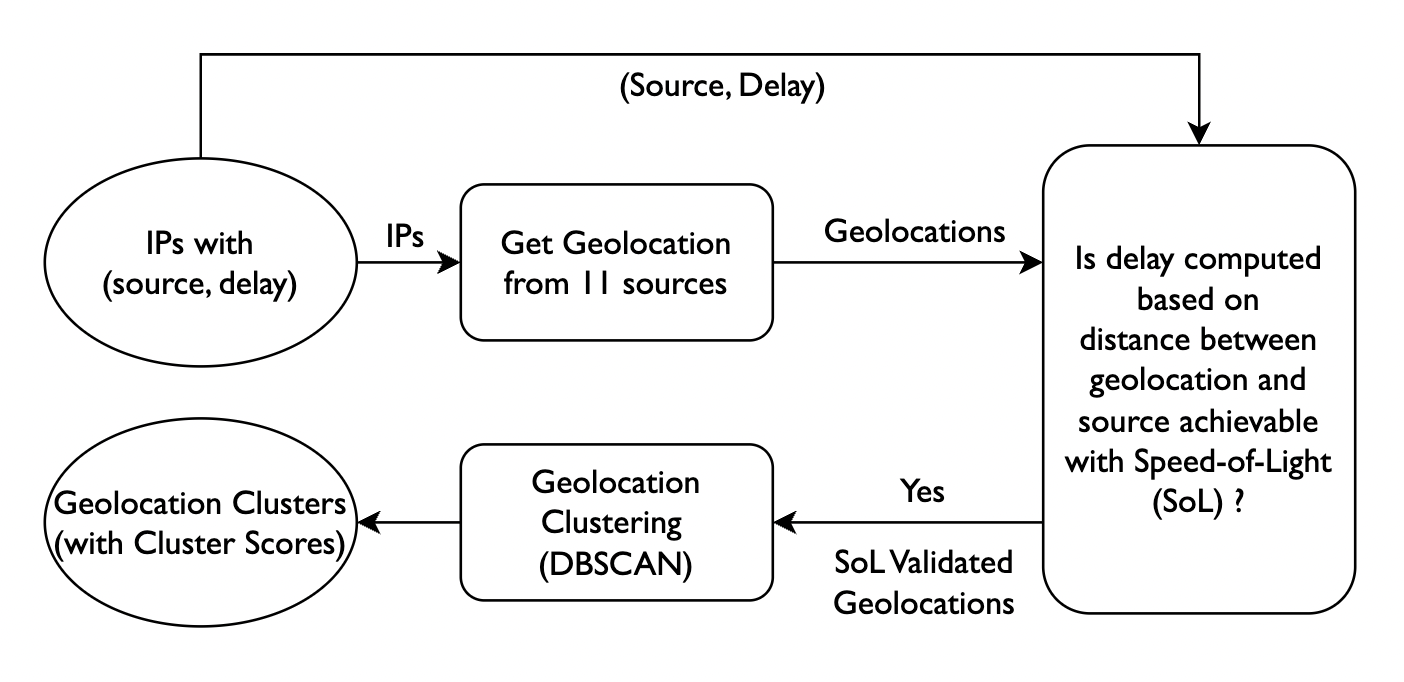}
    \caption{\ashwin{The workflow of generating geolocation of IPs and performing SoL validation.}}
    \label{fig:geolocation_and_sol_validation}
\end{figure}

\begin{figure}[h]
    \centering
    \includegraphics[width=\columnwidth]{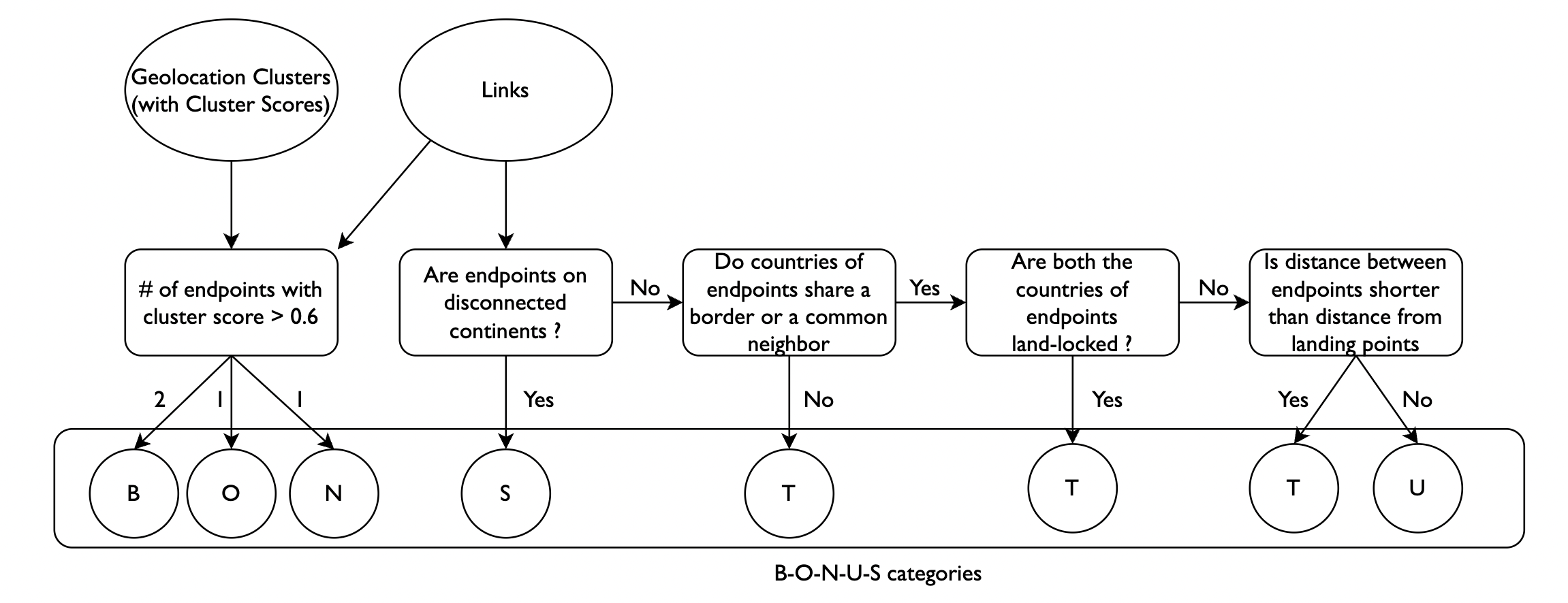}
    \caption{\ashwin{The workflow of B-O-N-U-S classification.}}
    \label{fig:bonus_classification_workflow}
\end{figure}

\begin{figure}
    \centering
    \includegraphics[width=0.7\columnwidth]{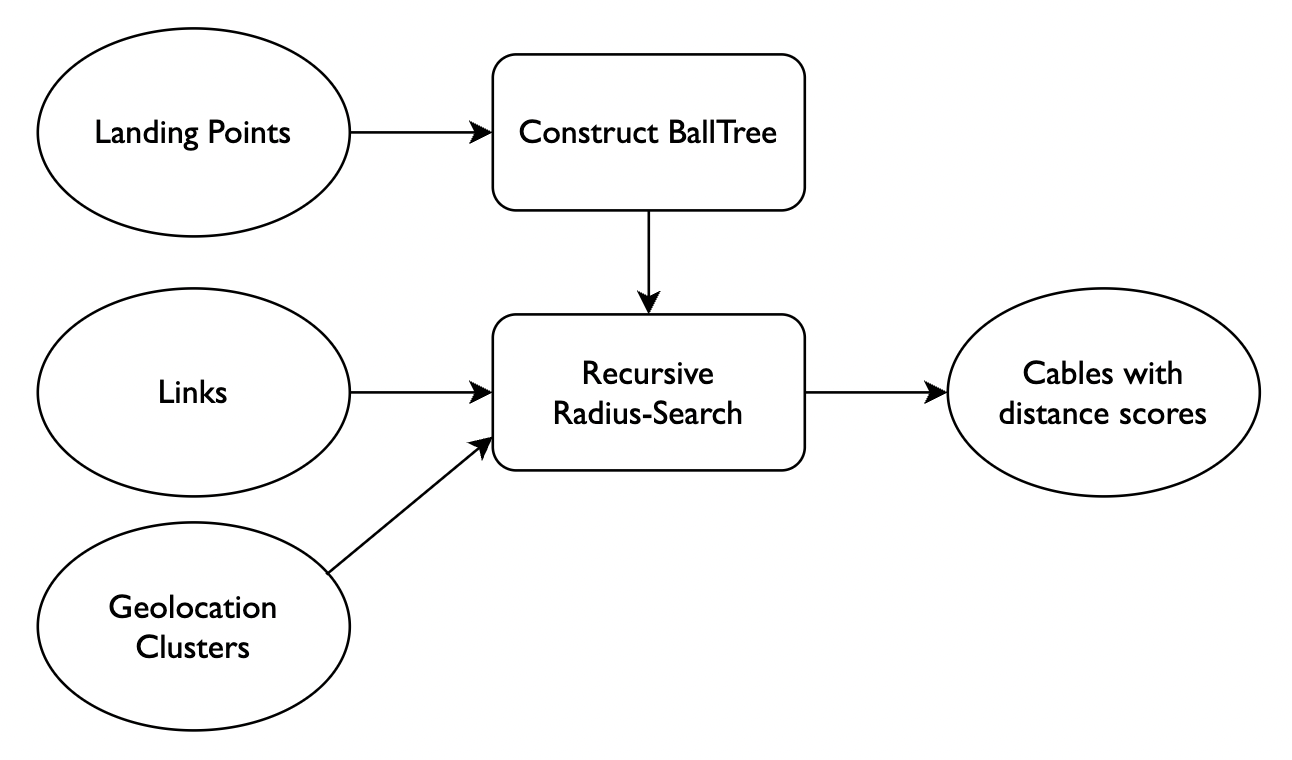}
    \caption{\ashwin{The workflow of geolocation-based cable mapping.}}
    \label{fig:geolocation_cable_mapping}
\end{figure}

\begin{figure}
    \centering
    \includegraphics[width=\columnwidth]{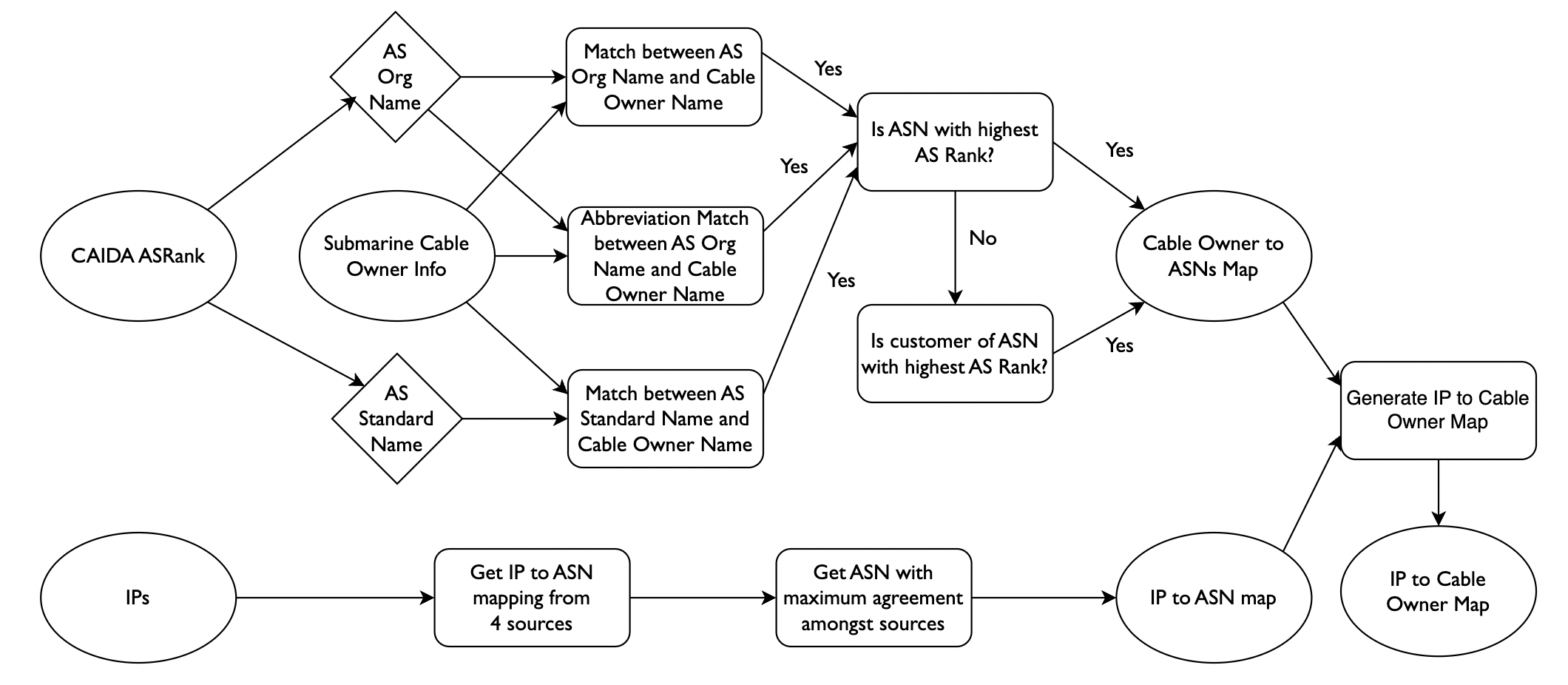}
    \caption{\ashwin{The workflow of IP to Cable Owner Mapping. For the processing steps with unlisted 'No', it indicates that those ASNs are dropped.}}
    \label{fig:ip_to_cable_owner_mapping}
\end{figure}

\begin{figure}
    \centering
    \includegraphics[width=0.7\columnwidth]{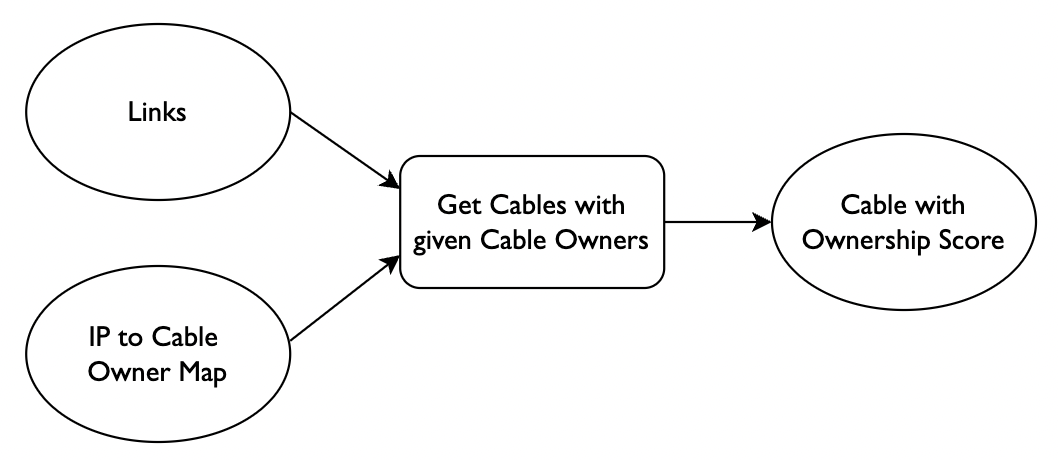}
    \caption{\ashwin{The workflow of ownership-based cable mapping.}}
    \label{fig:owner_cable_mapping}
\end{figure}

\begin{figure}
    \centering
    \includegraphics[width=\columnwidth]{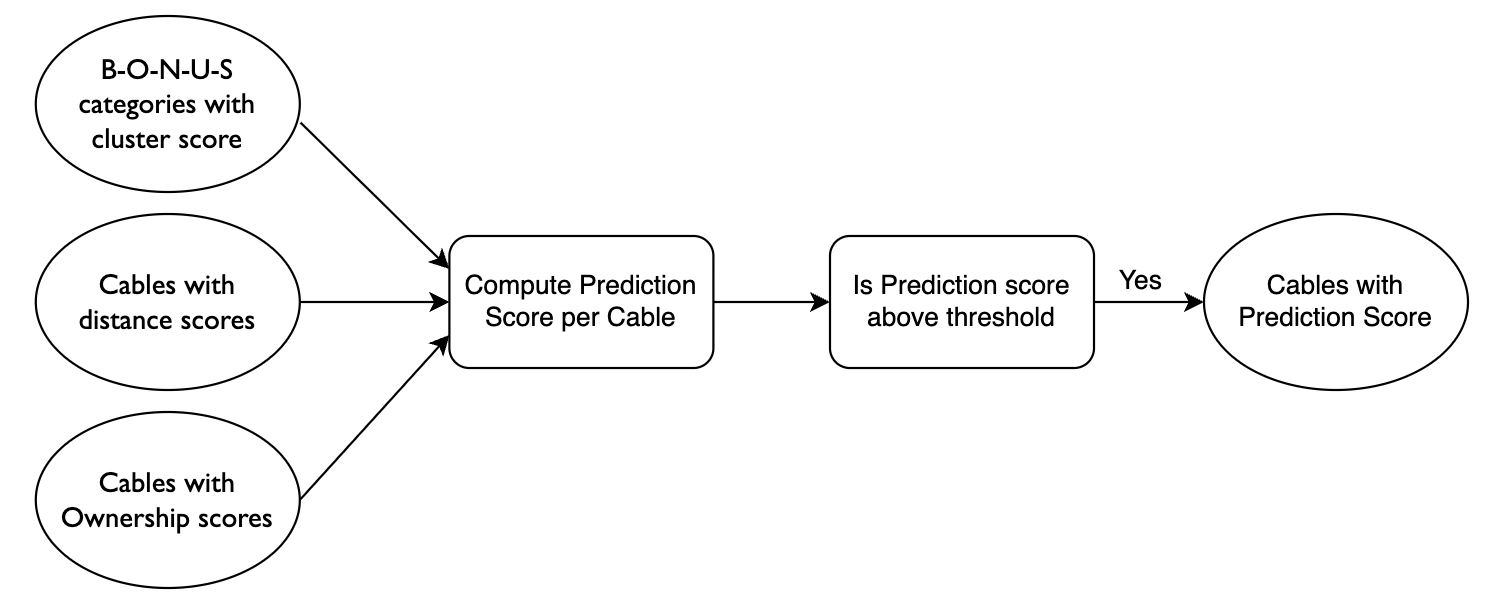}
    \caption{\ashwin{The workflow of aggregation and final mapping.}}
    \label{fig:final_mapping}
\end{figure}

\ashwin{Figures~\ref{fig:ip_and_link_extraction},~\ref{fig:geolocation_and_sol_validation},~\ref{fig:bonus_classification_workflow},~\ref{fig:geolocation_cable_mapping},~\ref{fig:ip_to_cable_owner_mapping},~\ref{fig:owner_cable_mapping},~\ref{fig:final_mapping} correspond to the 7 stages of the Nautilus pipeline in sequential order.}

\vspace{10mm}

\vspace{20mm}

\vspace{10mm}

\vspace{10mm}

\vspace{20mm}
\section{Clustering in Multi-Source Geolocation Aggregation} ~\label{need_for_all_clusters_retention}

\ashwin{The outputs from the Multi-Source Geolocation Aggregation step are clusters that meet the SoL constraints. While one might argue for selecting the cluster with the higher cluster score, we illustrate the need to maintain all clusters for downstream processing.}

\ashwin{Imagine we are geolocating an IP link endpoint using multiple sources, with 5 sources producing valid geolocation outcomes. Of these, 3 sources agree on location A, and 2 sources agree on location B. We will call the cable near location A Cable-A and the one near location B Cable-B. These cables have cluster scores of 0.6 and 0.4, respectively.}

\ashwin{In some downstream modules, such as cable owner matching, Cable-B might receive a higher score. For example, Cable-A may have a cable owner score of 0, while it is 1 for Cable-B, with all other scores (indicated by Y) being the same. In this case, the final prediction scores for both cables will be:}

$$Cable-A: (0.5 * 0.6 + 0.1 * 0) + Y = 0.3 + Y$$
$$Cable-B: (0.5 * 0.4 + 0.1 * 1) + Y = 0.3 + Y$$

\ashwin{Now, Cable-A and Cable-B are equally likely choices for mapping the IP link. Changes to scores from other modules could favor Cable-B over Cable-A. To address such scenarios, it is essential to maintain all clusters for further analysis and decision-making.}

\section{Understanding the need for unconfirmed submarine category} ~\label{unconfirmed_submarine_explanation}

\begin{figure}[ht]
    \centering
    \begin{subfigure}{0.475\textwidth}
        \centering
        \includegraphics[width=\textwidth]{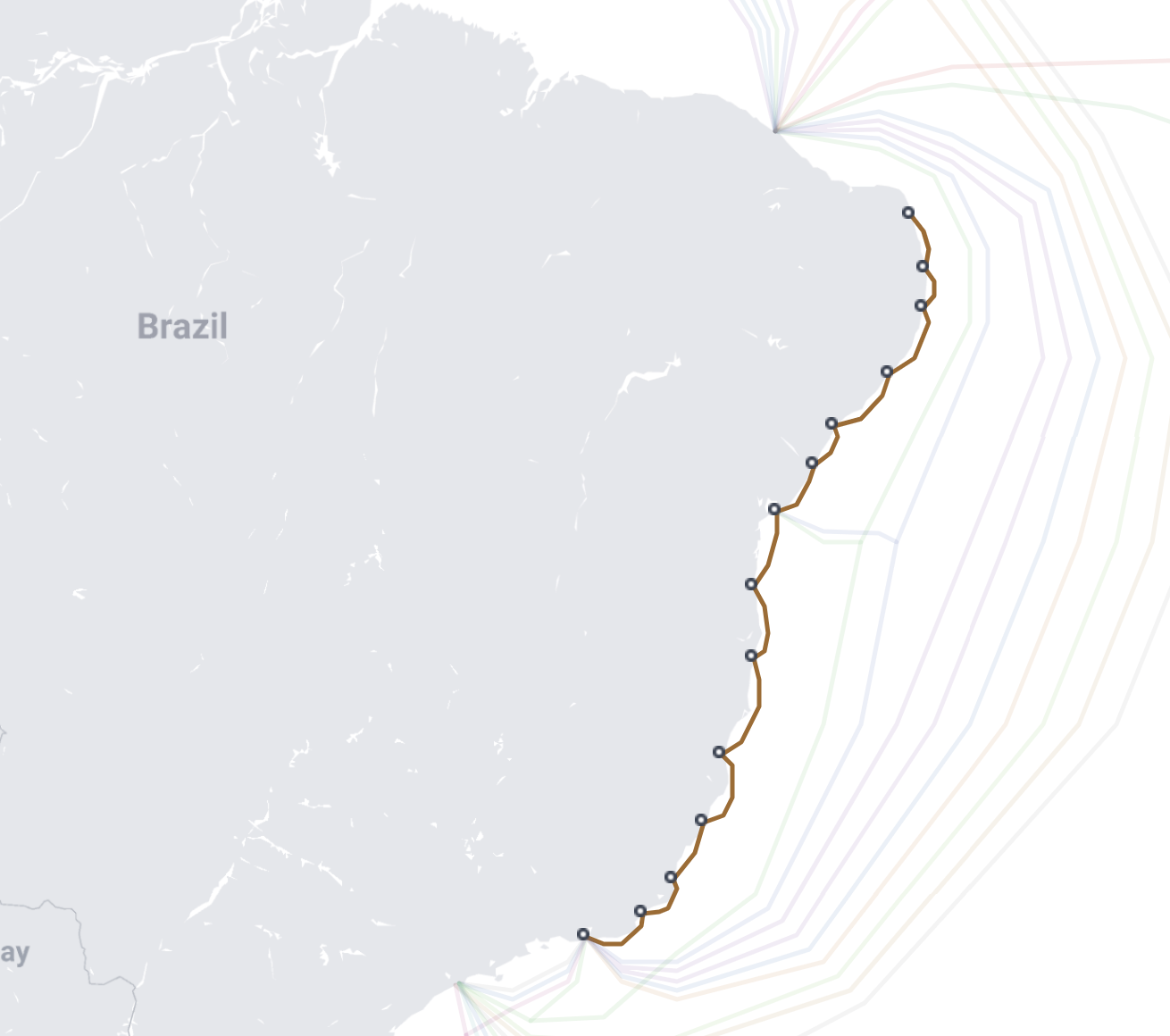}
        \caption[]%
        {{\small Brazilian Festoon is an entirely domestic submarine cable that connects various locations in Brazil's coastline (Image Source: Telegeography~\cite{tele_website})}}
    \end{subfigure}
    \hfill
    \begin{subfigure}{0.475\textwidth}
        \centering
        \includegraphics[width=\textwidth]{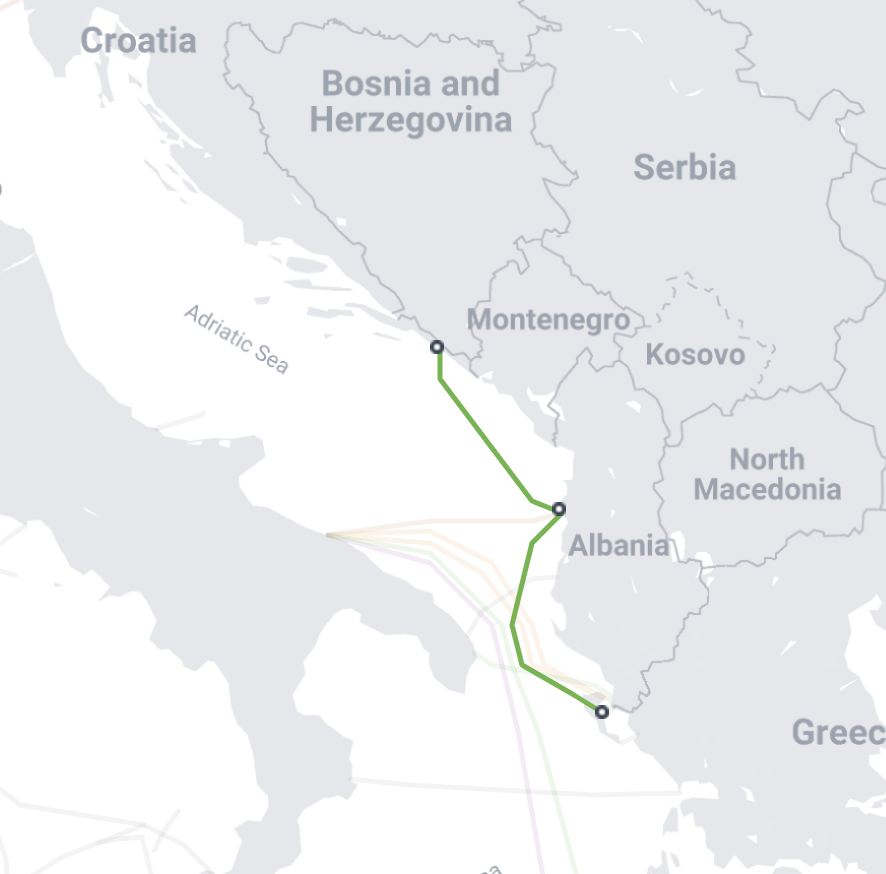}
        \caption[]%
        {{\small Adria-1 is a cable that connects Albania, Croatia, and Greece. (Image Source: Telegeography~\cite{tele_website})}}
    \end{subfigure}
    \caption{Examples of submarine cables where potentially a terrestrial cable could be used to connect the landing points in the cable }
    \label{potential_land_submarine_cables}
\end{figure}

Submarine cables tend to be viewed as long-haul cables that connect disjoint (non-connected) landmasses in the world. But it is important to note that this is not always the case. There are many domestic and small inter-country submarine cables and Figure~\ref{potential_land_submarine_cables} shows examples of 2 such submarine cables. As a terrestrial cable could also cover the path taken by these submarine cables, to account for the possibility that it could be a submarine cable, Nautilus generates a category called Unconfirmed submarine. This category thus captures such submarine cables and assigns a lower prediction score as there is still a potential for a terrestrial cable along that path.

\section{Examples from the Manual Validation for Cable Owner to AS Mappings} ~\label{cable_owner_to_as_examples}

\ashwin{The cable owner to AS mapping process involves a search across three dimensions and utilizes a pruning mechanism to select the ASN with the highest AS rank and its associated customers from among the matches. For instance, if the search yields 10 matches, Nautilus chooses the AS with the highest AS Rank as the primary match. Subsequently, the matches that are not customers of the primary choice are eliminated from the remaining nine matches.}

\ashwin{To validate this mapping approach, we manually examined mappings for several cable owners. Here are a few examples: For cable owner "KT," potential matches include Korea Telecom and Kazakh Telecom. As Korea Telecom has a higher AS rank and Kazakh Telecom is not a customer of Korea Telecom, only Korea Telecom is considered a valid match. Similarly, for cable owner "Orange," potential matches encompass Orange, Orange Cote d'Ivoire, and Orange Mali, among others. Orange is selected as the primary choice due to its AS rank of 14, and since Orange Cote d'Ivoire and Orange Mali are customers of Orange, they are also recognized as valid choices. This approach effectively handles cases where the submarine cable owner or AS name varies in regional segments or due to internal operations.}

\section{Parallel Cables} \label{parallel_cables_analysis}

\begin{figure}
    \centering
    \includegraphics[width=0.8\columnwidth]{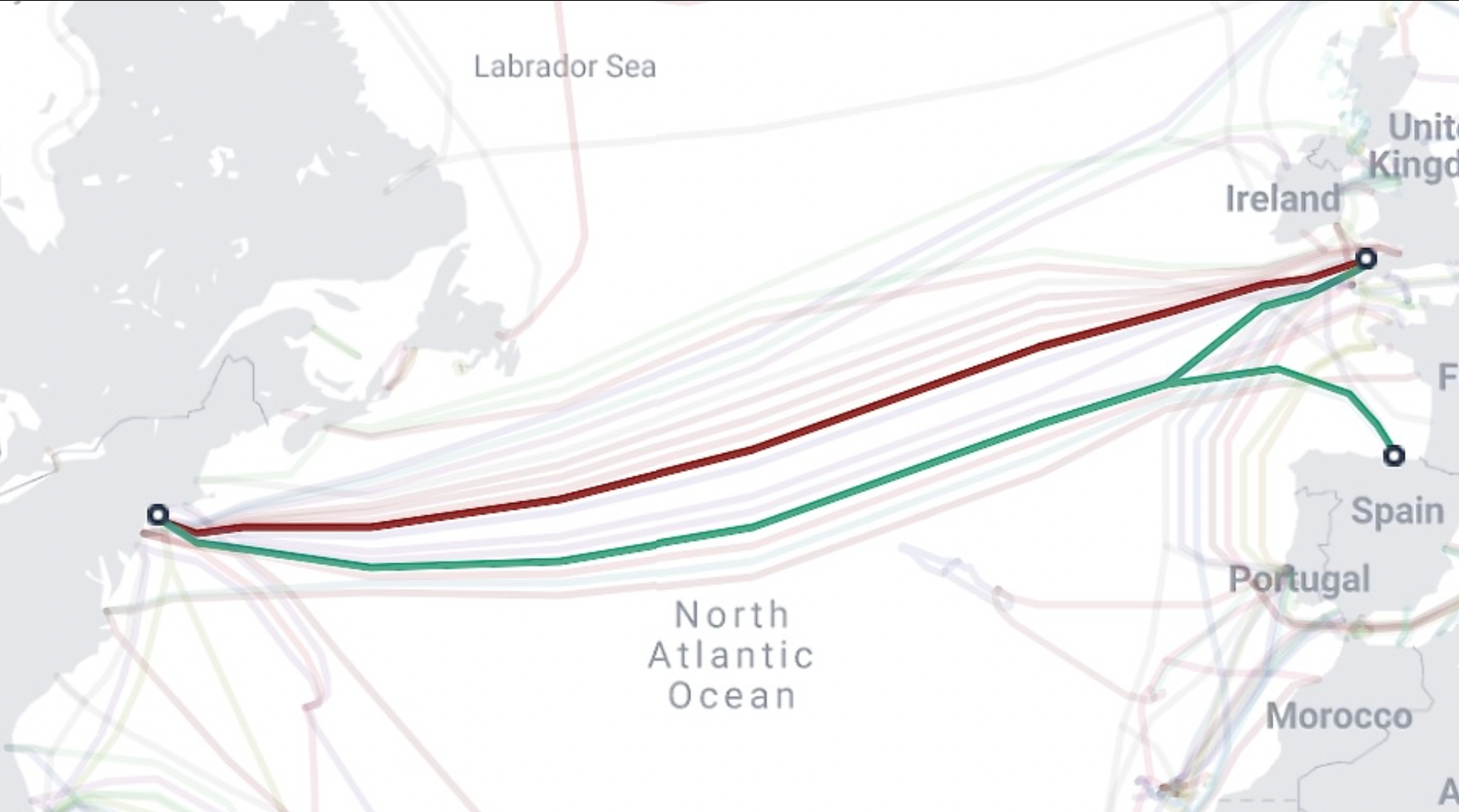}
    \caption{An example of a parallel cable - Yellow and Grace Hopper - which are essentially parallel for the path between Bude, UK, and Bellport, NY, US. While these cables share the exact same landing points on both ends, to be a parallel cable, the landing points on both ends need to be just proximal to each other (Image Source: Telegeography~\cite{tele_website})}
    \label{fig:parallel_cables}
\end{figure}

We define parallel cables as cables whose landing points on either end are in close proximity to each other. As cables tend to be on the shortest path that avoids any obstacles or hazards like undersea mountains or regions with high seismic activity, in regions with high demand and capacity requirements, cables tend to be laid in parallel. An example of a parallel cable is shown in Figure~\ref{fig:parallel_cables}. Hence to identify the exact cable a particular IP link takes, Nautilus relies on the idea that links of an AS are more likely to be on a submarine cable that the same AS owns. Using this, Nautilus is able to significantly reduce the number of cable predictions it makes finally.

\begin{table}[ht]
  \centering
  \begin{tabularx}{\linewidth}{p{3cm} p{4cm} X} %
    \toprule
    \textbf{Category} & \textbf{Source} & \textbf{Description} \\
    \midrule
    Geolocation & RIPE IPMap & RIPE geolocation service generated by using active measurement and crowd-sourced data\\
                & CAIDA geolocation & CAIDA geolocation collected using Hoiho, IXP information\\
                & Maxmind & MaxMind's geolocation database \\
                & IP2Location & IP2Location geolocation service using custom geolocation mechanism\\
                & IPinfo & IPinfo geolocation service using custom geolocation mechanism\\
                & DB-IP & DB-IP geolocation database using custom geolocation mechanism\\
                & IPregistry & IPregistry geolocation service using custom geolocation mechanism\\
                & IPGeolocation & IPGeolocation geolocation service using custom geolocation mechanism\\
                & IPapi & IPapi geolocation service using custom geolocation mechanism\\
                & IPapi.co & IPapi.co geolocation service using custom geolocation mechanism\\
                & IPdata & IPdata geolocation service using custom geolocation mechanism\\
    \midrule
    IP to ASN & RADB servers & Routing Assets Database (RADB) servers \\
              & Routinator RPKI validator & Routinator RPKI validator by NLnet labs\\
              & Cymru WhoIS & Team Cymru WhoIS service based on cymru.whois servers\\
              & CAIDA AS2Org & CAIDA AS2Org using WHOIS and PCH\\
    \midrule
    Traceroute & RIPE 5051 & RIPE Atlas traceroute measurement 5051\\
              & RIPE 5151 & RIPE Atlas traceroute measurement 5151\\
              & RIPE 6052 & RIPE Atlas traceroute measurement 6052\\
              & RIPE 6152 & RIPE Atlas traceroute measurement 6152\\
              & CAIDA /24 & CAIDA traceroute /24 probing \\
              & CAIDA /48 & CAIDA traceroute /48 probing \\
    \midrule
    AS Information & CAIDA ASRank & CAIDA ASRank derived from Ark, RouteViews, and RIPE NCC\\
    \midrule
    IXP & PeeringDB & PeeringDB Internet Exchange Point (IXP) information \\
    \midrule
    Submarine Cable & Telegeography & Telegeography's submarine cable data source \\
    \bottomrule
  \end{tabularx}
  \caption{\ashwin{List of data sources used by Nautilus}}
  \label{nautilus_data_sources}
\end{table}

\section{SoL validation performance analysis} \label{sol_validation_analysis}

\begin{figure}
    \centering
    \includegraphics[width=0.7\columnwidth]{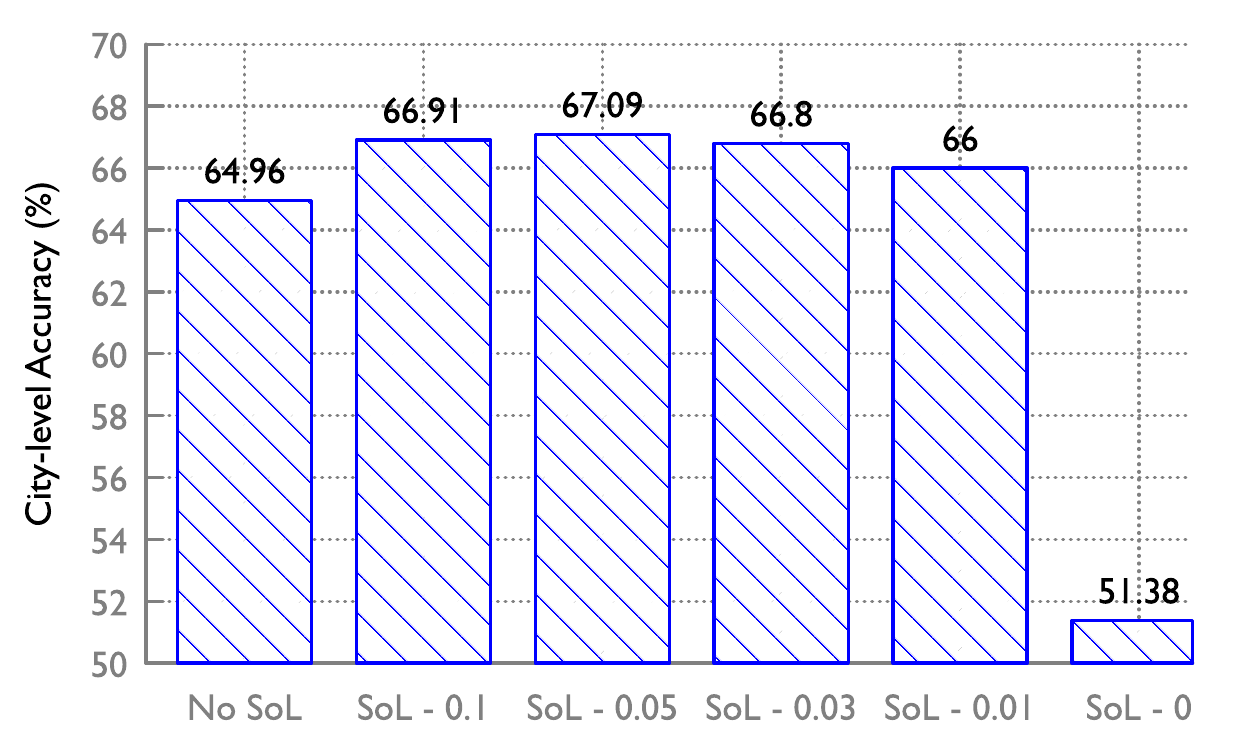}
    \caption{Comparison of normalized geolocation accuracy without SoL and with different SoL thresholds. SoL validation with a 0.05 threshold gives the best-normalized geolocation accuracy}
    \label{fig:sol_validation}
\end{figure}

To analyze the gains with SoL validation, we again utilize the ground truth data generated by Gharaibeh et. al.~\cite{geolocation_problem} and normalize the results based on the total number of IPs. Figure~\ref{fig:sol_validation} contains the results of this experiment. As observed from this figure, though with SoL validation, some fraction of IPs tend to have no valid geolocation output, the overall geolocation fidelity improves up to 2\%. This validates the overall performance gains with SoL validation. Additionally, to identify the best SoL threshold, we examine the normalized geolocation accuracy and observe that the 0.05 threshold has the best performance. With a lower SoL threshold, we impose a harder constraint for the margin of error with traceroute delay annotations, and by increasing the threshold, we loosen this constraint. As observed with a threshold of 0, the accuracy drops drastically due to more geolocation sources not producing output for a given IP. On the other hand with a threshold of 0.1, we start observing a decrease in normalized geolocation accuracy as more IPs might have erroneous geolocation.

\section{Cable Failures} \label{cable_failures_appendix}

\begin{figure}
    \centering
    \includegraphics[width=0.8\columnwidth]{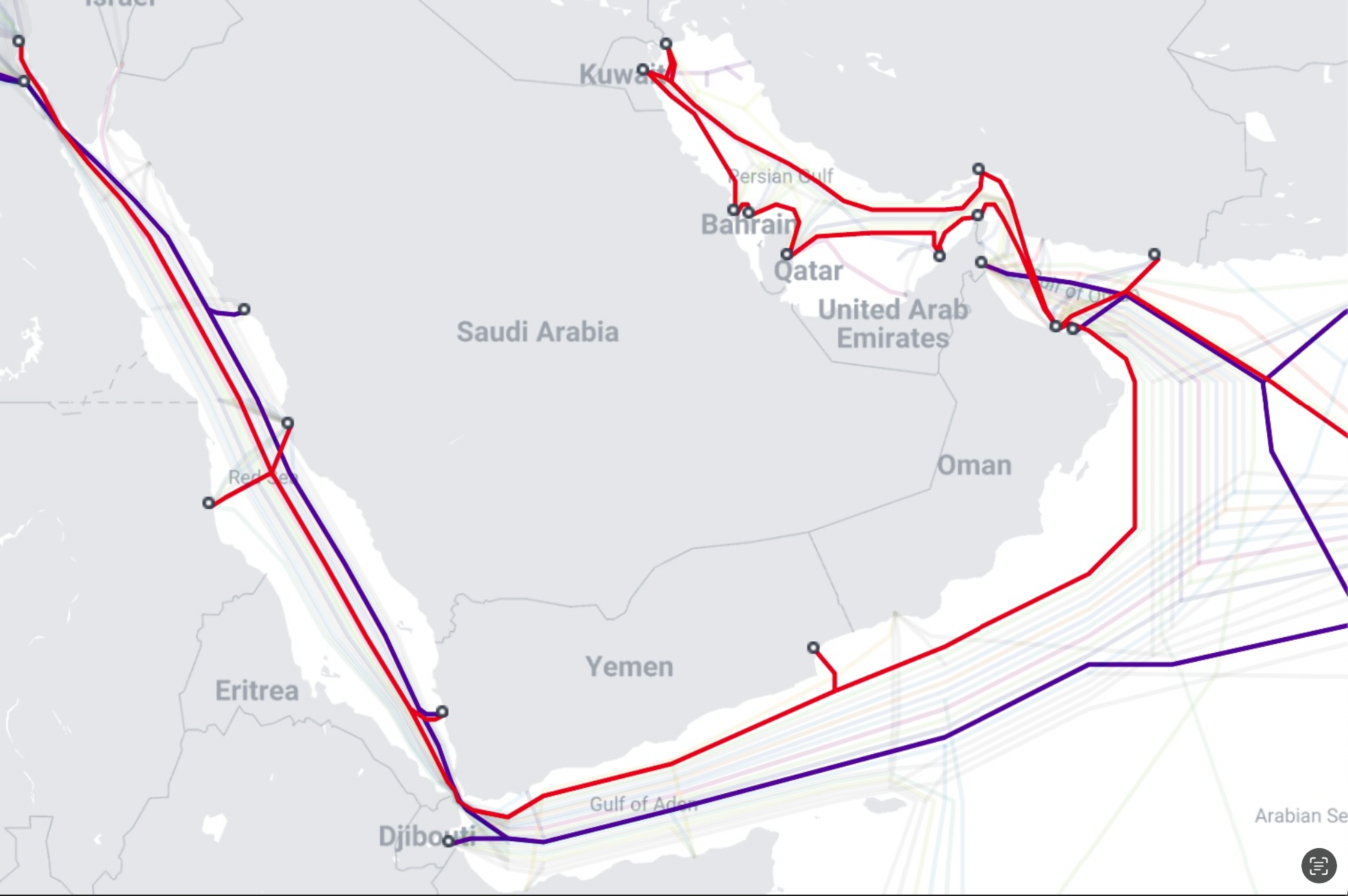}
    \caption{The 2 cables (FALCON shown in red and SeaMeWe-5 shown in blue) with a landing point in Al Hudaydah which was damaged by airstrike (Image Source: Telegeography~\cite{tele_website})}
    \label{fig:yemen_cables}
\end{figure}

\begin{figure}
    \centering
    \includegraphics[width=0.8\columnwidth]{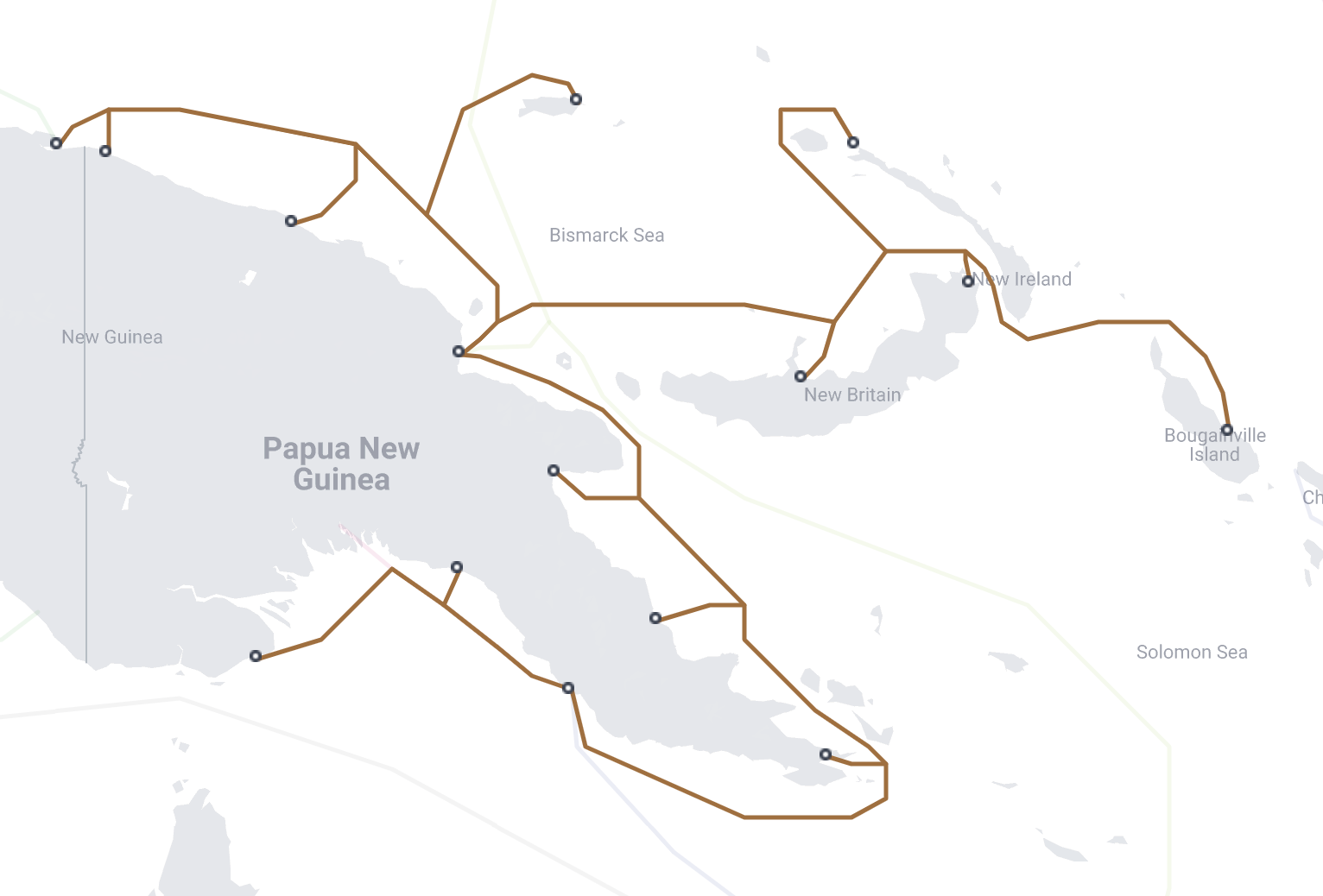}
    \caption{The Kumul domestic cable in Papua New Guinea that was damaged by an earthquake in the region near Madang (Image Source: Telegeography~\cite{tele_website})}
    \label{fig:kdcn}
\end{figure}

Figures~\ref{fig:yemen_cables} and ~\ref{fig:kdcn} show the submarine cables that experienced an outage during the Yemen outage and Papua New Guinea earthquake respectively. 

\section{Validating prior work using submarine cable failures} \label{validating_prior_work}

In this section, we present a comprehensive evaluation of iGDB's ability to identify mapping changes observed during submarine cable failures. As part of this assessment, we examined iGDB's performance against the Yemen failure, a validation experiment conducted for Nautilus. The findings underscore the limitations of iGDB for submarine cable mapping, as it failed to generate any mapping for the damaged cable segment, highlighting its inapplicability in this context.

For the evaluation, we used CAIDA's geolocation as a proxy for iGDB, as one of its services employs a DNS-based inference for geolocation due to the absence of geolocation scripts in iGDB's project repository. This approach provided us with 45 Yemen IPs. Upon excluding IPs geolocated by Maxmind (non-DNS-based method), we were left with 19 IPs. Standardizing these geolocations to the nearest city using iGDB's method yielded 5 IPs mapped to Al Hudaydah in Yemen, from which a total of 21 links were derived. However, out of these 21 links, 15 links lacked geolocation for the other endpoint (not in Al Hudaydah), leaving us with only 6 links where both ends had geolocation, and at least one end was in Al Hudaydah.

Upon examining the city pair combinations for these 6 links, we observed that the other endpoint locations were in Great Britain, the US, or China. Unfortunately, no cables connected these exact standardized locations, as they were not port cities (for instance, London). Conversely, Nautilus successfully tackled these challenges by utilizing multiple geolocation services, enabling a larger number of IPs to be accurately geolocated. Additionally, Nautilus conducted searches for cables beyond the immediate location, making diligent efforts to find suitable cable matches.

Importantly, even when we replaced the DNS-based geolocation with the best geolocation results from Nautilus, we encountered a similar outcome, reaffirming the concerns regarding iGDB's suitability for submarine cable mapping. These comparative results further highlight the strengths of Nautilus in the context of submarine cable mapping, showcasing its effectiveness in overcoming geolocation-related challenges and providing more accurate and reliable mapping outcomes.

%% file: main.bbl

\begin{thebibliography}{92}


\ifx \showCODEN    \undefined \def \showCODEN     #1{\unskip}     \fi
\ifx \showDOI      \undefined \def \showDOI       #1{#1}\fi
\ifx \showISBNx    \undefined \def \showISBNx     #1{\unskip}     \fi
\ifx \showISBNxiii \undefined \def \showISBNxiii  #1{\unskip}     \fi
\ifx \showISSN     \undefined \def \showISSN      #1{\unskip}     \fi
\ifx \showLCCN     \undefined \def \showLCCN      #1{\unskip}     \fi
\ifx \shownote     \undefined \def \shownote      #1{#1}          \fi
\ifx \showarticletitle \undefined \def \showarticletitle #1{#1}   \fi
\ifx \showURL      \undefined \def \showURL       {\relax}        \fi
\providecommand\bibfield[2]{#2}
\providecommand\bibinfo[2]{#2}
\providecommand\natexlab[1]{#1}
\providecommand\showeprint[2][]{arXiv:#2}

\bibitem[\protect\citeauthoryear{??}{ton}{[n. d.]a}]%
        {tonga_cable_issue}
 \bibinfo{year}{[n. d.]}\natexlab{a}.
\newblock \bibinfo{title}{{80km Stretch of Tonga Cable In Pieces From Eruption - Submarine Telecoms Forum}}.
\newblock \bibinfo{howpublished}{\url{https://subtelforum.com/80km-stretch-of-tonga-cable-in-pieces-from-eruption/}}.   (\bibinfo{year}{[n. d.]}).
\newblock


\bibitem[\protect\citeauthoryear{??}{99_}{[n. d.]}]%
        {99_traffic}
 \bibinfo{year}{[n. d.]}\natexlab{}.
\newblock \bibinfo{title}{{A map of all the underwater cables that connect the internet - Vox}}.
\newblock \bibinfo{howpublished}{\url{https://www.vox.com/2015/3/13/8204655/submarine-cables-internet}}.   (\bibinfo{year}{[n. d.]}).
\newblock


\bibitem[\protect\citeauthoryear{??}{mul}{[n. d.]}]%
        {multiple_providers_aae1_outage}
 \bibinfo{year}{[n. d.]}\natexlab{}.
\newblock \bibinfo{title}{{AAE-1 cable cut causes widespread outages in Europe, East Africa, Middle East, and South Asia - DCD}}.
\newblock \bibinfo{howpublished}{\url{https://www.datacenterdynamics.com/en/news/aae-1-cable-cut-causes-widespread-outages-in-europe-east-africa-middle-east-and-south-asia/}}.   (\bibinfo{year}{[n. d.]}).
\newblock


\bibitem[\protect\citeauthoryear{??}{ASR}{[n. d.]}]%
        {ASRank}
 \bibinfo{year}{[n. d.]}\natexlab{}.
\newblock \bibinfo{title}{{AS Rank: A ranking of the largest Autonomous Systems (AS) in the Internet.}}
\newblock \bibinfo{howpublished}{\url{https://asrank.caida.org/}}.   (\bibinfo{year}{[n. d.]}).
\newblock


\bibitem[\protect\citeauthoryear{??}{cai}{[n. d.]}]%
        {caida_as_to_org}
 \bibinfo{year}{[n. d.]}\natexlab{}.
\newblock \bibinfo{title}{{AS2org API Doc}}.
\newblock \bibinfo{howpublished}{\url{https://api.data.caida.org/as2org/v1/doc}}.   (\bibinfo{year}{[n. d.]}).
\newblock


\bibitem[\protect\citeauthoryear{??}{cab}{[n. d.]a}]%
        {cable-security}
 \bibinfo{year}{[n. d.]}\natexlab{a}.
\newblock \bibinfo{title}{{Beyond Triple Invisibility: Do Submarine Data Cables Require Better Security? | IPI Global Observatory}}.
\newblock \bibinfo{howpublished}{\url{https://theglobalobservatory.org/2021/09/beyond-triple-invisibility-do-submarine-data-cables-require-better-security/}}.   (\bibinfo{year}{[n. d.]}).
\newblock
\newblock
\shownote{(Accessed on 02/11/2023).}


\bibitem[\protect\citeauthoryear{??}{CAI}{[n. d.]}]%
        {CAIDA}
 \bibinfo{year}{[n. d.]}\natexlab{}.
\newblock \bibinfo{title}{{CAIDA}}.
\newblock \bibinfo{howpublished}{\url{https://www.caida.org/}}.   (\bibinfo{year}{[n. d.]}).
\newblock


\bibitem[\protect\citeauthoryear{??}{ipi}{[n. d.]}]%
        {ipinfo}
 \bibinfo{year}{[n. d.]}\natexlab{}.
\newblock \bibinfo{title}{{Comprehensive IP address data, IP geolocation API and database - IPinfo.io}}.
\newblock \bibinfo{howpublished}{\url{https://ipinfo.io/}}.   (\bibinfo{year}{[n. d.]}).
\newblock


\bibitem[\protect\citeauthoryear{??}{cab}{[n. d.]b}]%
        {cable_failures_1}
 \bibinfo{year}{[n. d.]}\natexlab{b}.
\newblock \bibinfo{title}{{Coral Sea Cable System Shunt Fault Repair Rescheduled - Submarine Telecoms Forum}}.
\newblock \bibinfo{howpublished}{\url{https://subtelforum.com/coral-sea-cable-system-shunt-fault-repair-rescheduled/}}.   (\bibinfo{year}{[n. d.]}).
\newblock


\bibitem[\protect\citeauthoryear{??}{cab}{[n. d.]c}]%
        {cable_failures_3}
 \bibinfo{year}{[n. d.]}\natexlab{c}.
\newblock \bibinfo{title}{{Fault Reported on Sea-MeWe-4 - Submarine Telecoms Forum}}.
\newblock \bibinfo{howpublished}{\url{https://subtelforum.com/fault-reported-on-sea-mewe-4/}}.   (\bibinfo{year}{[n. d.]}).
\newblock


\bibitem[\protect\citeauthoryear{??}{ipg}{[n. d.]}]%
        {ipgeolocation}
 \bibinfo{year}{[n. d.]}\natexlab{}.
\newblock \bibinfo{title}{{Free IP Geolocation API and Accurate IP Geolocation Database}}.
\newblock \bibinfo{howpublished}{\url{https://ipgeolocation.io/}}.   (\bibinfo{year}{[n. d.]}).
\newblock


\bibitem[\protect\citeauthoryear{??}{sub}{[n. d.]a}]%
        {submarine_cable_map}
 \bibinfo{year}{[n. d.]}\natexlab{a}.
\newblock \bibinfo{title}{{GitHub - telegeography/www.submarinecablemap.com: Comprehensive interactive map of the world's major operating and planned submarine cable systems and landing stations, updated frequently.}}
\newblock \bibinfo{howpublished}{\url{https://github.com/telegeography/www.submarinecablemap.com}}.   (\bibinfo{year}{[n. d.]}).
\newblock


\bibitem[\protect\citeauthoryear{??}{cab}{[n. d.]d}]%
        {cable_failures_5}
 \bibinfo{year}{[n. d.]}\natexlab{d}.
\newblock \bibinfo{title}{{Internet Outage in Indonesia Due to Submarine Cable Cut - SubTel Forum}}.
\newblock \bibinfo{howpublished}{\url{https://subtelforum.com/internet-outage-in-indonesia-due-to-submarine-cable-cut/}}.   (\bibinfo{year}{[n. d.]}).
\newblock


\bibitem[\protect\citeauthoryear{??}{ip2}{[n. d.]}]%
        {ip2location}
 \bibinfo{year}{[n. d.]}\natexlab{}.
\newblock \bibinfo{title}{{IP Address to IP Location and Proxy Information | IP2Location}}.
\newblock \bibinfo{howpublished}{\url{https://www.ip2location.com/}}.   (\bibinfo{year}{[n. d.]}).
\newblock


\bibitem[\protect\citeauthoryear{??}{max}{[n. d.]}]%
        {maxmind}
 \bibinfo{year}{[n. d.]}\natexlab{}.
\newblock \bibinfo{title}{{IP Geolocation and Online Fraud Prevention | MaxMind}}.
\newblock \bibinfo{howpublished}{\url{https://www.maxmind.com/en/home}}.   (\bibinfo{year}{[n. d.]}).
\newblock


\bibitem[\protect\citeauthoryear{??}{ipd}{[n. d.]}]%
        {ipdata}
 \bibinfo{year}{[n. d.]}\natexlab{}.
\newblock \bibinfo{title}{{IP Geolocation API | 20B+ Requests Served - ipdata}}.
\newblock \bibinfo{howpublished}{\url{https://ipdata.co/}}.   (\bibinfo{year}{[n. d.]}).
\newblock


\bibitem[\protect\citeauthoryear{??}{db-}{[n. d.]}]%
        {db-ip}
 \bibinfo{year}{[n. d.]}\natexlab{}.
\newblock \bibinfo{title}{{IP Geolocation API \& Free Address Database | DB-IP}}.
\newblock \bibinfo{howpublished}{\url{https://db-ip.com/}}.   (\bibinfo{year}{[n. d.]}).
\newblock


\bibitem[\protect\citeauthoryear{??}{cym}{[n. d.]}]%
        {cymru_whois}
 \bibinfo{year}{[n. d.]}\natexlab{}.
\newblock \bibinfo{title}{{IP to ASN mapping - Team Cymru}}.
\newblock \bibinfo{howpublished}{\url{https://team-cymru.com/community-services/ip-asn-mapping/\#whois}}.   (\bibinfo{year}{[n. d.]}).
\newblock


\bibitem[\protect\citeauthoryear{??}{ipa}{[n. d.]a}]%
        {ipapi}
 \bibinfo{year}{[n. d.]}\natexlab{a}.
\newblock \bibinfo{title}{{ipapi - IP Address Lookup and Geolocation API}}.
\newblock \bibinfo{howpublished}{\url{https://ipapi.com/}}.   (\bibinfo{year}{[n. d.]}).
\newblock


\bibitem[\protect\citeauthoryear{??}{ipa}{[n. d.]b}]%
        {ipapi_co}
 \bibinfo{year}{[n. d.]}\natexlab{b}.
\newblock \bibinfo{title}{{ipapi - IP Address Lookup and Geolocation API | No SignUp}}.
\newblock \bibinfo{howpublished}{\url{https://ipapi.co/}}.   (\bibinfo{year}{[n. d.]}).
\newblock


\bibitem[\protect\citeauthoryear{??}{tat}{[n. d.]}]%
        {tata_map}
 \bibinfo{year}{[n. d.]}\natexlab{}.
\newblock \bibinfo{title}{Map | Tata Communications}.
\newblock \bibinfo{howpublished}{\url{https://www.tatacommunications.com/map/}}.   (\bibinfo{year}{[n. d.]}).
\newblock


\bibitem[\protect\citeauthoryear{??}{rip}{[n. d.]a}]%
        {ripe_atlas}
 \bibinfo{year}{[n. d.]}\natexlab{a}.
\newblock \bibinfo{title}{{Measurements | RIPE Atlas}}.
\newblock \bibinfo{howpublished}{\url{https://atlas.ripe.net/measurements/\#tab-traceroute}}.   (\bibinfo{year}{[n. d.]}).
\newblock


\bibitem[\protect\citeauthoryear{??}{cog}{[n. d.]}]%
        {cogent_network_map}
 \bibinfo{year}{[n. d.]}\natexlab{}.
\newblock \bibinfo{title}{{Network Map}}.
\newblock \bibinfo{howpublished}{\url{https://www.cogentco.com/en/network/network-map}}.   (\bibinfo{year}{[n. d.]}).
\newblock


\bibitem[\protect\citeauthoryear{??}{Pee}{[n. d.]}]%
        {PeeringDB}
 \bibinfo{year}{[n. d.]}\natexlab{}.
\newblock \bibinfo{title}{PeeringDB}.
\newblock \bibinfo{howpublished}{\url{https://www.peeringdb.com/}}.   (\bibinfo{year}{[n. d.]}).
\newblock


\bibitem[\protect\citeauthoryear{??}{cab}{[n. d.]e}]%
        {cable-10-tril}
 \bibinfo{year}{[n. d.]}\natexlab{e}.
\newblock \bibinfo{title}{Protecting undersea cables must be made a national security priority}.
\newblock \bibinfo{howpublished}{\url{https://www.defensenews.com/opinion/commentary/2020/07/01/protecting-undersea-cables-must-be-made-a-national-security-priority/}}.   (\bibinfo{year}{[n. d.]}).
\newblock
\newblock
\shownote{(Accessed on 02/09/2023).}


\bibitem[\protect\citeauthoryear{??}{rip}{[n. d.]b}]%
        {ripe_ipmap}
 \bibinfo{year}{[n. d.]}\natexlab{b}.
\newblock \bibinfo{title}{{RIPE IPmap}}.
\newblock \bibinfo{howpublished}{\url{https://ipmap.ripe.net/}}.   (\bibinfo{year}{[n. d.]}).
\newblock


\bibitem[\protect\citeauthoryear{??}{rou}{[n. d.]}]%
        {routinator_whois}
 \bibinfo{year}{[n. d.]}\natexlab{}.
\newblock \bibinfo{title}{{Routinator}}.
\newblock \bibinfo{howpublished}{\url{https://rpki-validator.ripe.net/ui/}}.   (\bibinfo{year}{[n. d.]}).
\newblock


\bibitem[\protect\citeauthoryear{??}{anc}{[n. d.]a}]%
        {anchor-2}
 \bibinfo{year}{[n. d.]}\natexlab{a}.
\newblock \bibinfo{title}{{Ship's anchor caused cut in Internet cable}}.
\newblock \bibinfo{howpublished}{\url{https://www.nbcnews.com/id/wbna23068571}}.   (\bibinfo{year}{[n. d.]}).
\newblock
\newblock
\shownote{(Accessed on 02/11/2023).}


\bibitem[\protect\citeauthoryear{??}{anc}{[n. d.]b}]%
        {anchor-1}
 \bibinfo{year}{[n. d.]}\natexlab{b}.
\newblock \bibinfo{title}{{Ship's Anchor Slices Through Three Undersea Internet Cables}}.
\newblock \bibinfo{howpublished}{\url{https://www.popularmechanics.com/technology/infrastructure/a24064/anchor-cuts-undersea-internet-cables/}}.   (\bibinfo{year}{[n. d.]}).
\newblock
\newblock
\shownote{(Accessed on 02/11/2023).}


\bibitem[\protect\citeauthoryear{??}{sub}{[n. d.]b}]%
        {submarine_faq}
 \bibinfo{year}{[n. d.]}\natexlab{b}.
\newblock \bibinfo{title}{{Submarine Cable FAQs}}.
\newblock \bibinfo{howpublished}{\url{https://www2.telegeography.com/submarine-cable-faqs-frequently-asked-questions}}.   (\bibinfo{year}{[n. d.]}).
\newblock


\bibitem[\protect\citeauthoryear{??}{tel}{[n. d.]}]%
        {tele_website}
 \bibinfo{year}{[n. d.]}\natexlab{}.
\newblock \bibinfo{title}{Submarine Cable Map}.
\newblock \bibinfo{howpublished}{\url{https://www.submarinecablemap.com/}}.   (\bibinfo{year}{[n. d.]}).
\newblock


\bibitem[\protect\citeauthoryear{??}{cab}{[n. d.]f}]%
        {cable_failures_6}
 \bibinfo{year}{[n. d.]}\natexlab{f}.
\newblock \bibinfo{title}{{Svalbard Suffers Power Fault On Subsea Fiber Cable - Submarine Telecoms Forum}}.
\newblock \bibinfo{howpublished}{\url{https://subtelforum.com/svalbard-suffers-power-fault-on-subsea-fiber-cable/}}.   (\bibinfo{year}{[n. d.]}).
\newblock


\bibitem[\protect\citeauthoryear{??}{cai}{[n. d.]a}]%
        {caida_itdk}
 \bibinfo{year}{[n. d.]}\natexlab{a}.
\newblock \bibinfo{title}{{The CAIDA Macroscopic Internet Topology Data Kit - ITDK-2022-02}}.
\newblock \bibinfo{howpublished}{\url{https://www.caida.org/catalog/datasets/internet-topology-data-kit/}}.   (\bibinfo{year}{[n. d.]}).
\newblock


\bibitem[\protect\citeauthoryear{??}{cai}{[n. d.]b}]%
        {caida_24_probing}
 \bibinfo{year}{[n. d.]}\natexlab{b}.
\newblock \bibinfo{title}{{The CAIDA UCSD IPv4 Routed /24 Topology Dataset - 03-13-2022-03-23-2022}}.
\newblock \bibinfo{howpublished}{\url{https://www.caida.org/catalog/datasets/ipv4_routed_24_topology_dataset/}}.   (\bibinfo{year}{[n. d.]}).
\newblock


\bibitem[\protect\citeauthoryear{??}{cai}{[n. d.]c}]%
        {caida_48_probing}
 \bibinfo{year}{[n. d.]}\natexlab{c}.
\newblock \bibinfo{title}{{The CAIDA UCSD IPv6 Routed /48 Topology Dataset - 03-13-2022-03-23-2022}}.
\newblock \bibinfo{howpublished}{\url{https://www.caida.org/catalog/datasets/ipv6_routed_48_topology_dataset/}}.   (\bibinfo{year}{[n. d.]}).
\newblock


\bibitem[\protect\citeauthoryear{??}{rad}{[n. d.]}]%
        {radb_server}
 \bibinfo{year}{[n. d.]}\natexlab{}.
\newblock \bibinfo{title}{{The Internet Routing Registry - RADb}}.
\newblock \bibinfo{howpublished}{\url{https://www.radb.net/}}.   (\bibinfo{year}{[n. d.]}).
\newblock


\bibitem[\protect\citeauthoryear{??}{ipr}{[n. d.]}]%
        {ipregistry}
 \bibinfo{year}{[n. d.]}\natexlab{}.
\newblock \bibinfo{title}{{The Trusted Source for IP Address Data (geolocation and threat) - Ipregistry}}.
\newblock \bibinfo{howpublished}{\url{https://ipregistry.co/}}.   (\bibinfo{year}{[n. d.]}).
\newblock


\bibitem[\protect\citeauthoryear{??}{ton}{[n. d.]b}]%
        {tonga_cable_repair}
 \bibinfo{year}{[n. d.]}\natexlab{b}.
\newblock \bibinfo{title}{{Tonga Cable Successfully Repaired - Submarine Telecoms Forum}}.
\newblock \bibinfo{howpublished}{\url{https://subtelforum.com/tonga-cable-successfully-repaired/}}.   (\bibinfo{year}{[n. d.]}).
\newblock


\bibitem[\protect\citeauthoryear{??}{ear}{[n. d.]}]%
        {earthquake-3}
 \bibinfo{year}{[n. d.]}\natexlab{}.
\newblock \bibinfo{title}{{Tsunami, earthquake damage undersea fiber-optic cables in Japan | Laser Focus World}}.
\newblock \bibinfo{howpublished}{\url{https://www.laserfocusworld.com/fiber-optics/article/16562059/tsunami-earthquake-damage-undersea-fiberoptic-cables-in-japan}}.   (\bibinfo{year}{[n. d.]}).
\newblock
\newblock
\shownote{(Accessed on 02/11/2023).}


\bibitem[\protect\citeauthoryear{??}{typ}{[n. d.]a}]%
        {typhoon-1}
 \bibinfo{year}{[n. d.]}\natexlab{a}.
\newblock \bibinfo{title}{{Undersea Internet cables undergoing repairs after typhoon | Computerworld}}.
\newblock \bibinfo{howpublished}{\url{https://www.computerworld.com/article/2527025/undersea-internet-cables-undergoing-repairs-after-typhoon.html}}.   (\bibinfo{year}{[n. d.]}).
\newblock
\newblock
\shownote{(Accessed on 02/11/2023).}


\bibitem[\protect\citeauthoryear{??}{cab}{[n. d.]g}]%
        {cable_failures_4}
 \bibinfo{year}{[n. d.]}\natexlab{g}.
\newblock \bibinfo{title}{{Vietnam's AAG Repair Complete - SubTel Forum}}.
\newblock \bibinfo{howpublished}{\url{https://subtelforum.com/vietnams-aag-repair-complete/}}.   (\bibinfo{year}{[n. d.]}).
\newblock


\bibitem[\protect\citeauthoryear{??}{cab}{[n. d.]h}]%
        {cable_failures_2}
 \bibinfo{year}{[n. d.]}\natexlab{h}.
\newblock \bibinfo{title}{{Vietnam’s AAG Repair Complete - Submarine Telecoms Forum}}.
\newblock \bibinfo{howpublished}{\url{https://subtelforum.com/vietnams-aag-repair-complete/}}.   (\bibinfo{year}{[n. d.]}).
\newblock


\bibitem[\protect\citeauthoryear{??}{vod}{[n. d.]}]%
        {vodafone_map}
 \bibinfo{year}{[n. d.]}\natexlab{}.
\newblock \bibinfo{title}{Vodafone Global Infrastructure Map}.
\newblock \bibinfo{howpublished}{\url{http://globalnetworkmap.vodafone.com/}}.   (\bibinfo{year}{[n. d.]}).
\newblock


\bibitem[\protect\citeauthoryear{??}{typ}{[n. d.]b}]%
        {typhoon-2}
 \bibinfo{year}{[n. d.]}\natexlab{b}.
\newblock \bibinfo{title}{{Your slow Internet just got worse: Several undersea cables damaged after typhoons in Southeast Asia | Coconuts}}.
\newblock \bibinfo{howpublished}{\url{https://coconuts.co/manila/news/slow-internet-just-got-worse-several-undersea-cables-damaged-typhoons-southeast-asia/}}.   (\bibinfo{year}{[n. d.]}).
\newblock
\newblock
\shownote{(Accessed on 02/11/2023).}


\bibitem[\protect\citeauthoryear{??}{AS_}{2006}]%
        {AS_4}
 \bibinfo{year}{2006}\natexlab{}.
\newblock \showarticletitle{{Building an AS-topology model that captures route diversity}, author={M{\"u}hlbauer, Wolfgang and Feldmann, Anja and Maennel, Olaf and Roughan, Matthew and Uhlig, Steve}}.
\newblock \bibinfo{journal}{{\em ACM SIGCOMM Computer Communication Review\/}} \bibinfo{volume}{36}, \bibinfo{number}{4} (\bibinfo{year}{2006}), \bibinfo{pages}{195--206}.
\newblock


\bibitem[\protect\citeauthoryear{??}{AS_}{2010}]%
        {AS_1}
 \bibinfo{year}{2010}\natexlab{}.
\newblock \showarticletitle{{Towards an AS-to-organization Map}, author={Cai, Xue and Heidemann, John and Krishnamurthy, Balachander and Willinger, Walter}}. In \bibinfo{booktitle}{{\em Proceedings of the 10th ACM SIGCOMM conference on Internet measurement}}. \bibinfo{pages}{199--205}.
\newblock


\bibitem[\protect\citeauthoryear{Akgun and Gunes}{Akgun and Gunes}{2013}]%
        {subnet_3}
\bibfield{author}{\bibinfo{person}{Mehmet~Burak Akgun} {and} \bibinfo{person}{Mehmet~Hadi Gunes}.} \bibinfo{year}{2013}\natexlab{}.
\newblock \showarticletitle{{Bipartite internet topology at the subnet-level}}. In \bibinfo{booktitle}{{\em 2013 IEEE 2nd Network Science Workshop (NSW)}}. IEEE, \bibinfo{pages}{94--97}.
\newblock


\bibitem[\protect\citeauthoryear{Anderson, Salamatian, Bischof, Dainotti, and Barford}{Anderson et~al\mbox{.}}{2022}]%
        {igdb}
\bibfield{author}{\bibinfo{person}{Scott Anderson}, \bibinfo{person}{Loqman Salamatian}, \bibinfo{person}{Zachary~S Bischof}, \bibinfo{person}{Alberto Dainotti}, {and} \bibinfo{person}{Paul Barford}.} \bibinfo{year}{2022}\natexlab{}.
\newblock \showarticletitle{{iGDB: connecting the physical and logical layers of the internet}}. In \bibinfo{booktitle}{{\em Proceedings of the 22nd ACM Internet Measurement Conference}}. \bibinfo{pages}{433--448}.
\newblock


\bibitem[\protect\citeauthoryear{Bischof, Fontugne, and Bustamante}{Bischof et~al\mbox{.}}{2018}]%
        {submarine}
\bibfield{author}{\bibinfo{person}{Zachary~S Bischof}, \bibinfo{person}{Romain Fontugne}, {and} \bibinfo{person}{Fabi{\'a}n~E Bustamante}.} \bibinfo{year}{2018}\natexlab{}.
\newblock \showarticletitle{{Untangling the world-wide mesh of undersea cables}}. In \bibinfo{booktitle}{{\em Proceedings of the 17th ACM Workshop on Hot Topics in Networks}}. \bibinfo{pages}{78--84}.
\newblock


\bibitem[\protect\citeauthoryear{Bischof, Otto, and Bustamante}{Bischof et~al\mbox{.}}{2011}]%
        {earthquake_3}
\bibfield{author}{\bibinfo{person}{Zachary~S Bischof}, \bibinfo{person}{John~S Otto}, {and} \bibinfo{person}{Fabi{\'a}n~E Bustamante}.} \bibinfo{year}{2011}\natexlab{}.
\newblock \showarticletitle{{Distributed systems and natural disasters: BitTorrent as a global witness}}. In \bibinfo{booktitle}{{\em Proceedings of the Special Workshop on Internet and Disasters}}. \bibinfo{pages}{1--8}.
\newblock


\bibitem[\protect\citeauthoryear{Bozkurt, Aqeel, Bhattacherjee, Chandrasekaran, Godfrey, Laughlin, Maggs, and Singla}{Bozkurt et~al\mbox{.}}{2018}]%
        {mpls_city}
\bibfield{author}{\bibinfo{person}{Ilker~Nadi Bozkurt}, \bibinfo{person}{Waqar Aqeel}, \bibinfo{person}{Debopam Bhattacherjee}, \bibinfo{person}{Balakrishnan Chandrasekaran}, \bibinfo{person}{Philip~Brighten Godfrey}, \bibinfo{person}{Gregory Laughlin}, \bibinfo{person}{Bruce~M Maggs}, {and} \bibinfo{person}{Ankit Singla}.} \bibinfo{year}{2018}\natexlab{}.
\newblock \showarticletitle{{Dissecting Latency in the Internet's Fiber Infrastructure}}.
\newblock \bibinfo{journal}{{\em arXiv preprint arXiv:1811.10737\/}} (\bibinfo{year}{2018}).
\newblock


\bibitem[\protect\citeauthoryear{Bush, Maennel, Roughan, and Uhlig}{Bush et~al\mbox{.}}{2009}]%
        {AS_8}
\bibfield{author}{\bibinfo{person}{Randy Bush}, \bibinfo{person}{Olaf Maennel}, \bibinfo{person}{Matthew Roughan}, {and} \bibinfo{person}{Steve Uhlig}.} \bibinfo{year}{2009}\natexlab{}.
\newblock \showarticletitle{{Internet optometry: assessing the broken glasses in internet reachability}}. In \bibinfo{booktitle}{{\em Proceedings of the 9th ACM SIGCOMM conference on Internet measurement}}. \bibinfo{pages}{242--253}.
\newblock


\bibitem[\protect\citeauthoryear{Cho, Pelsser, Bush, and Won}{Cho et~al\mbox{.}}{2011}]%
        {earthquake_1}
\bibfield{author}{\bibinfo{person}{Kenjiro Cho}, \bibinfo{person}{Cristel Pelsser}, \bibinfo{person}{Randy Bush}, {and} \bibinfo{person}{Youngjoon Won}.} \bibinfo{year}{2011}\natexlab{}.
\newblock \showarticletitle{{The Japan Earthquake: the impact on traffic and routing observed by a local ISP}}. In \bibinfo{booktitle}{{\em Proceedings of the Special Workshop on Internet and Disasters}}. \bibinfo{pages}{1--8}.
\newblock


\bibitem[\protect\citeauthoryear{Donnet and Friedman}{Donnet and Friedman}{2007}]%
        {internet_mapping_survey_1}
\bibfield{author}{\bibinfo{person}{Benoit Donnet} {and} \bibinfo{person}{Timur Friedman}.} \bibinfo{year}{2007}\natexlab{}.
\newblock \showarticletitle{{Internet topology discovery: a survey}}.
\newblock \bibinfo{journal}{{\em IEEE Communications Surveys \& Tutorials\/}} \bibinfo{volume}{9}, \bibinfo{number}{4} (\bibinfo{year}{2007}), \bibinfo{pages}{56--69}.
\newblock


\bibitem[\protect\citeauthoryear{Donnet, Luckie, M{\'e}rindol, and Pansiot}{Donnet et~al\mbox{.}}{2012}]%
        {mpls_issues}
\bibfield{author}{\bibinfo{person}{Benoit Donnet}, \bibinfo{person}{Matthew Luckie}, \bibinfo{person}{Pascal M{\'e}rindol}, {and} \bibinfo{person}{Jean-Jacques Pansiot}.} \bibinfo{year}{2012}\natexlab{}.
\newblock \showarticletitle{{Revealing MPLS tunnels obscured from traceroute}}.
\newblock \bibinfo{journal}{{\em ACM SIGCOMM Computer Communication Review\/}} \bibinfo{volume}{42}, \bibinfo{number}{2} (\bibinfo{year}{2012}), \bibinfo{pages}{87--93}.
\newblock


\bibitem[\protect\citeauthoryear{Durairajan, Barford, Sommers, and Willinger}{Durairajan et~al\mbox{.}}{2015}]%
        {physical_1}
\bibfield{author}{\bibinfo{person}{Ramakrishnan Durairajan}, \bibinfo{person}{Paul Barford}, \bibinfo{person}{Joel Sommers}, {and} \bibinfo{person}{Walter Willinger}.} \bibinfo{year}{2015}\natexlab{}.
\newblock \showarticletitle{{InterTubes: A study of the US long-haul fiber-optic infrastructure}}. In \bibinfo{booktitle}{{\em Proceedings of the 2015 ACM Conference on Special Interest Group on Data Communication}}. \bibinfo{pages}{565--578}.
\newblock


\bibitem[\protect\citeauthoryear{Durairajan, Ghosh, Tang, Barford, and Eriksson}{Durairajan et~al\mbox{.}}{2013}]%
        {pop_2}
\bibfield{author}{\bibinfo{person}{Ramakrishnan Durairajan}, \bibinfo{person}{Subhadip Ghosh}, \bibinfo{person}{Xin Tang}, \bibinfo{person}{Paul Barford}, {and} \bibinfo{person}{Brian Eriksson}.} \bibinfo{year}{2013}\natexlab{}.
\newblock \showarticletitle{{Internet atlas: a geographic database of the internet}}. In \bibinfo{booktitle}{{\em Proceedings of the 5th ACM workshop on HotPlanet}}. \bibinfo{pages}{15--20}.
\newblock


\bibitem[\protect\citeauthoryear{Ester, Kriegel, Sander, Xu, et~al\mbox{.}}{Ester et~al\mbox{.}}{1996}]%
        {dbscan}
\bibfield{author}{\bibinfo{person}{Martin Ester}, \bibinfo{person}{Hans-Peter Kriegel}, \bibinfo{person}{J{\"o}rg Sander}, \bibinfo{person}{Xiaowei Xu}, {et~al\mbox{.}}} \bibinfo{year}{1996}\natexlab{}.
\newblock \showarticletitle{{A density-based algorithm for discovering clusters in large spatial databases with noise.}}. In \bibinfo{booktitle}{{\em kdd}}, Vol.~\bibinfo{volume}{96}. \bibinfo{pages}{226--231}.
\newblock


\bibitem[\protect\citeauthoryear{Feldman and Shavitt}{Feldman and Shavitt}{2008}]%
        {pop_5}
\bibfield{author}{\bibinfo{person}{Dima Feldman} {and} \bibinfo{person}{Yuval Shavitt}.} \bibinfo{year}{2008}\natexlab{}.
\newblock \showarticletitle{{Automatic large scale generation of internet pop level maps}}. In \bibinfo{booktitle}{{\em IEEE GLOBECOM 2008-2008 IEEE Global Telecommunications Conference}}. IEEE, \bibinfo{pages}{1--6}.
\newblock


\bibitem[\protect\citeauthoryear{Forum}{Forum}{[n. d.]}]%
        {earthquake-2}
\bibfield{author}{\bibinfo{person}{SubTel Forum}.} \bibinfo{year}{[n. d.]}\natexlab{}.
\newblock \bibinfo{title}{{PNG Earthquake Causing Multiple Cable Disruptions}}.
\newblock \bibinfo{howpublished}{\url{https://subtelforum.com/png-earthquake-causing-multiple-cable-disruptions/}}.   (\bibinfo{year}{[n. d.]}).
\newblock
\newblock
\shownote{(Accessed on 02/11/2023).}


\bibitem[\protect\citeauthoryear{Gharaibeh, Shah, Huffaker, Zhang, Ensafi, and Papadopoulos}{Gharaibeh et~al\mbox{.}}{2017}]%
        {geolocation_problem}
\bibfield{author}{\bibinfo{person}{Manaf Gharaibeh}, \bibinfo{person}{Anant Shah}, \bibinfo{person}{Bradley Huffaker}, \bibinfo{person}{Han Zhang}, \bibinfo{person}{Roya Ensafi}, {and} \bibinfo{person}{Christos Papadopoulos}.} \bibinfo{year}{2017}\natexlab{}.
\newblock \showarticletitle{{A look at router geolocation in public and commercial databases}}. In \bibinfo{booktitle}{{\em Proceedings of the 2017 Internet Measurement Conference}}. \bibinfo{pages}{463--469}.
\newblock


\bibitem[\protect\citeauthoryear{Govindan and Tangmunarunkit}{Govindan and Tangmunarunkit}{2000}]%
        {router_paper_1}
\bibfield{author}{\bibinfo{person}{Ramesh Govindan} {and} \bibinfo{person}{Hongsuda Tangmunarunkit}.} \bibinfo{year}{2000}\natexlab{}.
\newblock \showarticletitle{{Heuristics for Internet map discovery}}. In \bibinfo{booktitle}{{\em Proceedings IEEE INFOCOM 2000. Conference on Computer Communications. Nineteenth Annual Joint Conference of the IEEE Computer and Communications Societies (Cat. No. 00CH37064)}}, Vol.~\bibinfo{volume}{3}. IEEE, \bibinfo{pages}{1371--1380}.
\newblock


\bibitem[\protect\citeauthoryear{Grailet, Tarissan, and Donnet}{Grailet et~al\mbox{.}}{2016}]%
        {subnet_4}
\bibfield{author}{\bibinfo{person}{Jean-Francois Grailet}, \bibinfo{person}{Fabien Tarissan}, {and} \bibinfo{person}{Benoit Donnet}.} \bibinfo{year}{2016}\natexlab{}.
\newblock \showarticletitle{{TreeNET: Discovering and connecting subnets}}. In \bibinfo{booktitle}{{\em 8th International Workshop on Traffic Monitoring and Analysis (TMA)}}.
\newblock


\bibitem[\protect\citeauthoryear{Gunes and Sarac}{Gunes and Sarac}{2009}]%
        {router_paper_14}
\bibfield{author}{\bibinfo{person}{Mehmet~H Gunes} {and} \bibinfo{person}{Kamil Sarac}.} \bibinfo{year}{2009}\natexlab{}.
\newblock \showarticletitle{{Resolving IP aliases in building traceroute-based Internet maps}}.
\newblock \bibinfo{journal}{{\em IEEE/ACM Transactions on Networking\/}} \bibinfo{volume}{17}, \bibinfo{number}{6} (\bibinfo{year}{2009}), \bibinfo{pages}{1738--1751}.
\newblock


\bibitem[\protect\citeauthoryear{Jyothi}{Jyothi}{2021}]%
        {solarstorm-cable}
\bibfield{author}{\bibinfo{person}{Sangeetha~Abdu Jyothi}.} \bibinfo{year}{2021}\natexlab{}.
\newblock \showarticletitle{{Solar Superstorms: Planning for an Internet Apocalypse}}. In \bibinfo{booktitle}{{\em Proceedings of the 2021 ACM SIGCOMM Conference}}. \bibinfo{pages}{692--704}.
\newblock


\bibitem[\protect\citeauthoryear{Keys}{Keys}{2010}]%
        {router_paper_2}
\bibfield{author}{\bibinfo{person}{Ken Keys}.} \bibinfo{year}{2010}\natexlab{}.
\newblock \showarticletitle{{Internet-scale IP alias resolution techniques}}.
\newblock \bibinfo{journal}{{\em ACM SIGCOMM Computer Communication Review\/}} \bibinfo{volume}{40}, \bibinfo{number}{1} (\bibinfo{year}{2010}), \bibinfo{pages}{50--55}.
\newblock


\bibitem[\protect\citeauthoryear{Keys}{Keys}{2020}]%
        {router_paper_8}
\bibfield{author}{\bibinfo{person}{Ken Keys}.} \bibinfo{year}{2020}\natexlab{}.
\newblock \showarticletitle{{iffinder,” a tool for mapping interfaces to routers}}.
\newblock \bibinfo{journal}{{\em See http://www. caida. org/tools/measurement/iffinder\/}} (\bibinfo{year}{2020}).
\newblock


\bibitem[\protect\citeauthoryear{Keys, Hyun, Luckie, and Claffy}{Keys et~al\mbox{.}}{2012}]%
        {router_paper_6}
\bibfield{author}{\bibinfo{person}{Ken Keys}, \bibinfo{person}{Young Hyun}, \bibinfo{person}{Matthew Luckie}, {and} \bibinfo{person}{Kim Claffy}.} \bibinfo{year}{2012}\natexlab{}.
\newblock \showarticletitle{{Internet-scale IPv4 alias resolution with MIDAR}}.
\newblock \bibinfo{journal}{{\em IEEE/ACM Transactions on Networking\/}} \bibinfo{volume}{21}, \bibinfo{number}{2} (\bibinfo{year}{2012}), \bibinfo{pages}{383--399}.
\newblock


\bibitem[\protect\citeauthoryear{Knight, Nguyen, Falkner, Bowden, and Roughan}{Knight et~al\mbox{.}}{2011}]%
        {pop_3}
\bibfield{author}{\bibinfo{person}{Simon Knight}, \bibinfo{person}{Hung~X Nguyen}, \bibinfo{person}{Nickolas Falkner}, \bibinfo{person}{Rhys Bowden}, {and} \bibinfo{person}{Matthew Roughan}.} \bibinfo{year}{2011}\natexlab{}.
\newblock \showarticletitle{{The internet topology zoo}}.
\newblock \bibinfo{journal}{{\em IEEE Journal on Selected Areas in Communications\/}} \bibinfo{volume}{29}, \bibinfo{number}{9} (\bibinfo{year}{2011}), \bibinfo{pages}{1765--1775}.
\newblock


\bibitem[\protect\citeauthoryear{Liu, Bischof, Madan, Chan, and Bustamante}{Liu et~al\mbox{.}}{2020}]%
        {submarine_drivability}
\bibfield{author}{\bibinfo{person}{Shucheng Liu}, \bibinfo{person}{Zachary~S Bischof}, \bibinfo{person}{Ishaan Madan}, \bibinfo{person}{Peter~K Chan}, {and} \bibinfo{person}{Fabi{\'a}n~E Bustamante}.} \bibinfo{year}{2020}\natexlab{}.
\newblock \showarticletitle{{Out of sight, not out of mind: A user-view on the criticality of the submarine cable network}}. In \bibinfo{booktitle}{{\em Proceedings of the ACM Internet Measurement Conference}}. \bibinfo{pages}{194--200}.
\newblock


\bibitem[\protect\citeauthoryear{Liu, Moore, Gray, and Cardie}{Liu et~al\mbox{.}}{2006}]%
        {balltree}
\bibfield{author}{\bibinfo{person}{Ting Liu}, \bibinfo{person}{Andrew~W Moore}, \bibinfo{person}{Alexander Gray}, {and} \bibinfo{person}{Claire Cardie}.} \bibinfo{year}{2006}\natexlab{}.
\newblock \showarticletitle{{New algorithms for efficient high-dimensional nonparametric classification.}}
\newblock \bibinfo{journal}{{\em Journal of Machine Learning Research\/}} \bibinfo{volume}{7}, \bibinfo{number}{6} (\bibinfo{year}{2006}).
\newblock


\bibitem[\protect\citeauthoryear{Luckie, Dhamdhere, Huffaker, Clark, and Claffy}{Luckie et~al\mbox{.}}{2016}]%
        {router_paper_10}
\bibfield{author}{\bibinfo{person}{Matthew Luckie}, \bibinfo{person}{Amogh Dhamdhere}, \bibinfo{person}{Bradley Huffaker}, \bibinfo{person}{David Clark}, {and} \bibinfo{person}{KC Claffy}.} \bibinfo{year}{2016}\natexlab{}.
\newblock \showarticletitle{{Bdrmap: Inference of borders between IP networks}}. In \bibinfo{booktitle}{{\em Proceedings of the 2016 Internet Measurement Conference}}. \bibinfo{pages}{381--396}.
\newblock


\bibitem[\protect\citeauthoryear{Madhyastha, Isdal, Piatek, Dixon, Anderson, Krishnamurthy, and Venkataramani}{Madhyastha et~al\mbox{.}}{2006}]%
        {pop_4}
\bibfield{author}{\bibinfo{person}{Harsha~V Madhyastha}, \bibinfo{person}{Tomas Isdal}, \bibinfo{person}{Michael Piatek}, \bibinfo{person}{Colin Dixon}, \bibinfo{person}{Thomas Anderson}, \bibinfo{person}{Arvind Krishnamurthy}, {and} \bibinfo{person}{Arun Venkataramani}.} \bibinfo{year}{2006}\natexlab{}.
\newblock \showarticletitle{{iPlane: An information plane for distributed services}}. In \bibinfo{booktitle}{{\em Proceedings of the 7th symposium on Operating systems design and implementation}}. \bibinfo{pages}{367--380}.
\newblock


\bibitem[\protect\citeauthoryear{Mahadevan, Krioukov, Fomenkov, Dimitropoulos, Claffy, and Vahdat}{Mahadevan et~al\mbox{.}}{2006}]%
        {AS_7}
\bibfield{author}{\bibinfo{person}{Priya Mahadevan}, \bibinfo{person}{Dmitri Krioukov}, \bibinfo{person}{Marina Fomenkov}, \bibinfo{person}{Xenofontas Dimitropoulos}, \bibinfo{person}{KC Claffy}, {and} \bibinfo{person}{Amin Vahdat}.} \bibinfo{year}{2006}\natexlab{}.
\newblock \showarticletitle{{The Internet AS-level topology: three data sources and one definitive metric}}.
\newblock \bibinfo{journal}{{\em ACM SIGCOMM Computer Communication Review\/}} \bibinfo{volume}{36}, \bibinfo{number}{1} (\bibinfo{year}{2006}), \bibinfo{pages}{17--26}.
\newblock


\bibitem[\protect\citeauthoryear{Mani, Hall, Durairajan, and Barford}{Mani et~al\mbox{.}}{2020}]%
        {physical_3}
\bibfield{author}{\bibinfo{person}{Sathiya~Kumaran Mani}, \bibinfo{person}{Matthew~Nance Hall}, \bibinfo{person}{Ramakrishnan Durairajan}, {and} \bibinfo{person}{Paul Barford}.} \bibinfo{year}{2020}\natexlab{}.
\newblock \showarticletitle{{Characteristics of metro fiber deployments in the us}}. In \bibinfo{booktitle}{{\em Proceedings of the Network Traffic Measurement and Analysis Conference}}.
\newblock


\bibitem[\protect\citeauthoryear{Mao, Jamjoom, Tao, and Smith}{Mao et~al\mbox{.}}{2007}]%
        {physical_2}
\bibfield{author}{\bibinfo{person}{Yun Mao}, \bibinfo{person}{Hani Jamjoom}, \bibinfo{person}{Shu Tao}, {and} \bibinfo{person}{Jonathan~M Smith}.} \bibinfo{year}{2007}\natexlab{}.
\newblock \showarticletitle{{Networkmd: topology inference and failure diagnosis in the last mile}}. In \bibinfo{booktitle}{{\em Proceedings of the 7th ACM SIGCOMM conference on Internet measurement}}. \bibinfo{pages}{189--202}.
\newblock


\bibitem[\protect\citeauthoryear{Mao, Rexford, Wang, and Katz}{Mao et~al\mbox{.}}{2003}]%
        {AS_9}
\bibfield{author}{\bibinfo{person}{Zhuoqing~Morley Mao}, \bibinfo{person}{Jennifer Rexford}, \bibinfo{person}{Jia Wang}, {and} \bibinfo{person}{Randy~H Katz}.} \bibinfo{year}{2003}\natexlab{}.
\newblock \showarticletitle{{Towards an accurate AS-level traceroute tool}}. In \bibinfo{booktitle}{{\em Proceedings of the 2003 conference on Applications, technologies, architectures, and protocols for computer communications}}. \bibinfo{pages}{365--378}.
\newblock


\bibitem[\protect\citeauthoryear{Marder, Luckie, Dhamdhere, Huffaker, Claffy, and Smith}{Marder et~al\mbox{.}}{2018}]%
        {router_paper_4}
\bibfield{author}{\bibinfo{person}{Alexander Marder}, \bibinfo{person}{Matthew Luckie}, \bibinfo{person}{Amogh Dhamdhere}, \bibinfo{person}{Bradley Huffaker}, \bibinfo{person}{KC Claffy}, {and} \bibinfo{person}{Jonathan~M Smith}.} \bibinfo{year}{2018}\natexlab{}.
\newblock \showarticletitle{{Pushing the boundaries with bdrmapIT: mapping router ownership at internet scale}}. In \bibinfo{booktitle}{{\em Proceedings of the Internet Measurement Conference 2018}}. \bibinfo{pages}{56--69}.
\newblock


\bibitem[\protect\citeauthoryear{Marder and Smith}{Marder and Smith}{2016}]%
        {router_paper_11}
\bibfield{author}{\bibinfo{person}{Alexander Marder} {and} \bibinfo{person}{Jonathan~M Smith}.} \bibinfo{year}{2016}\natexlab{}.
\newblock \showarticletitle{{MAP-IT: Multipass accurate passive inferences from traceroute}}. In \bibinfo{booktitle}{{\em Proceedings of the 2016 Internet Measurement Conference}}. \bibinfo{pages}{397--411}.
\newblock


\bibitem[\protect\citeauthoryear{Motamedi, Rejaie, and Willinger}{Motamedi et~al\mbox{.}}{2014}]%
        {internet_mapping_survey_2}
\bibfield{author}{\bibinfo{person}{Reza Motamedi}, \bibinfo{person}{Reza Rejaie}, {and} \bibinfo{person}{Walter Willinger}.} \bibinfo{year}{2014}\natexlab{}.
\newblock \showarticletitle{{A survey of techniques for internet topology discovery}}.
\newblock \bibinfo{journal}{{\em IEEE Communications Surveys \& Tutorials\/}} \bibinfo{volume}{17}, \bibinfo{number}{2} (\bibinfo{year}{2014}), \bibinfo{pages}{1044--1065}.
\newblock


\bibitem[\protect\citeauthoryear{Oliveira, Pei, Willinger, Zhang, and Zhang}{Oliveira et~al\mbox{.}}{2008}]%
        {AS_2}
\bibfield{author}{\bibinfo{person}{Ricardo~V Oliveira}, \bibinfo{person}{Dan Pei}, \bibinfo{person}{Walter Willinger}, \bibinfo{person}{Beichuan Zhang}, {and} \bibinfo{person}{Lixia Zhang}.} \bibinfo{year}{2008}\natexlab{}.
\newblock \showarticletitle{{In search of the elusive ground truth: the Internet's AS-level connectivity structure}}.
\newblock \bibinfo{journal}{{\em ACM SIGMETRICS Performance Evaluation Review\/}} \bibinfo{volume}{36}, \bibinfo{number}{1} (\bibinfo{year}{2008}), \bibinfo{pages}{217--228}.
\newblock


\bibitem[\protect\citeauthoryear{Pansiot, M{\'e}rindol, Donnet, and Bonaventure}{Pansiot et~al\mbox{.}}{2010}]%
        {router_recursive_2}
\bibfield{author}{\bibinfo{person}{Jean-Jacques Pansiot}, \bibinfo{person}{Pascal M{\'e}rindol}, \bibinfo{person}{Benoit Donnet}, {and} \bibinfo{person}{Olivier Bonaventure}.} \bibinfo{year}{2010}\natexlab{}.
\newblock \showarticletitle{{Extracting intra-domain topology from mrinfo probing}}. In \bibinfo{booktitle}{{\em International Conference on Passive and Active Network Measurement}}. Springer, \bibinfo{pages}{81--90}.
\newblock


\bibitem[\protect\citeauthoryear{Roughan, Willinger, Maennel, Perouli, and Bush}{Roughan et~al\mbox{.}}{2011}]%
        {AS_3}
\bibfield{author}{\bibinfo{person}{Matthew Roughan}, \bibinfo{person}{Walter Willinger}, \bibinfo{person}{Olaf Maennel}, \bibinfo{person}{Debbie Perouli}, {and} \bibinfo{person}{Randy Bush}.} \bibinfo{year}{2011}\natexlab{}.
\newblock \showarticletitle{{10 lessons from 10 years of measuring and modeling the internet's autonomous systems}}.
\newblock \bibinfo{journal}{{\em IEEE Journal on Selected Areas in Communications\/}} \bibinfo{volume}{29}, \bibinfo{number}{9} (\bibinfo{year}{2011}), \bibinfo{pages}{1810--1821}.
\newblock


\bibitem[\protect\citeauthoryear{Shavitt and Zilberman}{Shavitt and Zilberman}{2010}]%
        {pop_7}
\bibfield{author}{\bibinfo{person}{Yuval Shavitt} {and} \bibinfo{person}{Noa Zilberman}.} \bibinfo{year}{2010}\natexlab{}.
\newblock \showarticletitle{{A structural approach for PoP geo-location}}. In \bibinfo{booktitle}{{\em 2010 INFOCOM IEEE Conference on Computer Communications Workshops}}. IEEE, \bibinfo{pages}{1--6}.
\newblock


\bibitem[\protect\citeauthoryear{Sinnott}{Sinnott}{1984}]%
        {haversine}
\bibfield{author}{\bibinfo{person}{Roger~W Sinnott}.} \bibinfo{year}{1984}\natexlab{}.
\newblock \showarticletitle{{Virtues of the Haversine}}.
\newblock \bibinfo{journal}{{\em Sky and telescope\/}} \bibinfo{volume}{68}, \bibinfo{number}{2} (\bibinfo{year}{1984}), \bibinfo{pages}{158}.
\newblock


\bibitem[\protect\citeauthoryear{Strasser, K{\"o}lling, Ferreira, Fink, Fujiwara, Henkel, Ikehara, Kanamatsu, Kawamura, Kodaira, et~al\mbox{.}}{Strasser et~al\mbox{.}}{2013}]%
        {earthquake_2}
\bibfield{author}{\bibinfo{person}{Michi Strasser}, \bibinfo{person}{Martin K{\"o}lling}, \bibinfo{person}{C~dos~Santos Ferreira}, \bibinfo{person}{Hiske~G Fink}, \bibinfo{person}{Toshyia Fujiwara}, \bibinfo{person}{Susann Henkel}, \bibinfo{person}{Ken Ikehara}, \bibinfo{person}{Toshyia Kanamatsu}, \bibinfo{person}{Kiichiro Kawamura}, \bibinfo{person}{Suichi Kodaira}, {et~al\mbox{.}}} \bibinfo{year}{2013}\natexlab{}.
\newblock \showarticletitle{{A slump in the trench: Tracking the impact of the 2011 Tohoku-Oki earthquake}}.
\newblock \bibinfo{journal}{{\em Geology\/}} \bibinfo{volume}{41}, \bibinfo{number}{8} (\bibinfo{year}{2013}), \bibinfo{pages}{935--938}.
\newblock


\bibitem[\protect\citeauthoryear{Tian, Dey, Liu, and Ross}{Tian et~al\mbox{.}}{2012}]%
        {pop_6}
\bibfield{author}{\bibinfo{person}{Ye Tian}, \bibinfo{person}{Ratan Dey}, \bibinfo{person}{Yong Liu}, {and} \bibinfo{person}{Keith~W Ross}.} \bibinfo{year}{2012}\natexlab{}.
\newblock \showarticletitle{{China's internet: Topology mapping and geolocating}}. In \bibinfo{booktitle}{{\em 2012 Proceedings IEEE INFOCOM}}. IEEE, \bibinfo{pages}{2531--2535}.
\newblock


\bibitem[\protect\citeauthoryear{Tozal and Sarac}{Tozal and Sarac}{2010}]%
        {subnet_2}
\bibfield{author}{\bibinfo{person}{M~Engin Tozal} {and} \bibinfo{person}{Kamil Sarac}.} \bibinfo{year}{2010}\natexlab{}.
\newblock \showarticletitle{{Tracenet: an internet topology data collector}}. In \bibinfo{booktitle}{{\em Proceedings of the 10th ACM SIGCOMM conference on Internet measurement}}. \bibinfo{pages}{356--368}.
\newblock


\bibitem[\protect\citeauthoryear{Tozal and Sarac}{Tozal and Sarac}{2011}]%
        {subnet_1}
\bibfield{author}{\bibinfo{person}{M~Engin Tozal} {and} \bibinfo{person}{Kamil Sarac}.} \bibinfo{year}{2011}\natexlab{}.
\newblock \showarticletitle{{Subnet level network topology mapping}}. In \bibinfo{booktitle}{{\em 30th IEEE International Performance Computing and Communications Conference}}. IEEE, \bibinfo{pages}{1--8}.
\newblock


\bibitem[\protect\citeauthoryear{Willinger and Roughan}{Willinger and Roughan}{2013}]%
        {router_hot}
\bibfield{author}{\bibinfo{person}{Walter Willinger} {and} \bibinfo{person}{Matthew Roughan}.} \bibinfo{year}{2013}\natexlab{}.
\newblock \showarticletitle{{Internet topology research redux}}.
\newblock \bibinfo{journal}{{\em ACM SIGCOMM eBook: Recent Advances in Networking\/}} (\bibinfo{year}{2013}).
\newblock


\bibitem[\protect\citeauthoryear{World}{World}{[n. d.]}]%
        {earthquake-1}
\bibfield{author}{\bibinfo{person}{Network World}.} \bibinfo{year}{[n. d.]}\natexlab{}.
\newblock \bibinfo{title}{{Taiwan earthquake damages undersea Internet cables}}.
\newblock \bibinfo{howpublished}{\url{https://www.networkworld.com/article/2203714/taiwan-earthquake-damages-undersea-internet-cables.html}}.   (\bibinfo{year}{[n. d.]}).
\newblock
\newblock
\shownote{(Accessed on 02/11/2023).}


\bibitem[\protect\citeauthoryear{Yoshida, Kikuchi, Yamamoto, Fujii, Nagami, Nakagawa, and Esaki}{Yoshida et~al\mbox{.}}{2009}]%
        {pop_1}
\bibfield{author}{\bibinfo{person}{Kaoru Yoshida}, \bibinfo{person}{Yutaka Kikuchi}, \bibinfo{person}{Masateru Yamamoto}, \bibinfo{person}{Yoriko Fujii}, \bibinfo{person}{Ken’ichi Nagami}, \bibinfo{person}{Ikuo Nakagawa}, {and} \bibinfo{person}{Hiroshi Esaki}.} \bibinfo{year}{2009}\natexlab{}.
\newblock \showarticletitle{{Inferring PoP-level ISP topology through end-to-end delay measurement}}. In \bibinfo{booktitle}{{\em International Conference on Passive and Active Network Measurement}}. Springer, \bibinfo{pages}{35--44}.
\newblock


\end{thebibliography}
